\pdfoutput=1
\def\tr{1} \def\ec{0} \def\old{0}
\ifnum\tr=1
\documentclass[11pt]{article}
\usepackage{geometry}                
\fi

\ifnum\ec=1
\documentclass{sig-alternate}
\fi

\usepackage{subfigure}
\usepackage{graphicx}
\usepackage{amssymb}
\usepackage{amsmath}
\usepackage{url}
\usepackage{epstopdf}
\usepackage{multirow}
\usepackage{cite}
\usepackage{pgf,pgfarrows,pgfnodes}
\newcommand{\skipthis}[1]{}

\ifnum\tr=1
\title{Information Asymmetries in Pay-Per-Bid Auctions\\
How Swoopo Makes Bank
}

\author{
John W. Byers
\thanks{John W. Byers is supported in part by NSF grant CNS-0520166. Computer Science Dept., Boston University, Boston, MA 02215 \& Adverplex Inc., Cambridge, MA 02139.} \\
       Boston University\\
       \texttt{byers@cs.bu.edu}
\and
Michael Mitzenmacher
\thanks{Supported in part by NSF grants CCF-0915922, CNS-0721491, and CCF-0634923, and in part by research grants from Yahoo!, Google, and Cisco Systems.  School of Engineering and Applied Sciences, Harvard University, Cambridge, MA 02138.}\\
       Harvard University\\
       \texttt{michaelm@eecs.harvard.edu}
\and
Georgios Zervas
\thanks{Computer Science Dept., Boston University, Boston, MA 02215 \& Adverplex Inc., Cambridge, MA 02139.} \\
       Boston University\\
       \texttt{zg@bu.edu}
}

\begin{document}
\maketitle

\fi

\ifnum\ec=1

\begin{document}
\conferenceinfo{Submitted to EC 2010,}{Cambridge, MA, 2010}
\title{Information Asymmetries in Pay-Per-Bid Auctions}
\conferenceinfo{EC'10,} {June 7--11, 2010, Cambridge, Massachusetts, USA.} 
\CopyrightYear{2010}
\crdata{978-1-60558-822-3/10/06}
\clubpenalty=10000
\widowpenalty = 10000 

\numberofauthors{3} 
\author{
\alignauthor John W. Byers\titlenote{E-mail: \{byers, zg\}@cs.bu.edu.  Supported in part by Adverplex, Inc. and by NSF grant 
  CNS-0520166.}\\
       \affaddr{Computer Science Dept.}\\
       \affaddr{Boston University}\\
\alignauthor Michael Mitzenmacher\titlenote{E-mail:  michaelm@eecs.harvard.edu.  Supported in part by NSF grants CCF-0915922, CNS-0721491, and CCF-0634923, and in part by research grants from Yahoo!, Google, and Cisco Systems.}\\
       \affaddr{School of Eng. Appl. Sci.}\\
       \affaddr{Harvard University}\\
 \alignauthor Georgios Zervas\raisebox{8pt}{$\ast$}\\
       \affaddr{Computer Science Dept.}\\
       \affaddr{Boston University}\\
}

\maketitle

\fi

\begin{abstract}
\ifnum\tr=1
Innovative auction methods can be exploited to increase profits, 
  with Shubik's famous ``dollar auction'' \cite{shubik1971} perhaps being
  the most widely known example.
Recently, some mainstream e-commerce web sites have apparently achieved the 
  same end on a much broader scale, by using ``pay-per-bid'' auctions to 
  sell items, from video games to bars of gold.
\fi
\ifnum\ec=1
Recently, some mainstream e-commerce web sites have 
  begun using ``pay-per-bid'' auctions to 
  sell items, from video games to bars of gold.
\fi
In these auctions, bidders incur a cost for placing each bid in addition to (or 
  sometimes in lieu of) the winner's final purchase cost.  
Thus even when a winner's purchase cost is a small fraction of the item's 
  intrinsic value, the auctioneer can still profit handsomely from the bid fees.
Our work provides novel analyses for these auctions, based on both modeling
  and datasets derived from auctions at Swoopo.com, the leading pay-per-bid auction site.
\ifnum\tr=1
While previous modeling work predicts profit-free equilibria, we analyze the 
  impact of {\em information asymmetry} broadly, as well as Swoopo features such 
  as bidpacks and the Swoop It Now option specifically, to quantify the
  effects of imperfect information in these auctions.
We find that even small asymmetries across players (cheaper bids, better estimates
  of other players' intent, different valuations of items, committed players 
  willing to play ``chicken'') can increase the auction duration well beyond that 
  predicted by previous work and thus skew the auctioneer's profit disproportionately.  
Finally, we discuss our findings in the context of a dataset of
  thousands of live auctions we observed on Swoopo, which enables us
  also to examine behavioral factors, such as the power of aggressive
  bidding.
\fi
\ifnum\ec=1
While previous modeling work predicts profit-free equilibria, we analyze the 
  impact of {\em information asymmetry} broadly, as well as Swoopo features such 
  as bidpacks and the Swoop It Now option specifically.
We find that even small asymmetries across players (cheaper bids, better estimates
  of other players' intent, different valuations of items, committed players 
  willing to play ``chicken'') can increase the auction duration significantly
   and thus skew the auctioneer's profit disproportionately.  
We discuss our findings in the context of a dataset of
  thousands of live auctions we observed on Swoopo, which enables us
  also to examine behavioral factors, such as the power of aggressive
  bidding.
\fi
Ultimately, our findings show that even with fully rational players, if
  players overlook or are unaware any of these factors, the result is 
  outsized profits for pay-per-bid auctioneers.
\end{abstract}

\ifnum\tr=1
\pagebreak
\tableofcontents
\pagebreak
\fi

\ifnum\ec=1


\fi

\ifnum\ec=1
\vspace{-2 mm}
\fi
\section{Introduction}

One of the more interesting commercial web sites to appear recently
from the standpoint of computational economics is Swoopo.  Swoopo runs
an auction website, using a nontraditional ``pay-per-bid'' auction
format.  Although we provide a more formal description later, the
basic framework is easy to describe.  As with standard eBay auctions,
pay-per-bid auctions for items begin at a reserve price (generally 0),
and have an associated countdown clock.  When a player places a bid,
the current auction price is incremented by a fixed amount, and some
additional time is added to the clock.  When the clock expires, the
last bidder must purchase the item at the final auction price.  The
\ifnum\tr=1
pay-per-bid enhancement is that each time a player increments the
price and becomes the current leader of the auction, they must pay a
bid fee.  At Swoopo.com, the price increment typically ranges from 
1 cent to 24 cents, and placing a bid typically costs 60 cents.
An important variation is a fixed-price auction, where the winner
obtains the right to buy the item at a fixed price $p$.  When $p = 0$,
such an auction is referred to as a 100\% off auction; in this case
Swoopo derives all of its revenue from the bids.
\fi
\ifnum\ec=1
pay-per-bid twist is that each time a player increments the
price and becomes the current leader of the auction, they must pay a
bid fee.  On Swoopo, the typical bid fee is 60 cents and the price
increment ranges from 1 cent to 24 cents.
\fi

While there are other web sites using similar auctions, Swoopo
has become the leader in this area, and recently has inspired multiple
papers that attempt to analyze the characteristics of the Swoopo
auction \cite{augenblick2009, hinnosaar2009, platt2009}.  These models 
share the same basic framework, based on assuming players decide whether or not 
to bid in a risk-neutral fashion, which we explain in detail in
Section~\ref{sec:model}.  Some of these papers then go further, and attempt to
justify their model by analyzing data from monitoring Swoopo auctions.

\begin{figure}[t]
\centering
\ifnum\tr=1
   \includegraphics{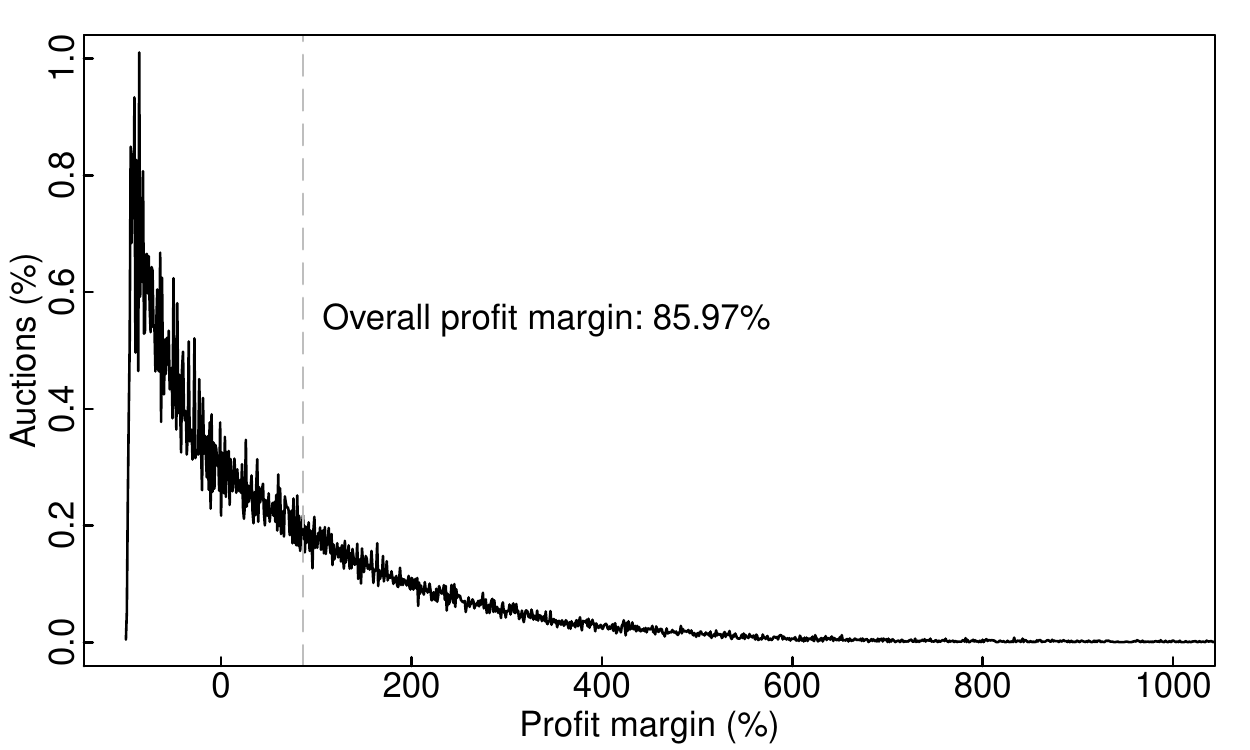} 
\fi
\ifnum\ec=1
   \includegraphics[scale=0.55]{figures/profit-margin.pdf} 
\fi
   \caption{Empirical estimate of profit margins for 114,628 Swoopo ascending-price auctions. 
\ifnum\tr=1
The overall profit margin is computed by summing the profit across all auctions and dividing it by the total cost.
\fi
}
\label{fig:profit-margin}
\end{figure}

One of the most interesting things about the nearly identical analyses
undertaken thus far is that the simple versions of the model predict 
negligible profits for Swoopo, in that the expected revenue 
matches the value of the item sold.  This fails to match the results from
datasets studied in these papers, other anecdotal evidence \cite{nyt1, nyt2}, as 
well as hard evidence we compiled from a dataset comprising over one hundred thousand 
auction outcomes that we collected, which show Swoopo making dramatic 
profits (see Figure~\ref{fig:profit-margin}).\footnote{We estimate Swoopo's 
net profits for an auction by summing up estimated bid fees plus the final purchase 
price and subtracting the stated retail value for the item.  
\ifnum\tr=1
While we know
the final purchase price exactly, we can only estimate bid fees, as some bidders
have access to discounted bids, for reasons we discuss in detail in Section~\ref{sec:vary-b}.
We therefore overestimate bid fees by assuming all bidders pay the standard bid fee. 
On the other hand, Swoopo's stated retail value for the item tends to be above
market rate, so by using the stated retail value for our calculation we underestimate
Swoopo's profit.  We do not suggest these effects simply cancel each other out, but we
believe our estimate provides a suitable ballpark figure.
\fi
}
Some suggestions in previous work have been made to account for this,
including the relaxation of the assumption that 
players are risk-neutral \cite{platt2009}, or the addition of a 
regret cost to model the impact of sunk costs \cite{augenblick2009}.

%
%
%

In this paper, we take the previous analysis as a starting point, but
  we focus on whether there are intrinsic aspects of the
  pay-per-bid auction framework that can derive profit from even
  rational, risk-neutral players who correctly model sunk costs.
Specifically, previous work has modeled the game as inherently
symmetric, with all players adopting identical randomized strategies.
However, there are natural asymmetries that can arise in the Swoopo
auction, particularly {\em asymmetries in information}.  
A rational player's strategy revolves around his
  assessment of the probability of winning the auction outright by bidding, based on 
  the current bid, the number of bidders, the bid fee, and the value of the item. 
Let us focus on one of these parameters, the number of players $n$.
Although previous models assume that $n$ is known to all players in advance,
 in practice, there is no way to know exactly how many players
are actively participating or monitoring the auction at any time.
In Section~\ref{sec:vary-n}, we show that 
even small asymmetries in beliefs about the number of active players can lead
to dramatic changes in overall auction revenue, and these changes can grow sharply
as the estimates vary from the true number of players.
We also quantify a similar effect due to {\em uncertain} beliefs, as opposed to 
  asymmetry across beliefs.


\ifnum\tr=1
As a related example, previous analyses assume that all players both pay the
  same fee to place a bid in an auction and ascribe an identical value
  to an item.
The latter is clearly not the case, and we discuss the implications in Section \ref{sec:vary-v}.
Less obviously, not all bidders on Swoopo are paying the same price per bid, since
  one item available for auction at Swoopo is a bidpack, which is effectively an
  option to make a fixed number of bids in future auctions for free (``freebids").
Players that win bidpacks at a discount on face value therefore have
the power to make bids at a cheaper price than other players.  As many
players may not factor in this effect (and are unlikely in any case to be able to
accurately estimate either the number of players using freebids, or the nominal 
bid fee those players pay), this again creates an information gap that tends to lead to 
increased profits for Swoopo.  In this case, 
players using less expensive bids 
have a decided advantage, as we show in Section~\ref{sec:vary-b}.  
Pushing this to the extreme, we have the case of
  shill bidders, who bid on behalf of the auctioneer, and can be modeled as bidders who 
  incur no cost to bid (but also never claim an item).
While we do not suggest shill bidders are present in online pay-per-bid auctions, 
our analysis in Section~\ref{sec:shill} nevertheless shows that they would have a striking impact
  on profitability.  
\fi

\ifnum\ec=1
As a related example, previous analyses assume that all players both pay the
  same fee to place a bid in an auction and ascribe an identical value
  to an item.
The latter is generally not the case.
Less obviously, not all bidders on Swoopo are paying the same price per bid, 
  for reasons we discuss in Section~\ref{sec:vary-b}.  
In this case, players using less expensive bids 
  have both a decided information advantage and a tactical advantage.
Pushing this to the extreme, we have the case of
  shill bidders, who bid on behalf of the auctioneer, and can be modeled as bidders who 
  incur no cost to bid (but also never claim an item).
While we do not suggest shill bidders are present in online pay-per-bid auctions, 
our analysis in Section~\ref{sec:shill} nevertheless shows that they would have a striking impact
  on profitability.  
\fi

\ifnum\tr=1
Previous analyses also assume that no players have
  any available side information that they can exploit.
However, two ways in which this side information may be present are when coalitions
  of bidders form, a case we address in Section~\ref{sec:coalitions}, and the setting 
  in which a Swoopo bidder is determined to buy the item, which we analyze in Section~\ref{sec:chicken}.
This interesting second case is facilitated by Swoopo's Swoop It Now feature, introduced
  in the US around July 2009, which enables any bidder to use all of their bid fees in
  an auction as a down payment for the item from Swoopo at full retail value, up until
  one hour after the auction concludes.
Auctions containing one or more {\em committed} players who will use the Swoop It
  Now feature if needed ultimately reduce to games of chicken \cite{rapoport1966}, an
  exceptionally profitable outcome for Swoopo that closely resembles auctions involving shill
  bidders.
\fi

\skipthis{
\ifnum\tr=1
Previous analyses also assume that no players have any available side information 
  that they can exploit.
However, when bidders collude, or when a bidder is determined to buy the item, 
  (as facilitated by Swoopo's Swoop It Now feature, which enables any bidder to use all 
  of their bid fees in a losing auction as a down payment for the item at full retail 
  value), this information can markedly impact strategic behavior.
In Sections~\ref{sec:coalitions} and \ref{sec:chicken}, we analyze these settings,
  and show that auctions containing one or more {\em committed} players who will use 
  the Swoop It Now feature as a last resort ultimately reduce to games of chicken \cite{rapoport1966}, 
  an exceptionally profitable outcome for Swoopo that closely resembles auctions involving shill
  bidders.
\fi
}

\ifnum\tr=1
Finally, our framework allows us to examine other interesting aspects of these 
  types of auctions that are difficult to model analytically, but which can be 
  studied via empirical observations.
One question that we are particularly interested in is whether certain bidder 
  behavior, such as aggressive bidding and bullying, is effective, as earlier
  work speculates~\cite{augenblick2009}.
In Section~\ref{sec:aggression}, we formulate a new definition of bidder aggression, 
  and demonstrate that bidders range widely across the aggression spectrum.
While aggressive bidders win more often, analysis of our dataset shows that 
  the most aggressive bidders contribute the lion's share of profits to Swoopo, and 
  successful strategies are most frequently associated with below-average aggression. 
\fi

\ifnum\ec=1
Finally, our framework allows us to examine other interesting aspects of these 
  auctions that are difficult to model analytically, but which can be 
  studied via empirical observations.
One question that we are particularly interested in is whether certain bidder 
  behavior, such as aggressive bidding, is effective, as earlier
  work speculates~\cite{augenblick2009}.
In Section~\ref{sec:aggression}, we formulate a new definition of bidder aggression, 
  and demonstrate that bidders range widely across the aggression spectrum.
While aggressive bidders win more often, analysis of our dataset shows that 
  the most aggressive bidders contribute the lion's share of profits to Swoopo, and 
  successful strategies are most frequently associated with below-average aggression. 
\fi

We believe that modeling and analyzing these information asymmetries
are interesting in their own right, although we also argue that they
provide a more realistic framework and possible explanation for Swoopo
profits than previous work.  Indeed, our work reveals the previously
hidden complexity of this auction process in the real-world setting.

We emphasize that while we provide data in an attempt to justify these
additions to the model, in contrast to previous work, we eschew
efforts to fit existing data to our model to parameterize and validate
it.  
\ifnum\tr=1
At a high level, we feel that at this stage validation attempts
based on data fitting are unwarranted.  Indeed, the attempt is
reminiscent of similar attempts in the area of power laws in computer networks,
where after many initial works attempted to justify their model of
power law growth by showing it fit the data, it has been widely argued
that fitting data is an improper approach for validation, as many
models with very different characteristics lead to power law
behaviors \cite{mitzenmacher2004brief, mitzenmacher2005editorial, willinger2009mathematics}.  
\fi
\ifnum\ec=1
We suggest that models at this stage can provide a high-level
understanding, but it may be difficult to disentangle
various effects through auction data alone without more detailed insight
into user behavior.  Moreover, current models appear as yet far from complete.
We therefore suggest future alternatives and directions
in the conclusion.

Finally, we note that, due to space limitations, we cannot fully describe all of our
results in this paper.  A much longer and more detailed version is available for
download on the arXiv \cite{bmz2010}. In particular, in many of our mathematical derivations here,
we focus on the simpler case of fixed-price auctions, described in Section~\ref{sec:model}, for space reasons.
\fi

\ifnum\ec=1
\fi
\subsection{Related Work}

Several recent working papers have studied pay-per-bid auctions
\cite{augenblick2009, hinnosaar2009, platt2009}.  While there are some
differences among the papers, they all utilize the same basic
framework, which is based on finding an equilibrium behavior for the
players of the auction.  We describe this framework in Section~\ref{sec:model},
and use it as a starting point.  The key feature of this framework
from our standpoint is that it treats the players as behaving
symmetrically, with full information.  Unsurprisingly, in such a
setting the expected profit for Swoopo is theoretically zero.

Our key deviation from past work is to consider asymmetries inherent
in such auctions, with a particular focus on information asymmetry.
Information asymmetry broadly refers to situations where one party has
better information than the others, and has become a key concept in
economics, with thousands of papers on the topic.  The pioneering work
of Akerlof \cite{akerlof1970}, Spence \cite{spence1973}, and Stiglitz
\cite{rothschild1976}, for which the authors received a Nobel Prize in
2001, established the area.  Typical examples of information asymmetry
include insider trading, used-car sales, and insurance.
Interestingly, the study of information asymmetry in auctions appears
significantly less studied.  We believe that our analysis of Swoopo
auctions provides a simple, natural example of the potential effects
of information asymmetry (as well as other asymmetries) in an auction
setting, and as such may be valuable beyond the analysis itself.
\ifnum\tr=1
Indeed, our first example shows how information asymmetry about a
basic parameter of an auction -- the number of participants -- can
significantly affect its profitability.
\fi

\ifnum\ec=1
\vspace{-2 mm}
\fi
\subsection{Datasets}

Where appropriate, we motivate our work or provide evidence for our
results via data from Swoopo auctions.  We have collected two
datasets.  One dataset is based on information published directly by
Swoopo, which contains limited information about an auction.
Information provided includes basic features such as the product
description, the retail price, the final auction price, the bid fee,
the price increment, and so on.  This dataset covers over 121,419
auctions.  We refer to this as the \textsc{Outcomes} dataset.  

\ifnum\tr=1
Our second dataset is based on traces of live auctions that we have
ourselves recorded using our own recording infrastructure.  Our traces
include the same information from the Swoopo auctions as well as detailed
bidding information for each auction, specifically the time and the
player associated with each bid.  This dataset spans 7,353 auctions and 2,541,332 bids.
We refer to this as the \textsc{Trace} dataset.  
Our methodology to collect bidding information entailed continuous
monitoring of Swoopo auctions.
We probe Swoopo according to a varying probing interval that is described in detail
in Appendix~\ref{sec:appendix-dataset};  
in particular, when the auction clock is at less than
2 minutes, we probe at least once a second.
Swoopo responds with a list of up to ten tuples of the form 
$(username, bid number)$ indicating the players that placed a bid since our 
previous probe and the order in which they did so. 
Often this list would include more than one tuple. In these cases
we ascribe the same timestamp to all of these bids; given
the high probing frequency, this is a reasonable approximation. A more
significant limitation from our methodology arises when more then ten players 
bid between successive probes. In these cases, Swoopo responded with just 
the ten latest bids. In particular this happened when more than ten players 
were using BidButlers, automatic bidding agents provided by the Swoopo 
interface, to bid at a given level.  Overall, we captured every bid from 4,328 auctions. 
The remaining 3,025 auctions had a total of 491,360 missing bids;  we did not consider
these in our study.
\fi

\ifnum\ec=1
Our second dataset is based on traces of live auctions that we have
ourselves recorded using our own recording infrastructure.  Our traces
include the same information from the Swoopo auctions as well as detailed
bidding information for each auction, specifically the time and the
player associated with each bid.  This dataset spans 7,353 auctions and 2,541,332 bids.
We refer to this as the \textsc{Trace} dataset.  
Our methodology to collect bidding information entailed continuous
monitoring of Swoopo auctions;  however, in some cases we could not obtain 
all of the bids.  In particular this happened when more than ten players 
were using BidButlers, automatic bidding agents provided by the Swoopo 
interface, to bid at a given level, as we collect at most ten bids with each
probe of Swoopo.  Overall, we captured every bid from 4,328 of the 7,353 auctions; 
only results from these complete auctions are included in our study.
Further details regarding our dataset and collection methods,
including how to download the data, can be found in the full version of our paper \cite{bmz2010}.  
\fi

\ifnum\tr=1
As two examples of the kind of information that can be derived from
our dataset we present Figures \ref{fig:profit-by-month-bidincrement}
and \ref{fig:profit-by-item}, derived from the \textsc{Outcomes} dataset. The
former presents Swoopo's monthly profit margin for a set of what we
refer to as regular auctions, grouped by price increment.  Regular
auctions exclude ``NailBiter'' auctions which do not allow the use of
BidButlers, beginner auctions which are for players who have not won
an auction previously, and fixed-price auctions. The latter figure
displays Swoopo's profit margin by item, for items that have been
auctioned off at least 200 times in regular auctions.

\begin{figure}[htbp]
   \centering
   \includegraphics{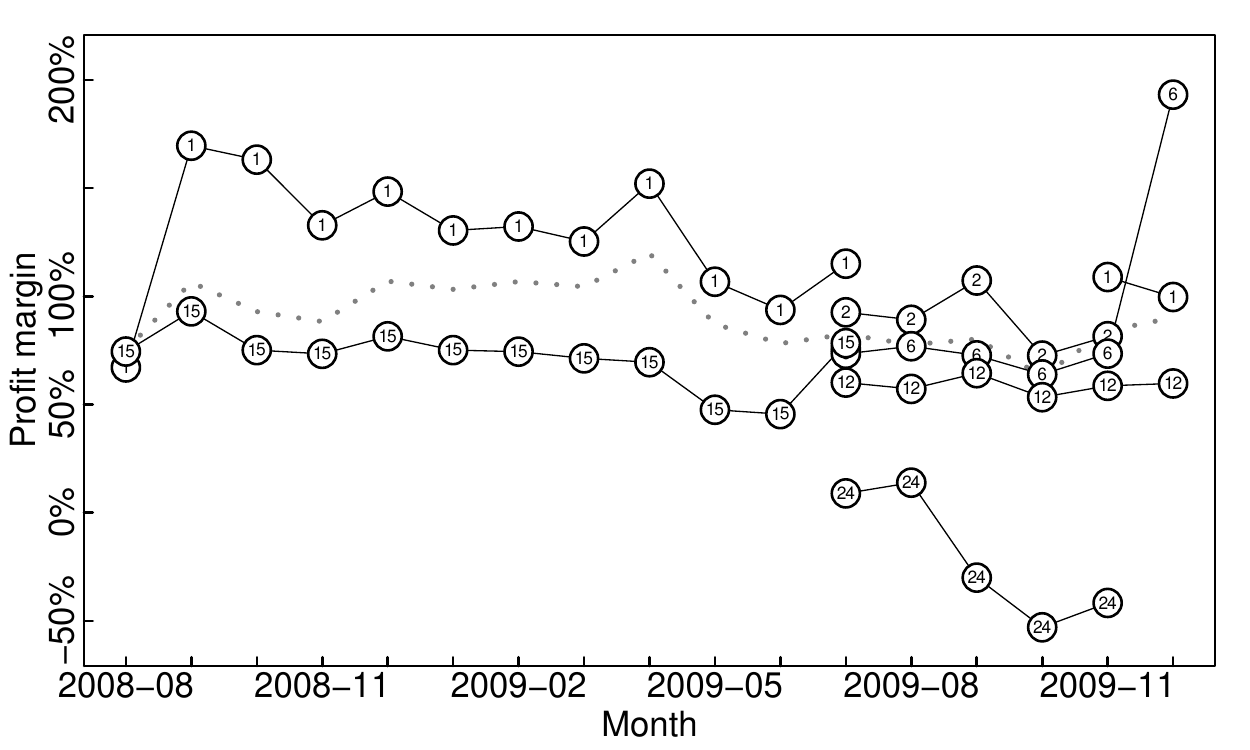} 
   \caption{Swoopo monthly profit margins by price increment for regular auctions from the \textsc{Outcomes} dataset. The dotted line represents the profit margin across all price increments.}
   \label{fig:profit-by-month-bidincrement}
\end{figure}

\begin{figure}[htbp]
   \centering
   \includegraphics{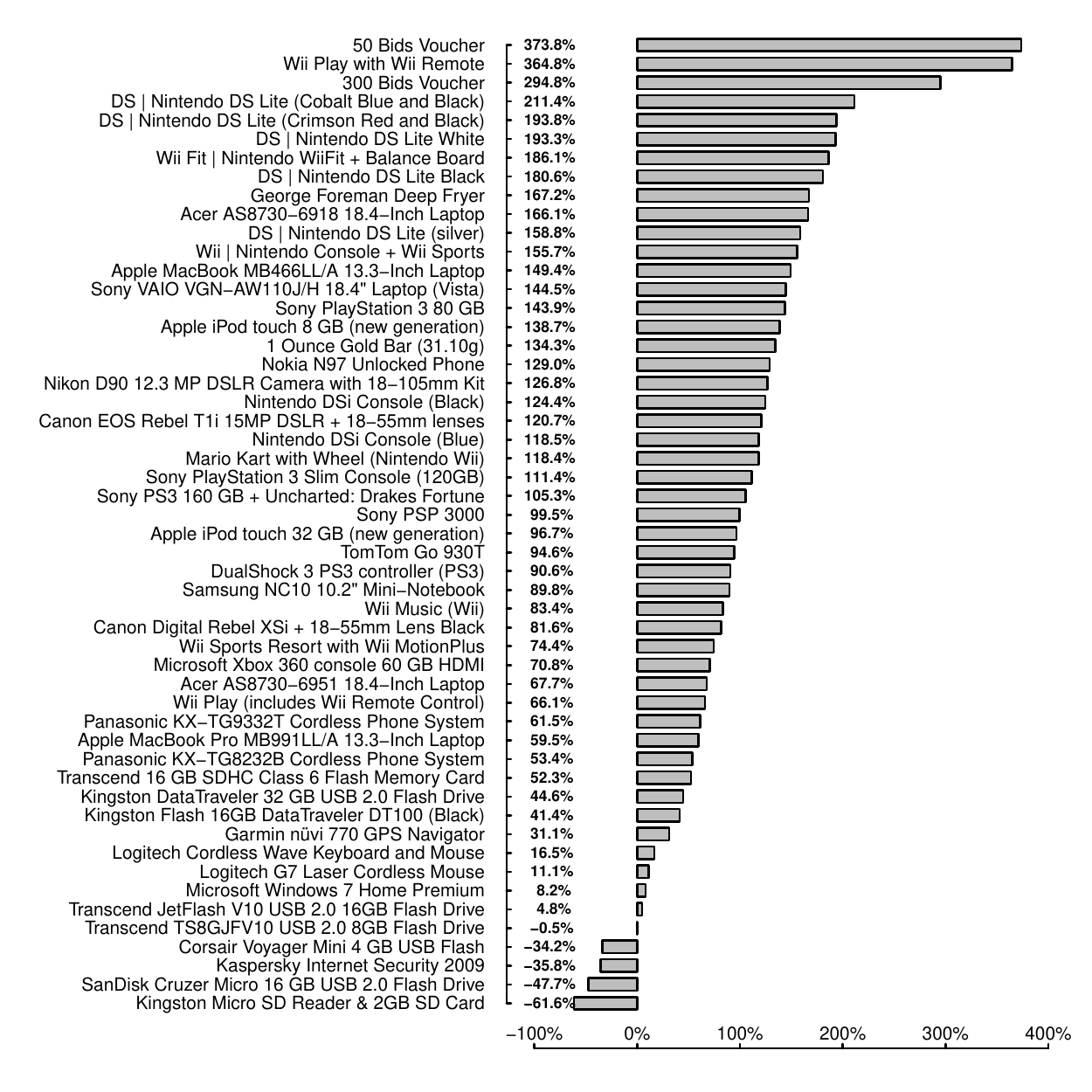} 
   \caption{Profit margin for items that have been auctioned off at least 200 times in regular auctions from the \textsc{Outcomes} dataset.}
   \label{fig:profit-by-item}
\end{figure}

More details regarding the datasets can be found in Appendix~\ref{sec:appendix-dataset}.
These datasets are publicly available.\footnote{Available at: \texttt{http://cs-people.bu.edu/zg/swoopo.html}. Please contact \texttt{zg@bu.edu} for questions regarding this dataset.}
\fi

\ifnum\ec=1
\vspace{-4 mm}
\fi
\section{A Symmetric Pay-Per-Bid Model}

\label{sec:model}

We start with a basic model and analysis of Swoopo auctions from
previous work, following the notation and framework of
\cite{platt2009}, although we note that essentially equivalent
analyses have also appeared in other work \cite{augenblick2009,
hinnosaar2009}.  This serves to provide background and context for our
work.

We consider an auction for an item with an objective value of $v$ to
all players.  There are $n$ players throughout the auction.  The
initial price of the item is 0.  In the {\em ascending-price} version
of the auction, when a player places a bid, he pays
an up-front cost of $b$ dollars, and the price is incremented by $s$
dollars.
\ifnum\tr=1
\footnote{The case of descending-price models could equally be
modeled and studied using our techniques.}
\fi
The auction has an associated countdown clock;  time is added
to the clock when a player bids to allow other players the opportunity
to bid again.
When an auction terminates, the last bidder pays
the current price of the item and receives the item.  In a variant
called a {\em fixed-price} auction, the winner
buys the item for a fixed price $p$;  bids still cost $b$ dollars 
but there is no price increment.  When $p=0$, this is called a {\em 100\%-off} 
auction.  
\ifnum\tr=1
In our analysis, 
we simplify players' strategies by removing the impact of timing.\footnote{We
study aspects of timing empirically in Section~\ref{sec:aggression}, where we
consider the repercussions of aggressive bidding in live auctions.}  
Instead of
bidding at a given time, players choose to bid based on the current
price; if multiple players choose to bid based on the current price,
we generally break ties by assuming a random bidder bids first.
A player that chooses not to bid at some price may bid later at
a higher price.
\fi
\ifnum\ec=1
In our analysis, we simplify players' strategies by removing the impact of timing
(but we do study this empirically in Section~\ref{sec:aggression}).
Instead of
bidding at a given time, players choose to bid based on the current
price, with ties broken at random.
A player that chooses not to bid at some price may bid later on.
\fi

The basic formulation for analyzing this game is that a player who
makes the $q$th bid is betting $b$ than no future player will bid.
Let $\mu_j$ be the probability that somebody makes the $j$th bid (given
that $j-1$ previous bids have been made).  Then the expected payoff
for the player that makes the $q$th bid is $(v - sq)(1-\mu_{q+1})$; a
player will only bid if this payoff is non-negative.  Note that when
$q > Q \equiv \lfloor \frac{v-b}{s} \rfloor$ it is clear that no
rational player will bid, as the item price plus bid fee exceeds the value.
For convenience in the analysis we will assume that $\frac{v-b}{s}$ is
an integer, to avoid technical issues when this does not hold (see
\cite{augenblick2009} for a discussion); this assumption ensures that a
player that makes the $Q$th bid is indifferent to the outcome (the
expected payoff is 0).  In the fixed-price variant, the payoff is $(v
- p)(1-\mu_{q+1})$; as long as $v > p$, bidding may occur.

\ifnum\tr=1
The equilibrium behavior is found by determining the probability that
a player should bid so that the expected payoff is zero 
whenever $q \leq Q$, leaving the players indifferent as to the choice of
whether to bid or not to bid.  (Alternative equilibria that are not
germane to our analyses are discussed in \cite{augenblick2009}.)  Hence 
the indifference condition is given by
$$b = (v - sq)(1-\mu_{q+1}),$$ or
$$\mu_{q+1} = 1 - \frac{b}{v-sq}$$
in the ascending-price auction, and 
$$\mu_{q} = 1 - \frac{b}{v-p}$$
at all steps in the fixed-price auction.  
\fi

\ifnum\ec=1
The equilibrium behavior is found by determining the probability that
a player should bid so that the expected payoff is zero 
whenever $q \leq Q$, leaving the players indifferent as to the choice of
whether to bid or not to bid.  
Hence 
the indifference condition is given by
$$b = (v - sq)(1-\mu_{q+1}),$$ or
$$\mu_{q+1} = 1 - b/(v-sq)$$
in the ascending-price auction, and 
$$\mu_{q} = 1 - b/(v-p)$$
at all steps in the fixed-price auction.  
\fi

In what follows it is helpful to let
$\beta_q$ be the probability that each player chooses to make
the $q$th bid given that the $(q-1)$st bid has been made and that the
player is not the current leader.  Note that by symmetry each player
bids with the same probability.  Hence, for $q > 1$, for ascending-price
auctions we must have
\ifnum\tr=1
\begin{eqnarray*}
1 - \mu_{q} & = & (1 - \beta_q)^{n-1} \\
\beta_q & = & 1 - (1 - \mu_{q})^{1/(n-1)} \\
\beta_q & = & 1 - \left(\frac{b}{v-s(q-1)}\right)^{1/(n-1)}.
\end{eqnarray*}
\fi
\ifnum\ec=1
\begin{eqnarray*}
1 - \mu_{q} & = & (1 - \beta_q)^{n-1} \\
\beta_q & = & 1 - \left(\frac{b}{v-s(q-1)}\right)^{1/(n-1)}.
\end{eqnarray*}
\fi
Similarly, we have 
$$\beta_q  = 1 - \left(\frac{b}{v-p}\right)^{1/(n-1)}$$
for the fixed-price auction.

We point out that the first bid is a special case, since at that point there
is no leader.  To maintain consistency, we want the indifference
condition to hold for the first bid; that is, players still bid such that their expected profit is zero. 
This requires a simple change, since at the first bid there are $n$ players who might
bid instead of $n-1$, giving for the ascending-price auction
\ifnum\tr=1
\begin{align}
\beta_{q} &= 
\begin{cases}
1 - \left(\frac{b}{v}\right)^{\frac{1}{n}} & \quad \textrm{for $q = 1$,} \\
1 - \left(\frac{b}{v-s(q-1)}\right)^{\frac{1}{n-1}} & \quad \textrm{for $q > 1$.}
\end{cases}
\end{align}
Similar equations hold for the fixed-price variant:
\begin{align}
\beta_{q} &= 
\begin{cases}
1 - \left(\frac{b}{v-p}\right)^{\frac{1}{n}} & \quad \textrm{for $q = 1$,} \\
1 - \left(\frac{b}{v-p}\right)^{\frac{1}{n-1}} & \quad \textrm{for $q > 1$.}
\end{cases}
\end{align}
\fi
\ifnum\ec=1
\begin{eqnarray*}
\beta_{1} = 1 - \left(\frac{b}{v}\right)^{\frac{1}{n}},
\end{eqnarray*}
and similarly $\beta_{1} = 1 - \left(b/(v-p)\right)^{\frac{1}{n}}$
for fixed-price auctions. 
\fi

The expected revenue for the auction can easily be calculated directly
using the above quantities.  However, we suggest a simple argument
(that can be formalized in various ways, such as by defining an appropriate
martingale) that demonstrates that Swoopo's expected revenue
is $v$ if there is at least one bid, and zero if no player bids.  
(A similar argument appears in \cite{augenblick2009}.)
First note that in auctions where there is at least one bid, an item of value 
$v$ is transferred to some player at the end of the auction.  Also,
by the indifference condition, the expected gain to the player that places
any bid is zero.  (Think of a bid $b$ as counterbalanced by the auctioneer
putting an expected value $b$ at risk.)  Therefore, by linearity of expectations, the auctioneer recoups
a sum of payments equal to $v$ in expectation over the course of the auction,
conditioned on there being at least one bid.
The probability that no player bids is $(1 - \beta_1)^n$ 
by definition of $\beta_1$, and thus the expected revenue is 
$v (1 - (1 - \beta_1)^n) = v - b$.

To be clear, in what follows, we will always consider revenue 
conditioned on the auction having had at least one bid, since otherwise, the
auction is essentially a non-operation for the auctioneer.  We call such
auctions {\em successful}.

\ifnum\tr=1
As we noted earlier, in practice the revenue Swoopo earns per auction is much 
larger than $v$ on average.  To help explain this, we consider the assumptions
behind this model.  This model assumes that the key parameters $n$,
$v$, and $b$ are the same for each player and known to all.  These
assumptions seem hard to justify in practice: players may misestimate
the total population, some may have access to cheaper bids, or they may
each value the auctioned object differently.  We provide some general 
extensions to the model for these types of variations in
Section~\ref{sec:model-ext}, but before that, we motivate our general
considerations by considering the specific example where the number of
players $n$ is not known.
\fi

\ifnum\ec=1
\fi

\ifnum\ec=1
\vspace{-3 mm}
\fi
\ifnum\tr=1
\section{Asymmetries in the Perceived Number of Bidders}

\label{sec:vary-n}

\subsection{Motivation}
\fi
\ifnum\ec=1
\section{Asymmetric Player Estimates} 

\label{sec:vary-n}

\fi

\ifnum\tr=1
The analysis of Section~\ref{sec:model} makes the assumption
that the number of players is fixed and known throughout.  This
assumption has been questioned in previous work; for example, in
\cite{platt2009}, they propose a simple alternative to the standard
model where participants enter and leave the auction over time.
However, even in this variation, the expected number of players at
each time step is known and the distribution is assumed to be Poisson,
so that the end result is a small variation on the previous
analysis.  Here we take a different approach and consider the
original model but without the assumption that every player
has the same estimate of $n$, the number of players in the game.
\fi

\ifnum\ec=1
The analysis of Section~\ref{sec:model} assumes
that the number of players is fixed and known throughout.  This
assumption has been questioned in previous work; for example, in
\cite{platt2009}, they propose a variation
where the expected number of players at
each time step is known and the distribution is assumed to be Poisson,
to model participants entering and leaving the auction over time.
The end result is a small variation on the previous
analysis.  Here we take a different approach and 
remove the assumption that every player
has the same estimate of $n$, the number of players in the game.
\fi

\ifnum\tr=1
Before diving into the analysis, we provide some motivating data from
our datasets.  During an auction, Swoopo provides some limited
information regarding the number of players participating in the
auction.  Specifically, it provides a list of the bidders that have
been active over the last 15 minute period.  Analysis of our
\textsc{Trace} dataset suggests this 
is insufficient information
for determining the number of players in the auction, and in fact we
suspect it can lead players to significantly underestimate the number of
other players in the auction.  
This will prove to have dramatic
impact on the analysis, and in particular on the expected revenue to Swoopo.
\fi
\ifnum\ec=1
Before diving into the analysis, we provide some motivating data from
our datasets.  During an auction, Swoopo provides 
a list of the bidders that have
been active over the last 15 minute period.  Analysis of our
\textsc{Trace} dataset indicate that this 
significantly underestimates the total number of participants in the auction, 
  so players who rely on this information to estimate the number of players 
  may be misled.
Our analysis shows that players underestimating $n$ can dramatically inflate Swoopo's expected revenue.
\fi

\ifnum\tr=1

\begin{figure}[tbp]
   \centering
\ifnum\tr=1
   \includegraphics{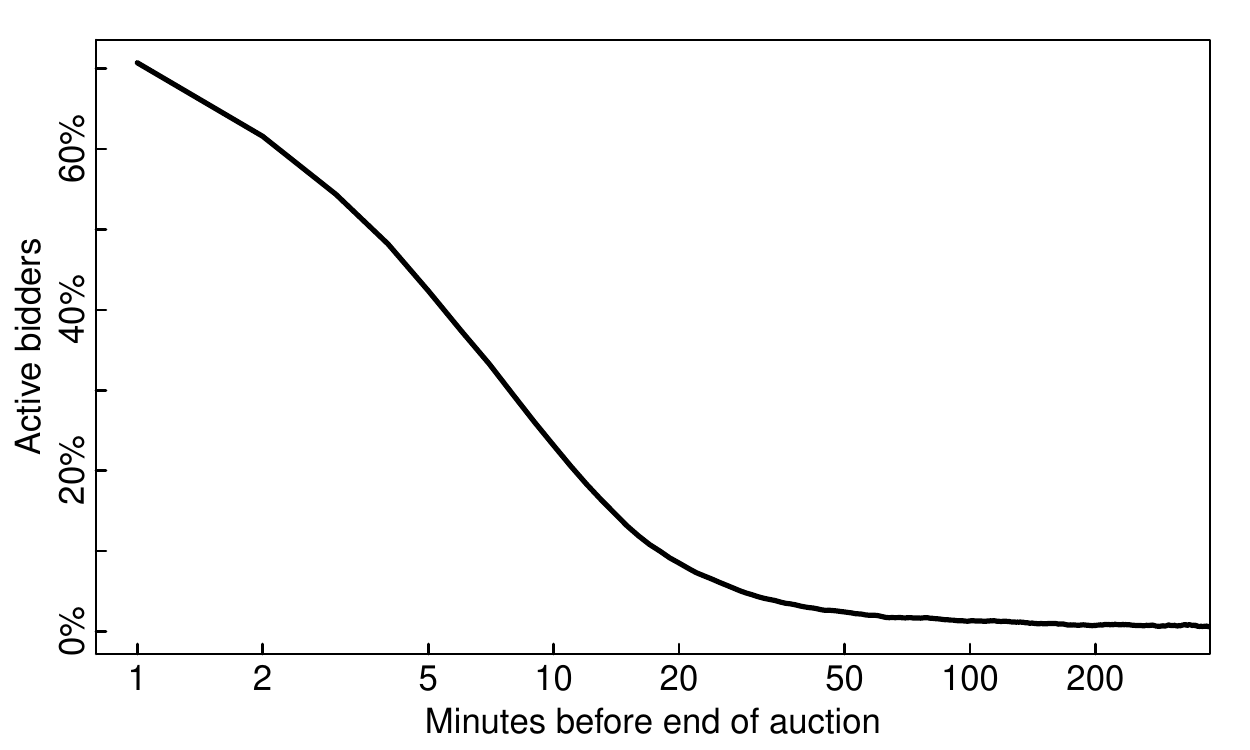} 
\fi
\ifnum\ec=1
   \includegraphics[scale=0.55]{figures/active-users.pdf} 
\fi
   \caption{Percentage of active bidders using a 15-minute sliding window. The $x$-axis is in log scale.}
   \label{fig:active-users}
\end{figure}

Following Swoopo, we define an {\em active bidder} as someone who has bid in the last fifteen minutes.
Using our \textsc{Trace} dataset, we observed each auction at one minute intervals, and at each time instant we computed the 
  number of active bidders as a percentage of the total number of players who 
  ultimately participated in the auction.
The points on the curve in our plot of these values in Figure \ref{fig:active-users}
  can be interpreted as the percentage of all participants accounted for 
  in Swoopo's active bidder list as a function of time. 
Our plot shows that auction participation builds to a crescendo at the end of 
  the auction, so a typical report ten minutes from the end of the auction
  only reports 20\% of all bidders, and when that number doubles in a five minute 
  span, it still reflects only 40\% of the population. 
Also, note that due to the nature of the auction, there is no fixed time at which the
  auction ends, so even bidders making predictions based on past observations are
  using a certain degree of guesswork.
We therefore suggest that players relying on active bidder information 
  may well be misguided about the size of the playing field. 
In the following sections we analytically quantify the effect such a misestimation has 
  on Swoopo's expected revenues.
\fi

\ifnum\ec=1
Following Swoopo, we define an {\em active bidder} as someone who has bid in the last fifteen minutes.
Using our \textsc{Trace} dataset, we observed each auction at one minute intervals, and at each time instant we computed the 
  number of active bidders as a percentage of the total number of players who 
  ultimately participated in the auction.
We found that auction participation typically builds to a crescendo at the end of 
  the auction;  on average, ten minutes from the end of the auction
  only 20\% of all bidders have participated, and five minutes from the end only
  40\% of all bidders have participated. 
Further, due to the nature of the auction, there is no fixed time at which the
  auction ends, so even bidders making predictions based on past observations are
  using a certain degree of guesswork.
\fi

\ifnum\tr=1
\subsection{Analysis for Fixed-Price Auctions}

For simplicity we begin with the case of fixed-price auctions.  To
initially frame the analysis, we further assume that the true number of players
is $n$, but {\em all} players perceive the number of players as $n
- k$ for some $k$ in the range $[1, n-2]$.  In this case, there is still 
symmetry among the players, but they choose to bid based on incorrect
information.  Following the previous analysis, to maintain the
indifference condition that the expected revenue for a player that
bids at each point should be equal to their bid fee, we have $(v -
p)(1 - \nu_{q}) = b$, where now $\nu_{q}$ is the perceived
probability that someone else will place the $q$th bid.  As before,
\begin{equation}
\nu_{q} = 1 - \frac{b}{v-p}.
\label{eqn:mq'}
\end{equation}
Again we let $\beta_q$ be the probability that a player chooses to make the $q$th bid.
For $q > 1$ we have $(1-\nu_{q}) = (1-\beta_{q})^{n-k-1}$, or
$\beta_{q} = 1 - (1 - \nu_{q})^{\frac{1}{n-k-1}}.$
\fi

\ifnum\ec=1

We now consider the analysis of fixed-price auctions.  To
initially frame the analysis, we further assume that the true number of players
is $n$, but {\em all} players perceive the number of players as $n
- k$ for some $k$ in the range $[1, n-2]$.  In this case, there is still 
symmetry among the players, but they choose to bid based on incorrect
information.  Following the previous analysis, to maintain the
indifference condition that the expected revenue for a player that
bids at each point should be equal to their bid fee, we have $(v -
p)(1 - \nu_{q}) = b$, where now $\nu_{q}$ is the perceived
probability that someone else will place the $q$th bid.  As before,
%
$\nu_{q} = 1 - \frac{b}{v-p}$.
%
Again we let $\beta_q$ be the probability that a player chooses to make the $q$th bid.
For $q > 1$ we have $(1-\nu_{q}) = (1-\beta_{q})^{n-k-1}$, or
$\beta_{q} = 1 - (1 - \nu_{q})^{\frac{1}{n-k-1}}.$
\fi

\ifnum\tr=1
Crucially, $\nu_{q}$ is not equal to $\mu_{q}$, the true probability that will 
make the $q$th bid.
Since $(1-\mu_{q})$ equals the probability that nobody makes the $q$th bid, we have
\begin{align}
1 - \mu_{q} &= (1 - \beta_{q})^{n-1} \notag \\
\mu_{q} &= 1 - \left((1 - \nu_{q})^{\frac{1}{n-k-1}}\right)^{n-1} \notag \\
\mu_{q} &= 1 - \left(\frac{b}{v-p}\right)^{\frac{n-1}{n-k-1}} \label{eqn:mq}.
\end{align}
Remember that the above holds for $q > 1$, as for the first bid the
bidding probabilities are slightly different, as explained in Section
\ref{sec:model}. To simplify the math, as an unsuccessful auction with zero bids is
uninteresting, we assume the first bid has been placed. Then $\mu_q$
is the same at all points, so we simply call the value $\mu$.  The
probability that the auction lasts another $r$ bids, after the
first, is given by
$\mu^{r} (1 - \mu)$.
If we let $R$ be the revenue from a successful auction, we calculate Swoopo's expected revenue as:
\begin{align}
E[R] &=  b + p + b \sum_{r=0}^{\infty} r \mu^{r} (1 - \mu).
\end{align}
In the simple case where $p = 0$ the expected revenue is:
\begin{equation}
E[R] = b \left(\frac{v}{b}\right)^{\frac{n-1}{n-k-1}}.
\end{equation}
\fi

\ifnum\ec=1
Crucially, $\nu_{q}$ is not equal to $\mu_{q}$, the true probability that someone will
make the $q$th bid.
Since $(1-\mu_{q})$ equals the probability that nobody makes the $q$th bid, we have
\begin{align}
1 - \mu_{q} &= (1 - \beta_{q})^{n-1} \notag \\
\mu_{q} &= 1 - \left((1 - \nu_{q})^{\frac{1}{n-k-1}}\right)^{n-1} \notag \\
\mu_{q} &= 1 - \left(\frac{b}{v-p}\right)^{\frac{n-1}{n-k-1}} \label{eqn:mq}.
\end{align}
Remember that the above holds for $q > 1$, as for the first bid the
bidding probabilities are slightly different, as explained in Section
\ref{sec:model}.  In a successful auction, $\mu_q$
is the same for all bids, so we simply call the value $\mu$.  The
probability that the auction lasts another $r$ bids, after the
first, is given by
$\mu^{r} (1 - \mu)$.
If we let $R$ be the revenue from a successful auction, we calculate Swoopo's expected revenue as:
\begin{align}
E[R] &=  b + p + b \sum_{r=0}^{\infty} r \mu^{r} (1 - \mu).
\end{align}
In the simple case where $p = 0$ the expected revenue is:
\begin{equation}
E[R] = b \left(\frac{v}{b}\right)^{\frac{n-1}{n-k-1}}.
\end{equation}
\fi

\begin{figure}[t]
\centering
\subfigure[All players underestimate the population by $k$. Negative values of $k$ stand for overestimates. ]
{
\ifnum\tr=1
   \includegraphics[scale=0.55]{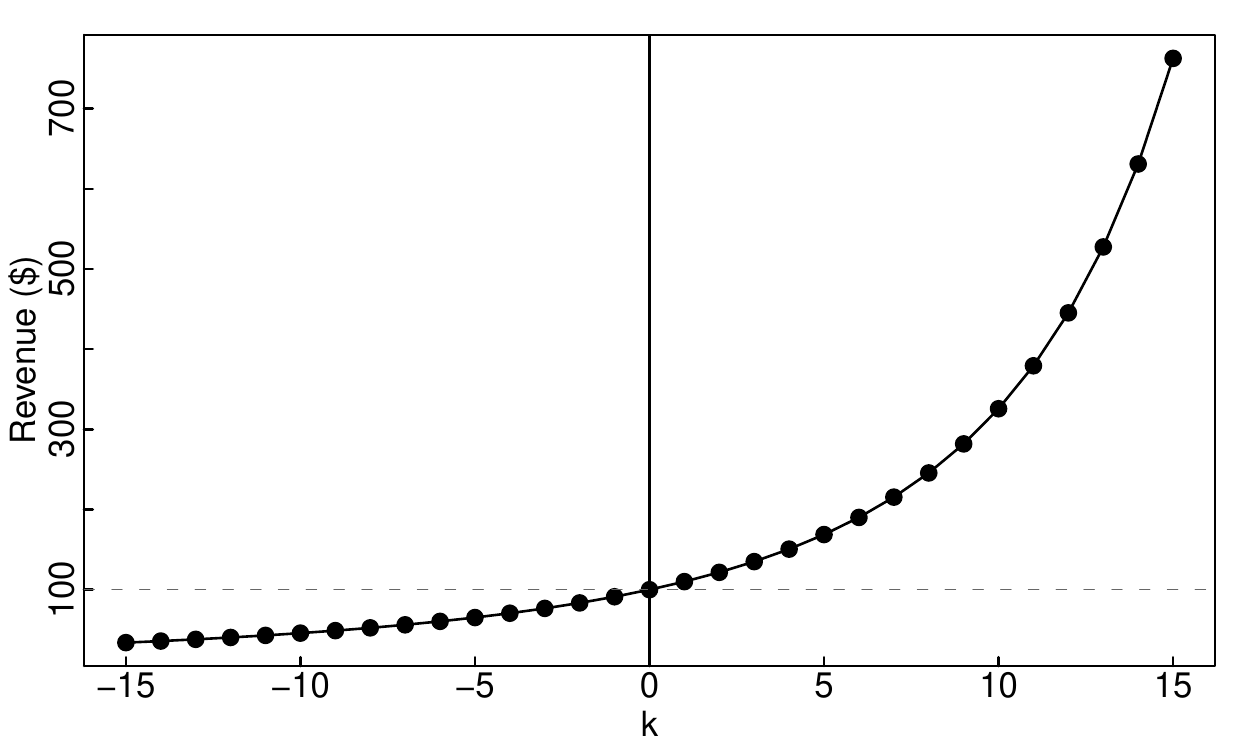} 
\fi
\ifnum\ec=1
   \includegraphics[scale=0.55]{figures/vary-n-fixed-revenue.pdf} 
\fi
   \label{fig:vary-k}
}
\subfigure[Half the players underestimate the population by $k$ and half overestimate it by an equal amount.]
{
\ifnum\tr=1
   \includegraphics[scale=0.55]{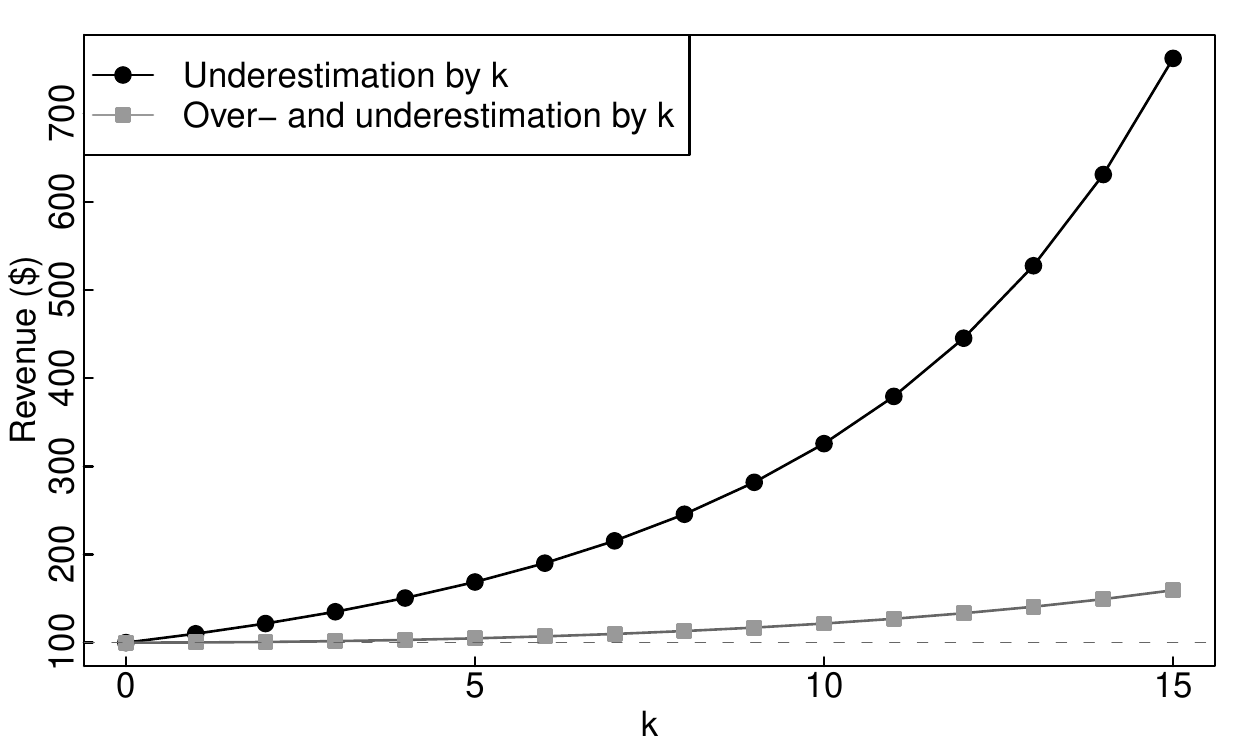} 
\fi
\ifnum\ec=1
   \includegraphics[scale=0.55]{figures/vary-pm-n-fixed-revenue.pdf} 
\fi
   \label{fig:vary-pm-k}
}
\caption{Expected revenue for Swoopo in a successful 100\% off 
  \ifnum\tr=1
auction for two different population misestimation settings;
  \fi
  \ifnum\ec=1
auction; 
  \fi
 $n=50$, $v=100$ and $b=1$.}
\label{fig:vary-n-revenue}
\end{figure}

\ifnum\tr=1
Notice that when $k=0$, the expected revenue from a successful auction is indeed $v$ as had been demonstrated in previous works.  Also, as $k$ appears in the 
exponent of the $v/b$ term, even small values of $k$ can have a significant effect on the revenue.
This dramatic impact on revenue as $k$ varies is depicted for a representative auction
for \$100 in cash with a bid fee of \$1 and 50 players in Figure~\ref{fig:vary-k}.
These will be our default parameters 
for fixed-price auctions throughout this work. 
\fi

\ifnum\ec=1
When $k = 0$, the expected revenue is $v$.
But as $k$ appears in the 
exponent of the $v/b$ term, even small values of $k$ can have a significant effect on the revenue.
This impact as $k$ varies is depicted for a representative auction
for \$100 in cash with a bid fee of \$1 and 50 players in Figure~\ref{fig:vary-k}.
These will be our default parameters 
for fixed-price auctions throughout this work. 
\fi

Conversely, one could consider what happens when players overestimate
the population, that is to say $k < 0$. As expected, the revenue for
Swoopo then shrinks, incurring an overall loss as demonstrated
in the left half of Figure \ref{fig:vary-k}.  
\ifnum\ec=1
Note the considerable asymmetry in the plot, however. 
\fi
\ifnum\tr=1
From our standpoint, the
key feature of this graph is the asymmetry:  underestimates of the player
population size have significantly larger revenue effects than overestimates.  
If players tend to underestimate the number of players, as our empirical
evidence suggests is likely, auction revenues increase dramatically.
\fi
Indeed, even if the average estimated number of players is correct, when
there is variation across estimates, Swoopo gains.    
For example, we can consider a simple case where half the players underestimate the population by $k$ and half overestimate it by the same amount. 
\ifnum\tr=1
Denote by $\beta_{q,-k}$ the probability with which the former bid and similarly by $\beta_{q,+k}$ the probability with which the latter do. Then we have:
\begin{align}
\beta_{q,-k} &= 1 - (1 - \nu_{q})^{\frac{1}{n-k-1}}, \\
\beta_{q,+k} &= 1 - (1 - \nu_{q})^{\frac{1}{n+k-1}}.
\end{align}

\fi
Computing the revenue for this mixed case involves more complicated machinery which we describe in detail in Section \ref{sec:model-ext}. 
For now, we observe that even though Swoopo has far more to gain by pure underestimation of the player population,
a mix of overestimation and underestimation in equal measures still yields
markedly increased revenues, as depicted in Figure \ref{fig:vary-pm-k}. 
(This can also be seen as a consequence of convexity
of the revenue curve as the estimate of the number of players varies.)

\ifnum\tr=1 Similar analyses can be made for different settings. \fi

\ifnum\ec=1 Similar analyses can be made for different settings, such as ascending-price auctions and other mixtures of estimates. \fi

\ifnum\tr=1
\subsection{Analysis for Ascending-Price Auctions}

Recall that in an ascending-price auction 
an auction lasts at most $Q = \lfloor
\frac{v-b}{s} \rfloor$ bids subsequent to the first. Following reasoning similar to that for
fixed-price auctions, we can express the expected revenue of a successful
ascending-price auction where the population is underestimated by
$k$ by
\begin{equation}
E[R] = (b+s) + \sum_{q=1}^{Q} (b + s) q (1 - \mu_{q+1}) \prod_{j=1}^{q} \mu_{j}
\label{eqn:rev-ascending-underestimate}
\end{equation}
where $\mu_{j} = 1 - \left( \frac{b}{v - s (j - 1)} \right)^{\frac{n-1}{n-k-1}}$. 
Following the same steps as \cite{platt2009}, 
  Equation \ref{eqn:rev-ascending-underestimate} can be simplified to yield:
\begin{align}
E[R]  &= 
(b+s)+\sum_{q=0}^{Q-1}\left( (b + s) \prod_{j=1}^{q+1} \mu_{j}\right).
\end{align}

In an ascending-price auction, the revenue is
trivially upper bounded by $(Q+1)(b+s)$, regardless of $k$.  
Figure \ref{fig:vary-pm-k-ascending}
displays the expected revenue for Swoopo in successful auctions
as the population estimation error $k$ varies.  
Here our auction is for \$100 in cash with a bid fee of \$1, 50 players, and a price increment
of \$0.25.  Again, these will be our default parameters 
for ascending-price auctions throughout this work. 
The two dashed lines correspond to the true value of the auctioned item
and the trivial upper bound on revenue of $(Q+1)(b+s)$.  
The plot confirms that for large enough $n$ and $k$,
when $\mu_{q} \approx 1$ throughout most of the auction, this upper
bound is nearly tight.  
Therefore, despite the asymptotic leveling off of revenues, Swoopo can still
capitalize significantly from misinformed players.

\begin{figure}[t]
   \centering
   \includegraphics[scale=0.55]{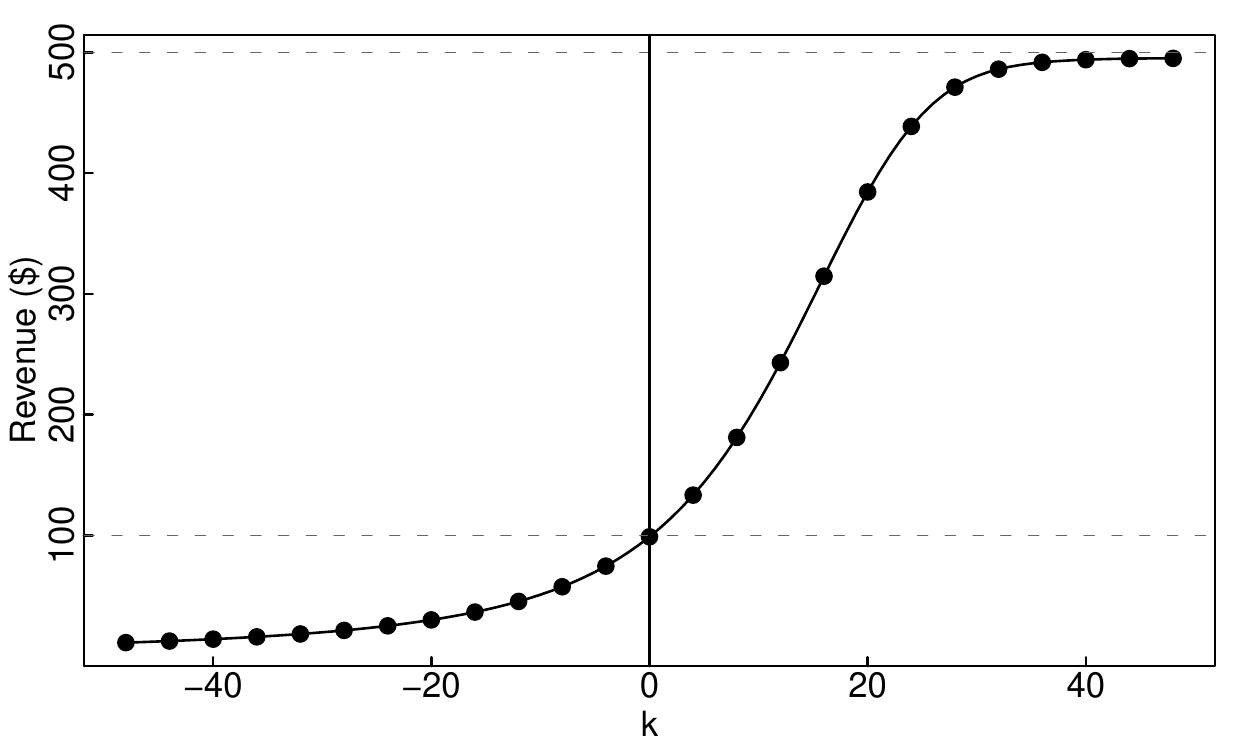} 
   \caption{Expected revenue as the underestimation error $k$ varies in an ascending price auction; $n=50$, $v=100$, $b=1$, and $s=0.25$.}
   \label{fig:vary-pm-k-ascending}
\end{figure}

\fi

\subsection{Incorporating Uncertainty Into Population Estimates}
Throughout the paper, we focus on the case where players have fixed
beliefs about relevant quantities, such as the number of players
participating in the game.  
However, it is also straightforward to extend our techniques
to settings where there is
underlying uncertainty within beliefs, either instead of or in addition to 
asymmetry across beliefs.
As a concrete example, we consider the case where
players are symmetric and all perceive the population size as being drawn 
from a distribution such that there are $i$ players with probability $p_i$.  
We do not focus on beliefs governed by distributions in the rest of the
paper, because our main thematic points can be made using simpler point
beliefs.  Moreover, beliefs governed by distributions also raise challenging
questions, such as how players determine an initial belief
distribution and whether they can update their beliefs as the auction
proceeds, that we do not consider in this work.  However, it should
intutitively be clear that uncertainty, as well as asymmmetry, can lead to 
situations that benefit the auctioneer.  We formalize one such situation now.

Consider a fixed-price auction auction with $n$ players,
where the players are symmetric and believe that the auction population size is
governed by a distribution where the number of players is $i$ with
probability $z_i$ such that $\sum_{i} i z_{i} = n$.  In other words,
the expectation of players' estimates is correct, but they do not know
the exact number of players.  We demonstrate that players bid more
frequently because of this uncertainty, leading to extra revenue for
the auctioneer.

If all players think that there are $n$ players, 
we have the indifference condition
$$b = (v - p)(1-\nu_q),$$ 
where here $\nu_q$ is the the perceived probability that any other player bids.
Let $\beta_1$ be the probability that a specific player who is not the leader bids.
As $1 - \nu_q = (1-\beta_1)^{n-1}$, we have $\beta_1$ is the solution to  
$$(1-\beta_1)^{n-1} = \frac{b}{v-p}.$$
In the setting where players believe the number of players is governed
by a distribution, we have the same indifference condition.
However, because of the uncertainty, if we let $\beta_2$ be the probability
that a specific player who is not the leader bids, we find 
$$1 - \nu_q = \sum_i z_i (1-\beta_2)^{i-1},$$
since the probability that no other player bids is now given by a mixture
based on the probability distribution.
Hence in this setting
$$\sum_i z_i (1-\beta_2)^{i-1} = \frac{b}{v-p}.$$

We now give a convexity argument to show that $\beta_2 \geq \beta_1$;  
that is, there is more bidding with uncertainty.  
Consider 
\[
f(\beta) = \sum_{i} z_{i} (1 - \beta)^{i-1}.
\]
and
\[
g(\beta) = (1 - \beta)^{n-1}.
\]
Note that both $f$ and $g$ are decreasing in $\beta$. Furthermore, $(1-\beta)^{x}$ is a
convex function in $x$ for $\beta \in [0,1]$.
Hence 
$$f(\beta) = \sum_{i} z_{i} (1 - \beta)^{i-1} \geq (1 - \beta)^{\sum_i z_i(i-1)}
= (1 - \beta)^{n-1} = g(\beta).$$
Now by definition of $\beta_2$ and $\beta_1$, 
$f(\beta_2) = b/(v-p) = g(\beta_1) \leq f(\beta_1)$. 
As $f$ is decreasing
in $\beta$, 
from $f(\beta_2) \leq f(\beta_1)$, we have $\beta_2 \geq
\beta_1$, as desired.

\ifnum\tr=1
\begin{figure}[t]
   \centering
   \includegraphics[scale=0.55]{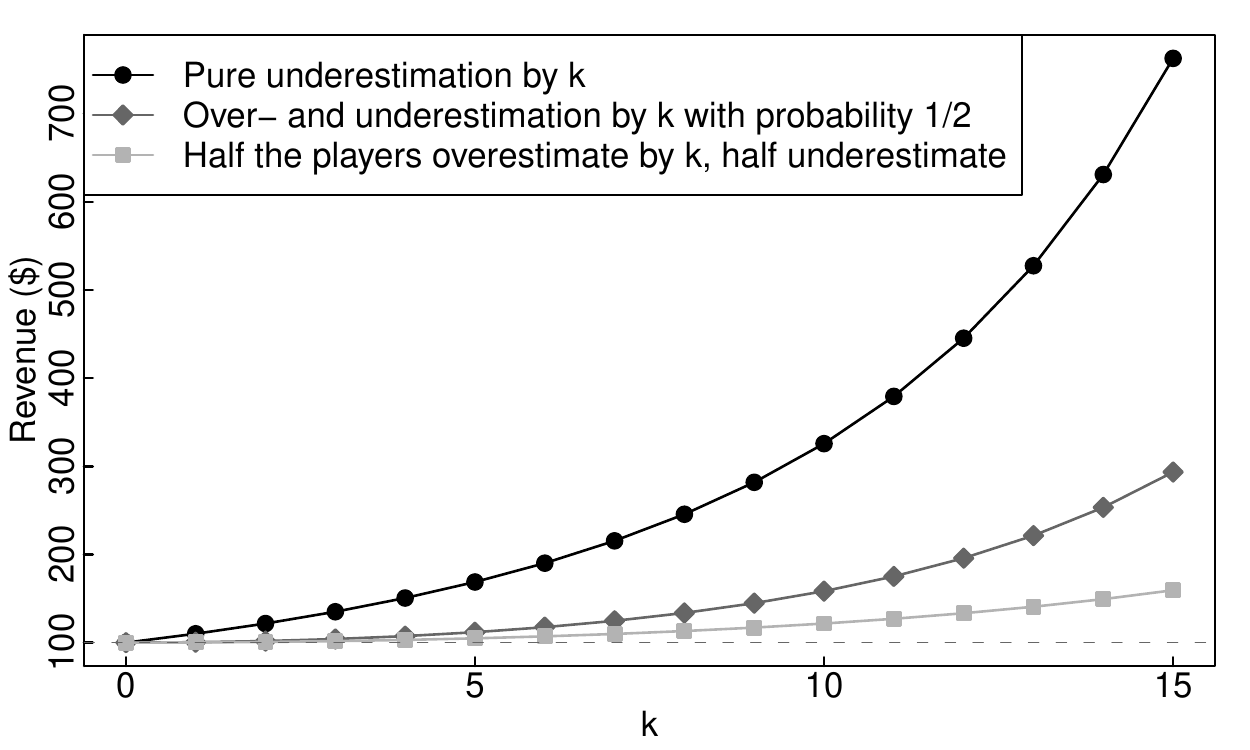} 
   \caption{Expected revenue for Swoopo when players are uncertain about the population.
}
   \label{fig:vary-pm-n-probabilistic}
\end{figure}

Figure~\ref{fig:vary-pm-n-probabilistic} provides a comparison of three settings, where in all
cases the number of players is $n=50$: 1) all players underestimate the
number of players by $k$;  2) with probability $1/2$, all players believe 
that there are $n-k$ players, otherwise all believe that there are $n+k$ players; and
3) half the players believe there are $n-k$ players and 
half believe there are $n+k$ players.  We see that for large
misestimates in this setting, uncertainty provides more benefit for
the auctioneer than asymmetry.
\fi

\ifnum\ec=1
\vspace{-3 mm}
\fi
\ifnum\tr=1
\section{Extended Models for Asymmetries}
\fi
\ifnum\ec=1
\section{Modeling General Asymmetries}
\fi

\label{sec:model-ext}

\ifnum\tr=1

We now consider a
general framework in which full information is not available to all
players and there are asymmetries.  We describe the underlying model
in Section~\ref{sec:model-ext2}, and in
Section~\ref{sec:asymmetric-markov}, we describe the analysis approach
based on Markov chains that we utilize throughout the paper.
We briefly discuss a full information model as an issue for future
work in the conclusion.

\subsection{Modeling Information Asymmetries}

\fi

\label{sec:model-ext2}

We now consider variations of the auction where there are asymmetries
in information.  For this we need to extend the symmetric model and
make a crucial distinction between the \emph{true} values of the
game's parameters -- $v$, $b$ and $n$ -- and the way players
\emph{perceive} them.  (We motivate the misperception of each these
parameters in the appropriate sections.)  For simplicity, we will
assume henceforth that there are two groups of players: $A$, of size
$k$, and $B$, of size $n-k$.  We can extend our approach to a larger
number of groups naturally, but with increased complexity.

Players in group $A$ perceive the value of the item as $v^{A}$, the
bid fee as $b^{A}$, and the number of participants in the game as
$n^{A}$. Define $v^{B}$, $b^{B}$ and $n^{B}$ similarly.  
Initially, we assume that each player is asymmetry-unaware, 
i.e. each player assumes all players have identical parameters and thus the 
groups are not aware of each other.  We will be also interested in cases 
where one group is aware of the split and therefore has an advantage 
over the other group.  That setting will utilize the same basic structure;
we develop it in later sections.  In both settings, except in the special
case of collusion (studied in Section~\ref{sec:coalitions}), members of the 
groups are not aware of the identities of individuals in either group.  

The parameters determine both the perceived and
the true probability of the $q$th bid being placed. So, for group
$A$, let $\nu_{q}^{A}$ be the perceived probability that anyone -- in
either group -- will place the $q^{th}$ bid. In other words,
$\nu_{q}^{A}$ is an estimate of $\mu_{q}$ from the perspective of
players in group $A$. Also, define $\mu_{q}^{A}$ as the true
probability that one or more players in group $A$ places the $q$th bid
and similarly define $\nu_{q}^{B}$ and $\mu_{q}^{B}$ for group $B$. If $\mu_{q}$ is the
true probability of the $q$th bid being placed then we have
$1-\mu_{q} = (1-\mu_{q}^{A})(1-\mu_{q}^{B})$.

Players in group $A$ will bid according to their perceived
indifference condition, which for ascending-price auctions is now $(v^{A} - s(q-1))(1 - \nu_{q}^{A}) =
b^{A}$, and similarly for group $B$. (Similar derivations hold for fixed-price auctions.)
Using the fact that $1 -
\nu_{q}^{A} = (1 - \beta_{q}^{A})^{n-1}$ we can easily derive the
individual bidding probability for group $A$ players:
\begin{equation}
\beta_{q}^{A} = 1 - \left(\frac{b^{A}}{v^{A}-s(q-1)}\right)^{\frac{1}{n^{A}-1}}.
\end{equation}
The derivation for group $B$ players is identical. Using the individual bidding probabilities we can compute the probability of a bid being placed by anyone in group $A$ as
\begin{align}
1 - \mu_{q}^{A} &= (1 - \beta_{q}^{A})^{k}\\
\mu_{q}^{A} &= 1 - \left(\frac{b^{A}}{v^{A} - s(q-1)}\right)^{\frac{k}{n^{A}-1}}.
\end{align}
Note that generally $\mu_{q} \ne \nu_{q}^{A} \ne \nu_{q}^{B}$.

\subsection{A Markov Chain Approach 
\ifnum\tr=1
for Analyzing Asymmetries
\fi
}
\label{sec:asymmetric-markov}

To compute various quantities of interest when we have asymmetric
behaviors requires a bit of work, primarily because the probability of
a bid at any given time depends in part on what group the current auction
leader belongs to.  In the models we have described,
however, the auction itself is memoryless, in that, given the leader
and the current number of bids, the history to reach the current state
is unimportant to the future of the auction.  Essentially all of our
models have this form.  Hence, we can place these auctions in the
setting of Markov chains in order to efficiently calculate the
distribution of auction outcomes.

\ifnum\tr=1
Specifically, the general case for two groups of players can be
captured by an absorbing, time-inhomogeneous 
Markov chain as shown in Figure \ref{fig:two-groups-markov}.  
(Recall that in a
time-inhomogeneous Markov chain, the transition probabilities can
depend on the current time as well as the state, but not on the
history of the chain.)  The chain contains four states: in state $A$ a
member of the first group is leading the auction, while in state
$W_{A}$ the auction has been won by a member of the first group. We
define states $B$ and $W_{B}$ similarly. 
Observe that $W_{A}$ and $W_{B}$ are absorbing states.  
Also observe we overload the notation
$A$ and $B$ to refer both to the sets of players and a state of the
Markov chain; this should not cause confusion as the meaning will be
clear by context.  Finally, note that there is no state corresponding
to the initial setting prior to the first bid.
While we could have a special initial state, we instead choose the
starting state probabilistically from $A$ or $B$ according to the appropriate
probabilities for the first bid, recalling that we assume our auction
is successful.
\fi

\ifnum\ec=1
Specifically, the general case for two groups of players can be
captured by an absorbing, time-inhomogeneous 
Markov chain as shown in Figure \ref{fig:two-groups-markov}.  
(Recall that in a
time-inhomogeneous Markov chain, the transition probabilities can
depend on the current time as well as the state, but not on the
history of the chain.)  The chain contains four states: in state $A$ a
member of the first group is leading the auction, while in absorbing state
$W_{A}$ the auction has been won by a member of the first group. We
define states $B$ and $W_{B}$ similarly. 
Note that we overload the notation
$A$ and $B$ to refer both to the sets of players and a state of the
Markov chain.
Finally, note that there is no state corresponding
to the initial setting prior to the first bid.
Instead we choose the 
starting state probabilistically from $A$ or $B$ according to the appropriate
probabilities for the first bid, recalling that we assume our auction
is successful.
\fi

\ifnum\tr=1

\begin{figure}[t]
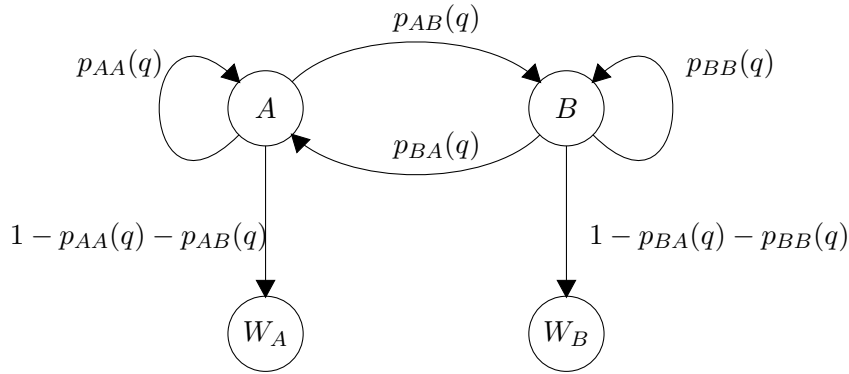

\begin{center}
\begin{pgfpicture}{0cm}{0cm}{8cm}{6cm}
\pgfnodecircle{A-leads}[stroke]{\pgfxy(2,4)}{0.5cm}
\pgfputat{\pgfrelative{\pgforigin}{\pgfnodecenter{A-leads}}}{\pgfbox[center,center]{$A$}}
\pgfnodecircle{B-leads}[stroke]{\pgfxy(6,4)}{0.5cm}
\pgfputat{\pgfrelative{\pgforigin}{\pgfnodecenter{B-leads}}}{\pgfbox[center,center]{$B$}}
\pgfnodecircle{A-wins}[stroke]{\pgfxy(2,1)}{0.5cm}
\pgfputat{\pgfrelative{\pgforigin}{\pgfnodecenter{A-wins}}}{\pgfbox[center,center]{$W_{A}$}}
\pgfnodecircle{B-wins}[stroke]{\pgfxy(6,1)}{0.5cm}
\pgfputat{\pgfrelative{\pgforigin}{\pgfnodecenter{B-wins}}}{\pgfbox[center,center]{$W_{B}$}}

\pgfsetendarrow{\pgfarrowtriangle{6pt}}
\pgfnodeconncurve{A-leads}{B-leads}{45}{135}{1cm}{1cm}
\pgfnodeconncurve{B-leads}{A-leads}{225}{315}{1cm}{1cm}
\pgfnodeconncurve{A-leads}{A-leads}{225}{135}{2cm}{2cm}
\pgfnodeconncurve{B-leads}{B-leads}{315}{45}{2cm}{2cm}
\pgfnodeconncurve{A-leads}{A-wins}{-90}{90}{1cm}{1cm}
\pgfnodeconncurve{B-leads}{B-wins}{-90}{90}{1cm}{1cm}

\pgfnodelabel{A-leads}{B-leads}[0.4][1.1cm]{$p_{AB}(q)$}
\pgfnodelabel{B-leads}{A-leads}[0.6][0.6cm]{$p_{BA}(q)$}
\pgfnodelabel{A-leads}{A-leads}[0.5][2.5cm]{$p_{AA}(q)$}
\pgfnodelabel{B-leads}{B-leads}[0.5][-1.6cm]{$p_{BB}(q)$}
\pgfnodelabel{A-leads}{A-wins}[0.65][-3.4cm]{$1-p_{AA}(q)-p_{AB}(q)$}
\pgfnodelabel{B-leads}{B-wins}[0.65][0.3cm]{$1-p_{BA}(q)-p_{BB}(q)$}

\end{pgfpicture}
\end{center}
\caption{A state-machine for an asymmetric game with two groups of players.}
\label{fig:two-groups-markov}
\end{figure}

\fi

\ifnum\ec=1

\begin{figure}[t]
\begin{center}
\begin{pgfpicture}{0cm}{0cm}{4cm}{3.2cm}
\pgfnodecircle{A-leads}[stroke]{\pgfxy(0,2)}{0.4cm}
\pgfputat{\pgfrelative{\pgforigin}{\pgfnodecenter{A-leads}}}{\pgfbox[center,center]{$A$}}
\pgfnodecircle{B-leads}[stroke]{\pgfxy(3.5,2)}{0.4cm}
\pgfputat{\pgfrelative{\pgforigin}{\pgfnodecenter{B-leads}}}{\pgfbox[center,center]{$B$}}
\pgfnodecircle{A-wins}[stroke]{\pgfxy(0,0)}{0.4cm}
\pgfputat{\pgfrelative{\pgforigin}{\pgfnodecenter{A-wins}}}{\pgfbox[center,center]{$W_{A}$}}
\pgfnodecircle{B-wins}[stroke]{\pgfxy(3.5,0)}{0.4cm}
\pgfputat{\pgfrelative{\pgforigin}{\pgfnodecenter{B-wins}}}{\pgfbox[center,center]{$W_{B}$}}

\pgfsetendarrow{\pgfarrowtriangle{6pt}}
\pgfnodeconncurve{A-leads}{B-leads}{45}{135}{1cm}{1cm}
\pgfnodeconncurve{B-leads}{A-leads}{225}{315}{1cm}{1cm}
\pgfnodeconncurve{A-leads}{A-leads}{225}{135}{1.5cm}{1.5cm}
\pgfnodeconncurve{B-leads}{B-leads}{315}{45}{1.5cm}{1.5cm}
\pgfnodeconncurve{A-leads}{A-wins}{-90}{90}{1cm}{1cm}
\pgfnodeconncurve{B-leads}{B-wins}{-90}{90}{1cm}{1cm}

\pgfnodelabel{A-leads}{B-leads}[0.4][1.1cm]{$\scriptstyle p_{AB}(q)$}
\pgfnodelabel{B-leads}{A-leads}[0.6][0.6cm]{$\scriptstyle p_{BA}(q)$}
\pgfnodelabel{A-leads}{A-leads}[0.6][2cm]{$\scriptstyle p_{AA}(q)$}
\pgfnodelabel{B-leads}{B-leads}[0.5][-1.2cm]{$\scriptstyle p_{BB}(q)$}
\pgfnodelabel{A-leads}{A-wins}[0.65][-2.4cm]{$\scriptstyle 1-p_{AA}(q)-p_{AB}(q)$}
\pgfnodelabel{B-leads}{B-wins}[0.65][0.2cm]{$\scriptstyle 1-p_{BA}(q)-p_{BB}(q)$}

\end{pgfpicture}
\end{center}
\caption{A state-machine for an asymmetric game with two groups of players.}
\label{fig:two-groups-markov}
\end{figure}

\fi

We use $p_{AB}(q)$ to denote the transition probability of going from state $A$
to state $B$ after the $q$th bid, and similarly we can define
$p_{BA}(q)$, $p_{AA}(q)$, and so on. For example, $p_{AB}(1)$ is the transition probability
from state $A$ to state $B$ when one bid has already been placed. When considering fixed-price
auctions, the bidding probabilities, and hence the state transition
probabilities, are invariant from bid to bid.  
In this special case
the Markov chain becomes time-homogeneous, and given the distribution
on the initial state we can derive analytical expressions for the
probability of terminating in state $A$ or $B$.  A good description of
this approach can be found in many standard texts; we provide a
summary based on \cite{grinstead1997} in 
\ifnum\tr=1 Appendix \ref{sec:appendix-model-ext}.  \fi
\ifnum\ec=1 the full version of the paper \cite{bmz2010}.  \fi

\ifnum\tr=1
For ascending-price auctions, which are time-inhomogeneous, 
we resort to numerical methods employing simple recurrence relations. 
This can also be useful to obtain more specific information in the case of
fixed-price auctions (or as an alternative approach for calculating various
quantities).  For example, let $P_{A}(q)$ be the probability
of being in state $A$ after $q$ bids;  here $P_{W_{A}}(q)$ represents
the probability that a player from $A$ has won the auction at some
point up to bid $q$, so that $P_{W_{A}}(q) + P_{W_{B}}(q)$ becomes 1
for an ascending-price auction when $q$ is sufficiently large and
converges to 1 for a fixed-price auction in the limit as $q$ goes to infinity.  Then we
have 
\begin{equation}
P_{A}(q+1) = P_{A}(q) p_{AA}(q) + P_{B}(q) p_{BA}(q),
\end{equation}
and other similar recurrences, including    
\begin{equation}
P_{W_A}(q+1) = P_{A}(q) p_{AW_A}(q) + P_{W_A}(q).
\end{equation}
\fi

\ifnum\ec=1
For ascending-price auctions, which are time-inhomogeneous, 
we resort to numerical methods employing simple recurrence relations. 
This can also be useful to obtain more specific information in the case of
fixed-price auctions (or as an alternative approach for calculating various
quantities).  For example, let $P_{A}(q)$ be the probability
of being in state $A$ after $q$ bids;  here $P_{W_{A}}(q)$ represents
the probability that a player from $A$ has won the auction at some
point up to bid $q$, so that $P_{W_{A}}(q) + P_{W_{B}}(q)$ becomes 1
for an ascending-price auction when $q$ is sufficiently large and
converges to 1 for a fixed-price auction as $q$ goes to infinity.  Then we
have 
\begin{equation}
P_{A}(q+1) = P_{A}(q) p_{AA}(q) + P_{B}(q) p_{BA}(q),
\end{equation}
and other similar recurrences, including    
\begin{equation}
P_{W_A}(q+1) = P_{A}(q) p_{AW_A}(q) + P_{W_A}(q).
\end{equation}
\fi

Given these various equations, it is easy to compute quantities such
as the expected revenue.  For example, in an ascending-price auction, assuming all players have a bid
fee of $b$, every time $A$ is in the
lead, he has paid a bid of $b$ for this, and the price has gone up by $s$.
Letting $R$ be the revenue, we easily find 
\begin{equation}
E[R] = (b + s) \left(1 + \sum_{i=1}^{Q} P_{A}(i) + \sum_{i=1}^{Q} P_{B}(i) \right).
\end{equation}

\ifnum\tr=1
Notice that the simple nature of the Markov chain framework allows us to derive 
all the important quantities, such as the expected revenue for Swoopo, directly
from the appropriate transition probabilities.  Hence, 
in the rest of the paper, we focus on finding these probabilities.  
Where suitable, we explicitly derive corresponding quantities, but in other cases we 
implicitly use this Markov chain characterization and leave the details to the reader.
\fi

\ifnum\ec=1
Notice that the simple nature of the Markov chain framework allows us to derive 
all the important quantities, such as the expected revenue for Swoopo, directly
from the appropriate transition probabilities.  Hence, 
in the rest of the paper, we focus on finding these probabilities, and leave
further details to the reader.
\fi

\ifnum\ec=1
\vspace{-1em}
\fi
\section{Asymmetries in Bid Fees}
\label{sec:vary-b}

\ifnum\tr=1
\subsection{Motivation}

\begin{figure}[htbp]
\centering
\subfigure[Winners' total cost of bidpacks as a percentage of retail price (\textsc{Outcomes} dataset).]
{
   \includegraphics[scale=0.55]{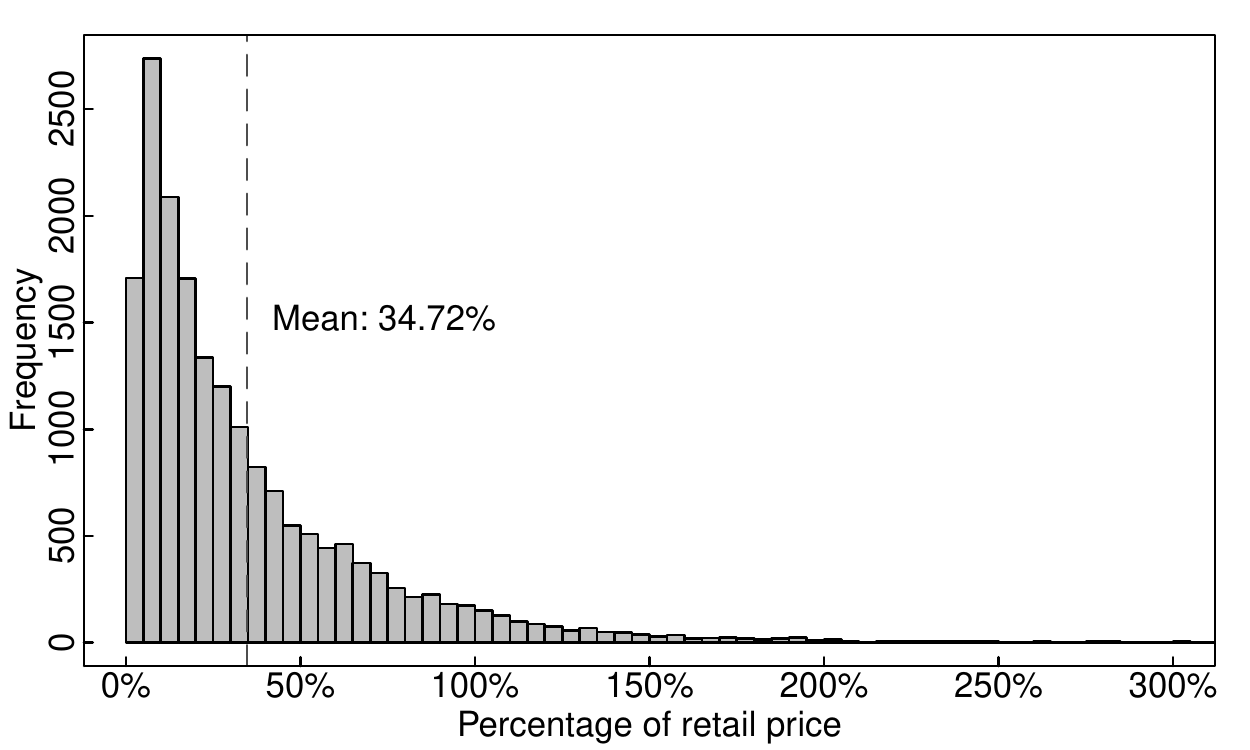} 
   \label{fig:bidpack-discount}
}
\subfigure[Winners' total cost of bidpacks as a percentage of retail price when accounting also for lost auctions
(\textsc{Trace} dataset).]
{
   \includegraphics[scale=0.55]{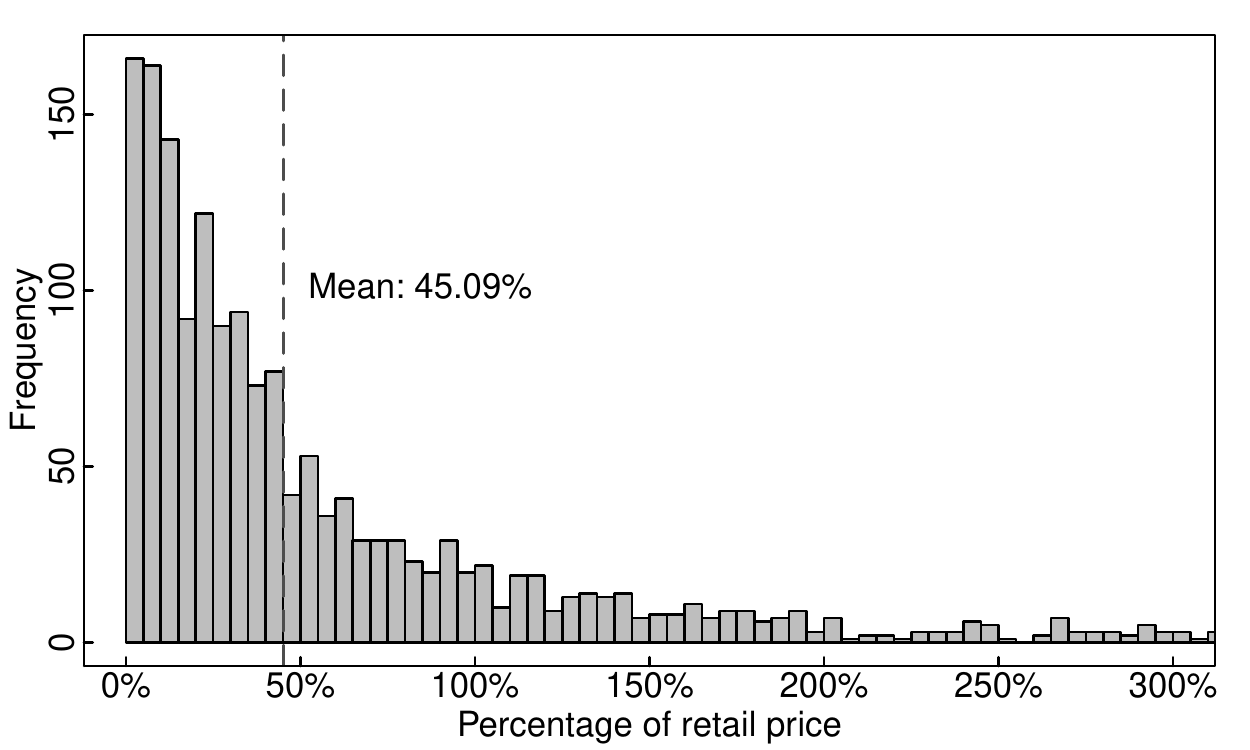} 
   \label{fig:bidpack-acquisition-cost}
}
\caption{Measuring the acquisition cost of bidpacks}
\label{fig:bidpack-cost}
\end{figure}
\fi

\ifnum\ec=1
\begin{figure}[t]
\centering
\includegraphics[scale=0.55]{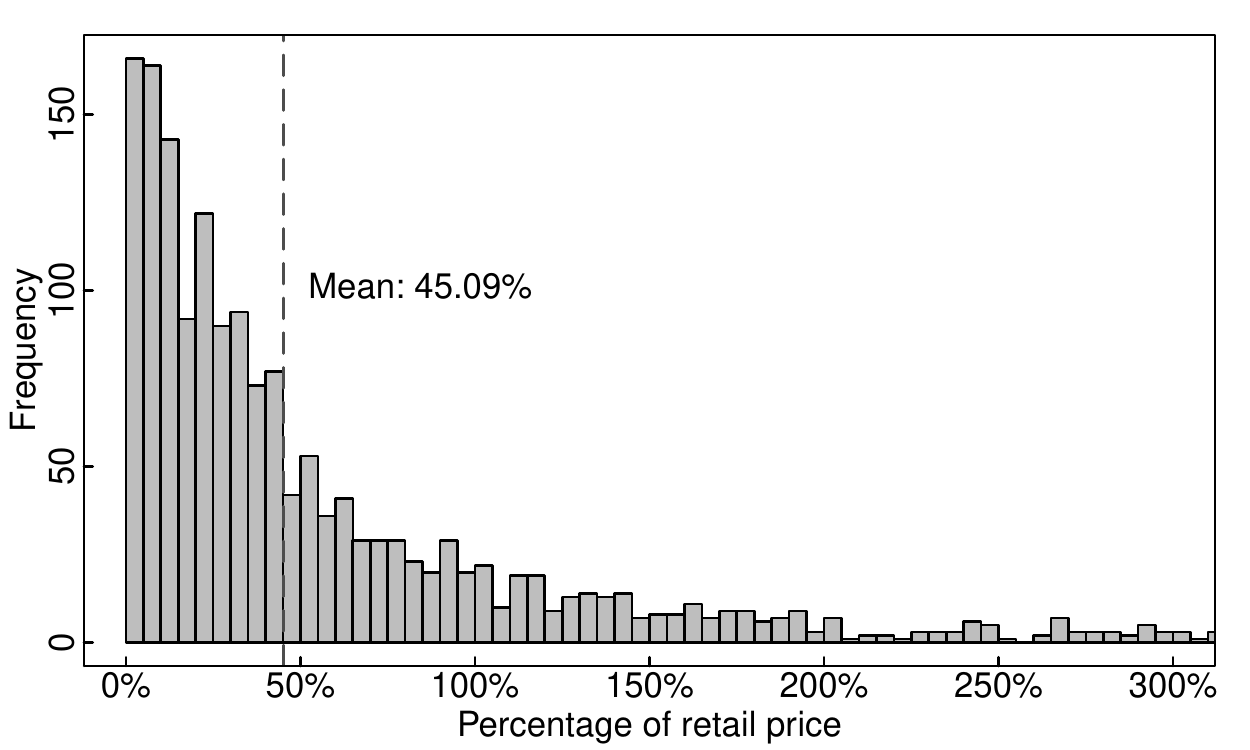} 
\caption{Winners' total cost of bidpacks as a percentage of retail price ({\sc
 Trace} dataset).}
\label{fig:bidpack-cost}
\end{figure}
\fi

\ifnum\tr=1
We now consider asymmetries that arise when players have different bid
fees.  While it seems clear that in pay-per-bid auctions players may
fail to properly estimate the population of bidders, it is less clear
why the cost per bid may vary among players, so we now motivate this direction. 

Among other items offered on auction at Swoopo are \emph{bidpacks}. A
bidpack, as its name suggests, is a set of prepaid bids. 
As of this writing they come in
five sizes: 40, 75, 150, 400, and 1000 bids. Their corresponding retail values are
\$24, \$45, \$90, \$240 and \$600, but if won in an auction they can potentially be had
for a substantial discount. 
Players who win bidpack auctions 
and
participate in later auctions can effectively enjoy lower bidding fees
compared to other participants, generally without the other
participants' knowledge.  
\fi

\ifnum\ec=1
We now consider asymmetries that arise when players have different bid
fees.  As one motivation, 
among other items offered on auction at Swoopo are \emph{bidpacks},
or sets of prepaid bids. 
Players who win bidpack auctions 
at a discount to face value
and
participate in later auctions can effectively enjoy lower bidding fees
compared to other participants, generally without the other
participants' knowledge.  
\fi

\ifnum\tr=1
To provide evidence that bidpacks can lead to varying bid fees, we
look to our data.  
Based on our \textsc{Outcomes} dataset, we derive the
average cost of a bidpack as a percentage of the nominal retail cost
for winners of bidpack auctions in Figure~\ref{fig:bidpack-discount};
this includes the winners' bid costs and the price they paid.  As can
be seen, bids can be had at a substantial discount, with the winner's total cost at just
over $1/3$ of the retail cost on average.  With the \textsc{Outcomes} dataset, however, we
can only determine the cost of bids made by the auction winners; this
clearly underestimates the costs to players who may also lose
auctions for bidpacks, raising their average cost per bid.  To attempt
to account for this, we similarly examine our \textsc{Trace} dataset, and
compute the average cost for any winner of a bidpack auction,
including the cost for bidpack auctions where the player has lost.
These results appear in Figure~\ref{fig:bidpack-acquisition-cost}.
Naturally, this leads to a smaller estimated average discount,
although the discount is still over $1/2$ of the retail cost.  While this may  
still be an underestimate of bidpack costs (as we cannot take into account auctions we have not
captured, and our results are biased towards winners) it strongly suggests that winners of bidpack auctions enjoy
a substantial discount in bid fees when applying those bids to other auctions.
\fi
\ifnum\ec=1
To provide evidence that bidpacks can lead to varying bid fees, we
estimate the total cost of bidpacks
for winners of bidpack auctions in our \textsc{Trace} dataset. 
Costs include the winners' bid costs and the prices they paid in winning auctions, 
  as well as the bid costs those winners incurred when {\em losing} other bidpack auctions 
  in our dataset.  We then plot the average cost 
of bidpacks as a percentage of the nominal retail cost in Figure~\ref{fig:bidpack-cost}.
This leads to an overall discount of over $1/2$ of the retail cost.  While this may  
still be an underestimate of bidpack costs (as we cannot take into account auctions we have not
captured, and our results are biased towards winners) it suggests that winners of bidpack auctions enjoy
a substantial discount in bid fees when applying those bids to other auctions.
\fi

\ifnum\tr=1
Cheaper bids are also available through seasonal promotions that
Swoopo conducts.  For example, Swoopo has had promotions offering 
more bids for the same price.  
A screenshot showing such a promotion
is given in Figure~\ref{fig:swoopo-bidpack-offer}. 
A further cause for variation in bid fees relates 
to the remarkable fact that Swoopo auctions take place with bidders bidding in different 
currencies.  An example of this is shown in Section~\ref{sec:vary-v}.
\fi

\ifnum\ec=1
Discounted bids are also available through seasonal promotions that
  Swoopo conducts.
Moreover, further variation in bid fees is 
 due to the remarkable fact that Swoopo auctions take place with bidders bidding in different 
currencies.  Further details are given in the full version \cite{bmz2010}.  Overall,
our evidence suggests varying bid fees are realistic in practice, and we turn to
quantifying their impact.
\fi

\ifnum\tr=1
\begin{figure}[t]
   \centering
   \includegraphics[scale=0.55]{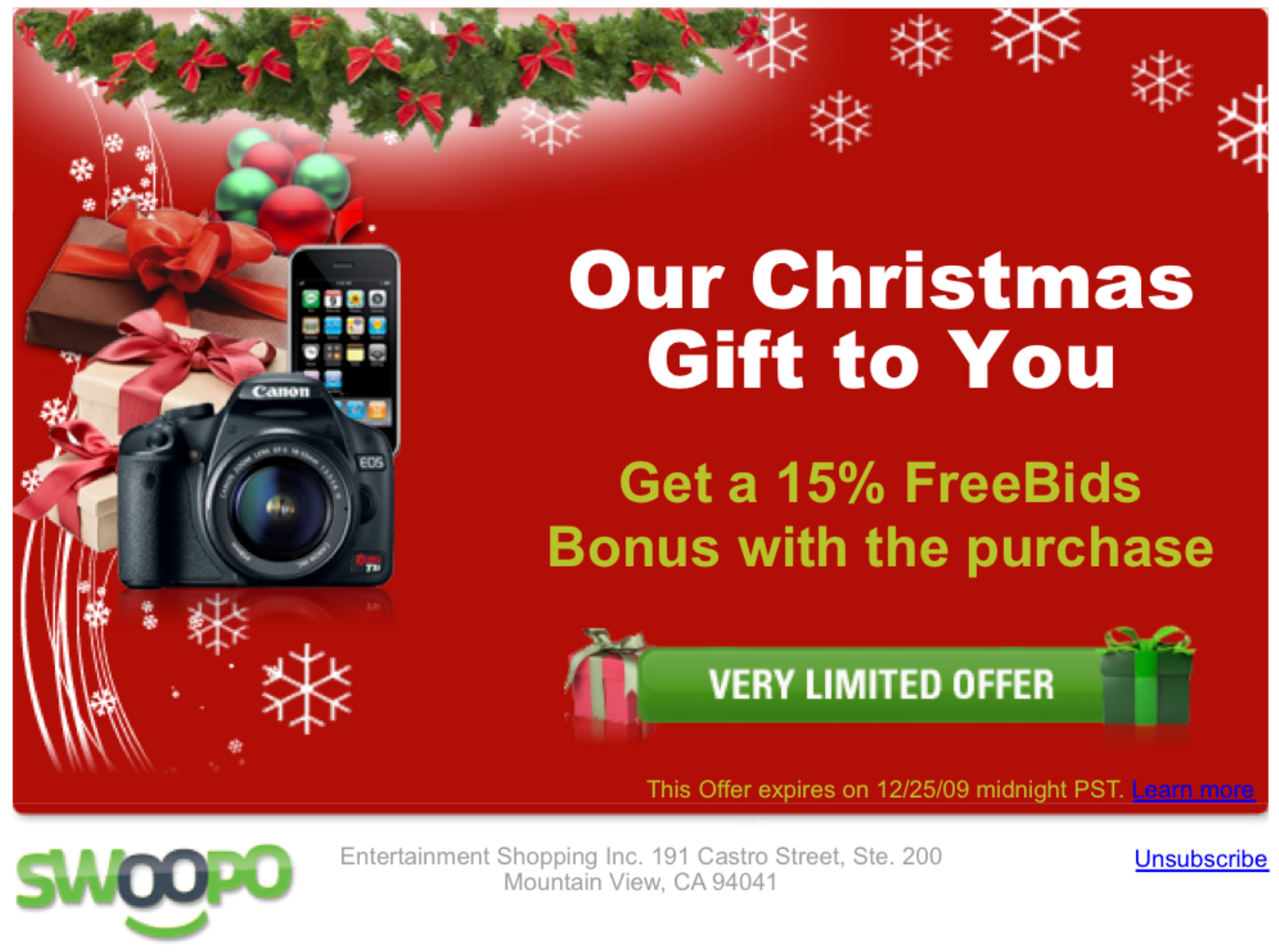} 
   \caption{A recent seasonal bidpack promotion.}
   \label{fig:swoopo-bidpack-offer}
\end{figure}
\fi

\ifnum\tr=1
Between bidpack auctions, seasonal promotions, and currency
differences, the possibility that players are paying varying bid fees 
becomes reality.  In the following we analyze and quantify the impact this has
on Swoopo auctions. 
\fi

\ifnum\tr=1
\subsection{Fixed-Price Auctions}

We first consider the simpler case of fixed-price auctions with price $p$.
We assume that the $n$ bidders are divided in two
groups $A$ and $B$, of size $k$ and $n-k$ respectively. We will assume that 
$k \ge 2$; the case where $k=1$ can be handled similarly but the case
structure of the analysis is slightly different.  Group $A$ incurs a bid fee of
$b^{A}$ while group $B$ incurs a bid fee of $b^{B}$ with $b^{A} <
b^{B}$. In context, we may presume that group $A$ is the set of
bidders who 
are bidding at a discount
whereas group $B$ is the
set of players who are charged regular bid fees.  In what follows we
also assume $A$ players are aware of the two groups while $B$ players
perceive everyone as belonging to the same group as themselves. This
creates an information asymmetry.  We believe this choice of model is
natural; we suspect many (less sophisticated) players may not recognize
that others are obtaining cheaper bids.  It also provides an example
of how our Markov chain approach of Section~\ref{sec:asymmetric-markov} applies to such a setting.

Intuitively, players in $A$, 
having access to lowered bid fees, would outlast players in group $B$ in 
expectation.  Much less obviously, it also turns out that the expected
length of the auction has {\em no} dependence on the bid fee of the $B$
players -- the auction length is solely determined by the bid fees incurred by the 
$A$ players.
\fi

\ifnum\ec=1

We consider the simpler case of fixed-price auctions with price $p$.
We assume that the $n$ bidders are divided in two
groups $A$ and $B$, of size $k$ and $n-k$ respectively. We will assume that 
$k \ge 2$; the case where $k=1$ can be handled similarly but the case
structure of the analysis is slightly different.  Group $A$ incurs a bid fee of
$b^{A}$ while group $B$ incurs a bid fee of $b^{B}$ with $b^{A} <
b^{B}$. In context, we may presume that group $A$ is the set of
bidders who 
are bidding at a discount
whereas group $B$ is the
set of players who are charged regular bid fees.  In what follows we
also assume $A$ players are aware of the two groups while $B$ players
perceive everyone as belonging to the same group as themselves. This
creates an information asymmetry.  We believe this choice of model is
natural; we suspect many (less sophisticated) players may not recognize
that others are obtaining cheaper bids.  It also provides an example
of how our Markov chain approach of Section~\ref{sec:asymmetric-markov} applies to such a setting.
\fi

Let $\mu_{q}^{A}$ be the collective probability that some player in group $A$ 
makes the $q$th bid, and similarly define $\mu_{q}^{B}$.
Then the probability that no player makes the $q$th bid is:
\begin{equation}
(1 - \mu_{q}) = (1 - \mu_{q}^{A}) (1 - \mu_{q}^{B}) \label{eqn:bid-fee-mq}
\end{equation}
where $\mu_{q}$ is defined to be the \emph{true} collective probability that anyone, in either group, bids.

Next, consider the game from the point of view of $B$ players. Remember that, according to them, everyone belongs to a single group incurring the same bid fee. Define $\nu_{q}^{B}$ as the \emph{perceived} probability that anyone, in either group, makes the $q$th bid according to the information available to $B$ players. From the indifference condition for $B$ players we have:
\begin{align}
(v - p)(1 - \nu_{q}^{B}) &= b^{B} \notag \\
\nu_{q}^{B} &= 1 - \frac{b^{B}}{v - p}.
\end{align}
We derive the \emph{true} 
probability $\beta_{q}^{B}$ that a $B$ player bids as:
\ifnum\tr=1
\begin{align}
(1 - \nu_{q}^{B}) &= (1 - \beta_{q}^{B})^{n-1} \notag \\
\beta_{q}^{B} &= 1 - (1 - \nu_{q}^{B})^{\frac{1}{n-1}} \notag \\
\beta_{q}^{B} &= 1 - \left(\frac{b^{B}}{v-p}\right)^{\frac{1}{n-1}}.
\end{align}
\fi
\ifnum\ec=1
\begin{align}
(1 - \nu_{q}^{B}) &= (1 - \beta_{q}^{B})^{n-1} \notag \\
\beta_{q}^{B} &= 1 - \left(\frac{b^{B}}{v-p}\right)^{\frac{1}{n-1}}.
\end{align}
\fi
We can then write the probability of group $B$ bidding as:
\begin{equation*}
1 - \mu_{q}^{B} = 
\begin{cases}
(1 - \beta_{q}^{B})^{n - k} & \quad \textrm{if group $A$ is leading,} \\
(1 - \beta_{q}^{B})^{n - k - 1} & \quad \textrm{if group $B$ is leading},
\end{cases}
\end{equation*}
which after manipulation becomes:
\begin{equation}
\mu_{q}^{B} = 
\begin{cases}
1 - \left(\frac{b^{B}}{v-p}\right)^{\frac{n - k}{n-1}} & \quad \textrm{if group $A$ is leading,} \\
1 - \left(\frac{b^{B}}{v-p}\right)^{\frac{n - k - 1}{n-1}} & \quad \textrm{if group $B$ is leading.} \\
\end{cases}
\label{eqn:mq-not-a}
\end{equation}
Remember that $\mu_{q}^{B}$ is the \emph{true} collective probability with which group $B$ players bid. Furthermore, notice that players in group $A$ are aware of this probability.

\begin{figure}[t]
\centering
\subfigure[Revenue as the price of cheap bids varies.]
{
   \includegraphics[scale=0.55]{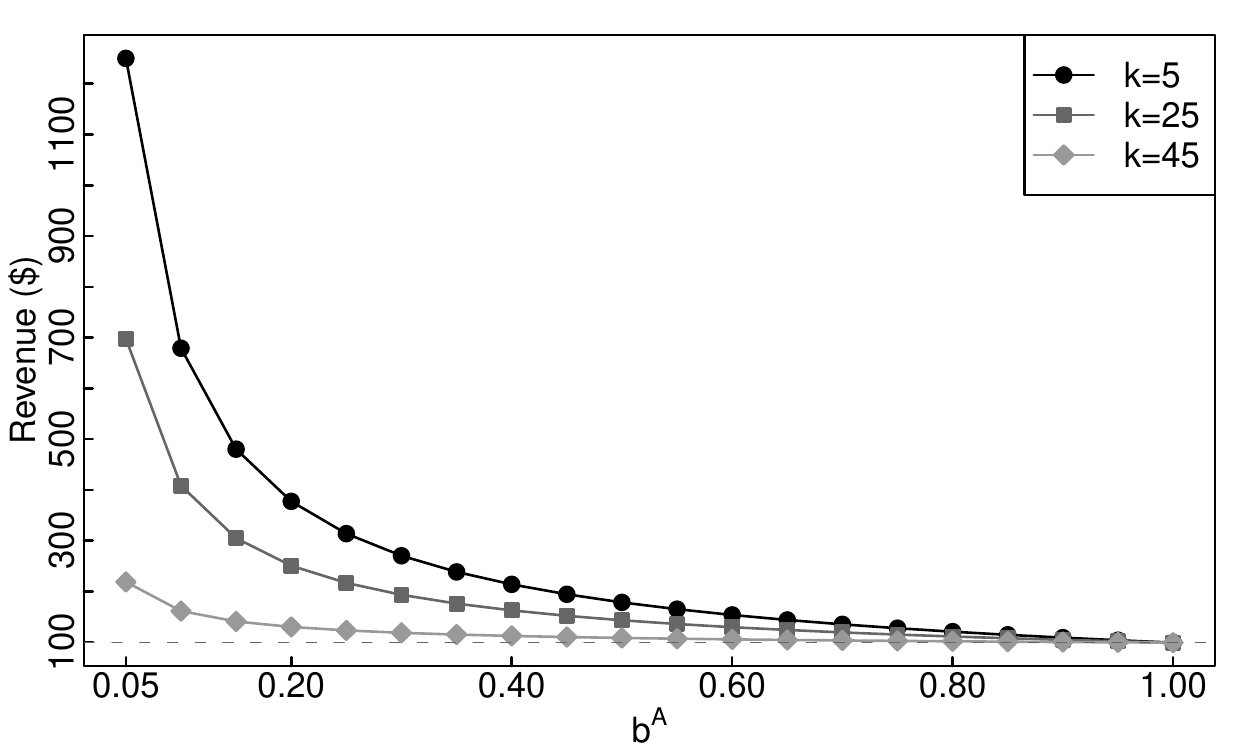} 
   \label{fig:vary-b-fixed-revenue}
}
\subfigure[Relative likelihood of a specific player in group $A$ winning the auction when compared with a specific player in group $B$.]
{
   \includegraphics[scale=0.55]{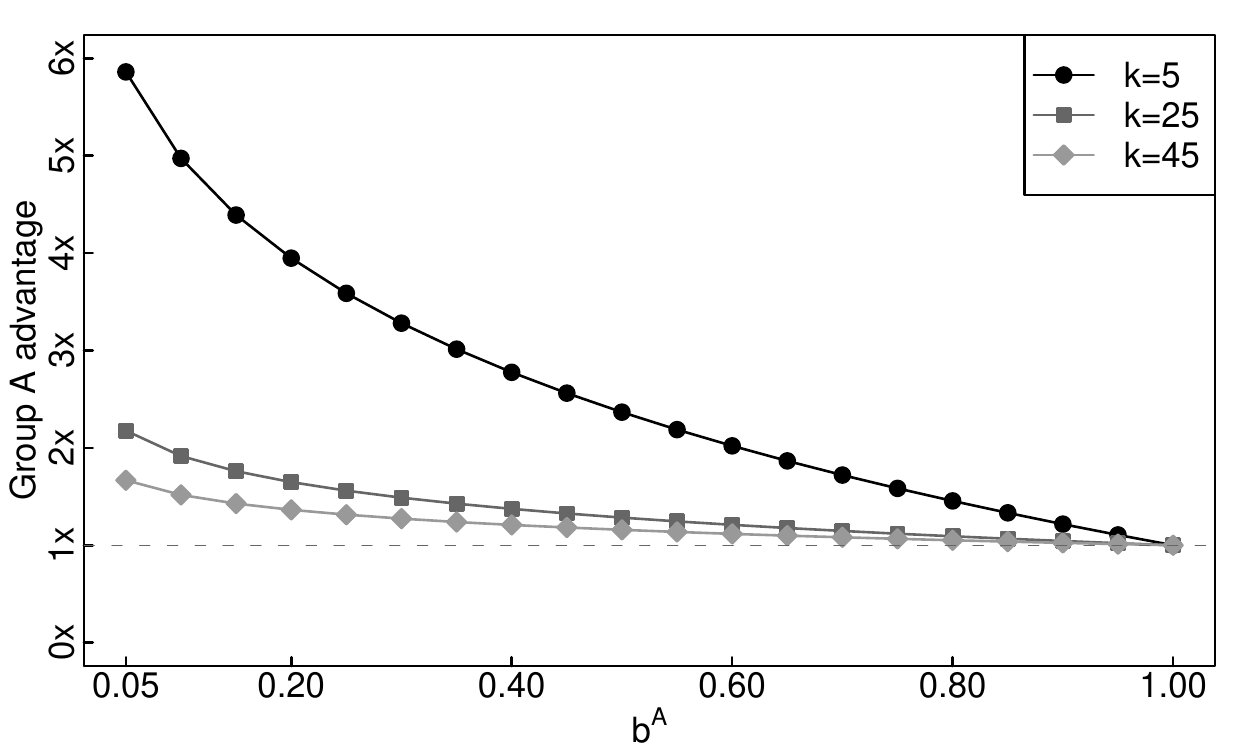} 
   \label{fig:vary-b-fixed-relative-p}
}
\caption{A fixed-priced auction with $k$ players provisioned with cheap bids; $n=50$, $v=100$ and $b^{B}=1$.}
\label{fig:vary-b-fixed}
\end{figure}

\ifnum\tr=1
Assuming the leader before the $q$th bid was from group $B$ and using the indifference condition for group $A$ we can derive an expression for $\mu_{q}^{A}$:
\begin{align}
(v - p) (1 - \mu_{q}^{A}) (1 - \mu_{q}^{B}) &= b^{A} \\
\mu_{q}^{A} &= 1 - \frac{b^{A}}{(v - p) (1 - \mu_{q}^{B})} \\
\mu_{q}^{A} &= 1 - \frac{b^{A}}{v-p} \left(\frac{v-p}{b^{B}}\right)^{\frac{n - k - 1}{n-1}} \\
\mu_{q}^{A} &= 1 - \frac{b^{A}}{b^{B}} \left(\frac{b^{B}}{v-p}\right)^{\frac{k}{n-1}}.
\end{align}
\fi
\ifnum\ec=1
Assuming the leader before the $q$th bid was from group $B$ and using the indifference condition for group $A$ we can derive an expression for $\mu_{q}^{A}$:
\begin{align}
(v - p) (1 - \mu_{q}^{A}) (1 - \mu_{q}^{B}) &= b^{A} \\
\mu_{q}^{A} &= 1 - \frac{b^{A}}{b^{B}} \left(\frac{b^{B}}{v-p}\right)^{\frac{k}{n-1}}.
\end{align}
\fi
The derivation for group $A$ leading is similar, leading to:
\begin{equation}
\mu_{q}^{A} = 
\begin{cases}
1-\frac{b^{A}}{b^{B}} \left(\frac{b^{B}}{v-p}\right)^{\frac{k-1}{n-1}} & \quad \textrm{if group $A$ is leading,} \\
1-\frac{b^{A}}{b^{B}} \left(\frac{b^{B}}{v-p}\right)^{\frac{k}{n-1}} & \quad \textrm{if group $B$ is leading.} \\ 
\end{cases}
\label{eqn:mq-a}
\end{equation}
Next, using Equation \ref{eqn:bid-fee-mq} we can derive an expression for $\mu_{q}$, the \emph{true} probability that a $q$th bid is placed:
\begin{equation}
\mu_{q} = 1 - \frac{b^{A}}{v-p},
\end{equation}
which holds irrespectively of who is the current leader. 
It seems counterintuitive that neither the number of $B$ players nor their
bid fee play any role in determining the probability $\mu_{q}$.
However, this is similar to the original setting where
all players pay the same bid fee, and $\mu_{q}$ was independent of $n$.

\ifnum\tr=1
We can also write an expression for $\beta_{q}^{A}$:
\begin{equation}
\beta_{q}^{A} = 
\begin{cases}
1 - \left(\frac{b^{A}}{b^{B}}\right)^{\frac{1}{k-1}} \left(\frac{b^{B}}{v-p}\right)^{\frac{1}{n-1}} & \quad \textrm{if group $A$ is leading,} \\
1 - \left(\frac{b^{A}}{b^{B}}\right)^{\frac{1}{k}} \left(\frac{b^{B}}{v-p}\right)^{\frac{1}{n-1}} & \quad \textrm{if group $B$ is leading.} \\
\end{cases}
\label{eqn:beta-q-a}
\end{equation}
\fi

\ifnum\ec=1
We can also write an expression for $\beta_{q}^{A}$:
\begin{equation*}
\beta_{q}^{A} = 
\begin{cases}
1 - \left(\frac{b^{A}}{b^{B}}\right)^{\frac{1}{k-1}} \left(\frac{b^{B}}{v-p}\right)^{\frac{1}{n-1}} & \quad \textrm{if group $A$ is leading,} \\
1 - \left(\frac{b^{A}}{b^{B}}\right)^{\frac{1}{k}} \left(\frac{b^{B}}{v-p}\right)^{\frac{1}{n-1}} & \quad \textrm{if group $B$ is leading.} \\
\end{cases}
\end{equation*}
\fi

\ifnum\tr=1
Having computed the individual bid probabilities for each player we can compute Swoopo's expected revenue in
successful auctions using the framework we developed in Section \ref{sec:asymmetric-markov}. 
Consider our usual fixed-price auction with $n=50$, $b=1$ and $p=0$.
Some bidders have access to a discounted bid fee $b^{A}$, while the rest pay the regular 
  rate of \$1 per bid. Figure \ref{fig:vary-b-fixed-revenue} displays Swoopo's excepted revenue 
  as the fee $b^{A}$ charged to group $A$ bidders for bidding varies. 
\fi
\ifnum\ec=1
With these bid probabilities in hand, we can apply 
the framework developed in Section \ref{sec:asymmetric-markov}. 
Consider our usual fixed-price auction with $n=50$, $b=1$ and $p=0$.
Some bidders have access to a discounted bid fee $b^{A}$, while the rest pay the regular 
  rate of \$1 per bid. Figure \ref{fig:vary-b-fixed-revenue} displays Swoopo's excepted revenue 
  as the fee $b^{A}$ charged to group $A$ bidders for bidding varies. 
\fi

\ifnum\tr=1
The expected revenue per successful Swoopo auction actually increases,
superlinearly, in the gap between bid fees.  This is somewhat
surprising, given that the amount of revenue from each bid from group
$A$ is {\em decreasing}.  We provide a high-level explanation of what
appears to be happening.  Group $B$ bidders not only pay full price
for their bids, but are also participating in an auction that
tends to last substantially longer than they expect. Furthermore, 
they bid with higher probability than they would if
they had complete information.  Consequently Swoopo's revenues
explode. Of course our analysis hinges on the assumption that group
$B$ bidders never realize that they have been dealt a losing hand;
recall for fixed-price auctions the underlying bidding behavior is
memoryless.  But, during an actual auction, Swoopo does not reveal
whether a player makes a discounted bid, making it hard for players to
assess how level the playing field is, and thus also making our model
plausible.  (We leave extensions to a setting where players change their
beliefs about other auction players as the auction proceeds, and
change their strategy accordingly, as future work.)

Also of interest is the advantage gained by a specific player having
access to cheap bids. Using the same example as above, Figure
\ref{fig:vary-b-fixed-relative-p} displays the relative likelihood of a
specific $A$ player winning the auction compared to a specific $B$ player winning the
auction as a function of the discounted bid fee. 
Observe that there is a clear synergy here: provisioning of
cheaper bids helps the players who receive them and the auctioneer
alike.
\fi

\ifnum\ec=1
The expected revenue per successful Swoopo auction actually increases,
superlinearly, in the gap between bid fees.  This is somewhat
surprising, given that the amount of revenue from each bid from group
$A$ is {\em decreasing}.  However, Group $B$ bidders not only pay full price
for their bids, but are also participating in an auction that
tends to last substantially longer than they expect. 
Consequently Swoopo's revenues increase as well. 
Our analysis hinges on the assumption that group
$B$ bidders never realize that they have been dealt a losing hand;
recall for fixed-price auctions the underlying bidding behavior is
memoryless.  During actual auctions, Swoopo does not reveal bid costs,
making our model plausible.  (Extending our model to a setting where 
players' beliefs about other players evolve as the auction proceeds, and
then adapt their strategies, is future work.)

Also of interest is the advantage gained by a specific player having
access to cheap bids. Using the same example as above, Figure
\ref{fig:vary-b-fixed-relative-p} displays the relative likelihood of a
specific $A$ player winning the auction compared to a specific $B$ player 
as a function of the discounted bid fee. 
There is a clear synergy here: provisioning of
cheaper bids helps the players who receive them and the auctioneer
alike.
\fi

\ifnum\tr=1
\subsection{A Single Player with Access to Cheap Bids $(k=1)$}

The analysis above depends on the size of the $A$ group of bidders being larger than one. If $k=1$, then Equations \ref{eqn:mq-not-a} and \ref{eqn:mq-a} do not hold any more. For completeness we provide the details for this case.  (The uninterested reader may skip ahead.)
In this case we have:
\begin{equation}
\mu_{q}^{B} = 
\begin{cases}
1 - \left(\frac{b^{B}}{v-p}\right) & \quad \textrm{if group $A$ is leading,} \\
1 - \left(\frac{b^{B}}{v-p}\right)^{\frac{n - 2}{n-1}} & \quad \textrm{if group $B$ is leading,} \\
\end{cases}
\label{eqn:mq-not-a-m1}
\end{equation}
and
\begin{equation}
\mu_{q}^{A} = 
\begin{cases}
0 & \quad \textrm{if group $A$ is leading,} \\
1-\frac{b^{A}}{b^{B}} \left(\frac{b^{B}}{v-p}\right)^{\frac{1}{n-1}} & \quad \textrm{if group $B$ is leading.} \\
\end{cases}
\label{eqn:mq-a-m1}
\end{equation}
%

Using equation \ref{eqn:bid-fee-mq} we can calculate the probability that a bid is placed as:
\begin{equation}
\mu_{q} = 
\begin{cases}
1 - \frac{b^{B}}{v-p} & \quad \textrm{if group $A$ is leading,} \\
1 - \frac{b^{A}}{v-p} & \quad \textrm{if group $B$ is leading.} \\
\end{cases}
\label{eqn:mq-m1}
\end{equation}
Notice that, unlike the case where $k > 1$, the probability of the
auction ending depends on which group is currently leading the
auction. This affects the duration of the auction and consequently
Swoopo's revenues. Auctions are 
substantially shorter when there is only one player with access to cheap bids,
and similarly yield much less revenue.  
Figure \ref{fig:vary-b-k1-revenue} displays Swoopo's revenue when
$k=1$, all other parameters being unchanged.
The relative advantage of a sole player with cheap bids increases
dramatically, however, an effect demonstrated in Figure \ref{fig:vary-b-k1-relative-p}.

\begin{figure}[t]
\centering
\subfigure[Swoopo's revenue as the price of cheap bids varies.]
{
   \includegraphics[scale=0.55]{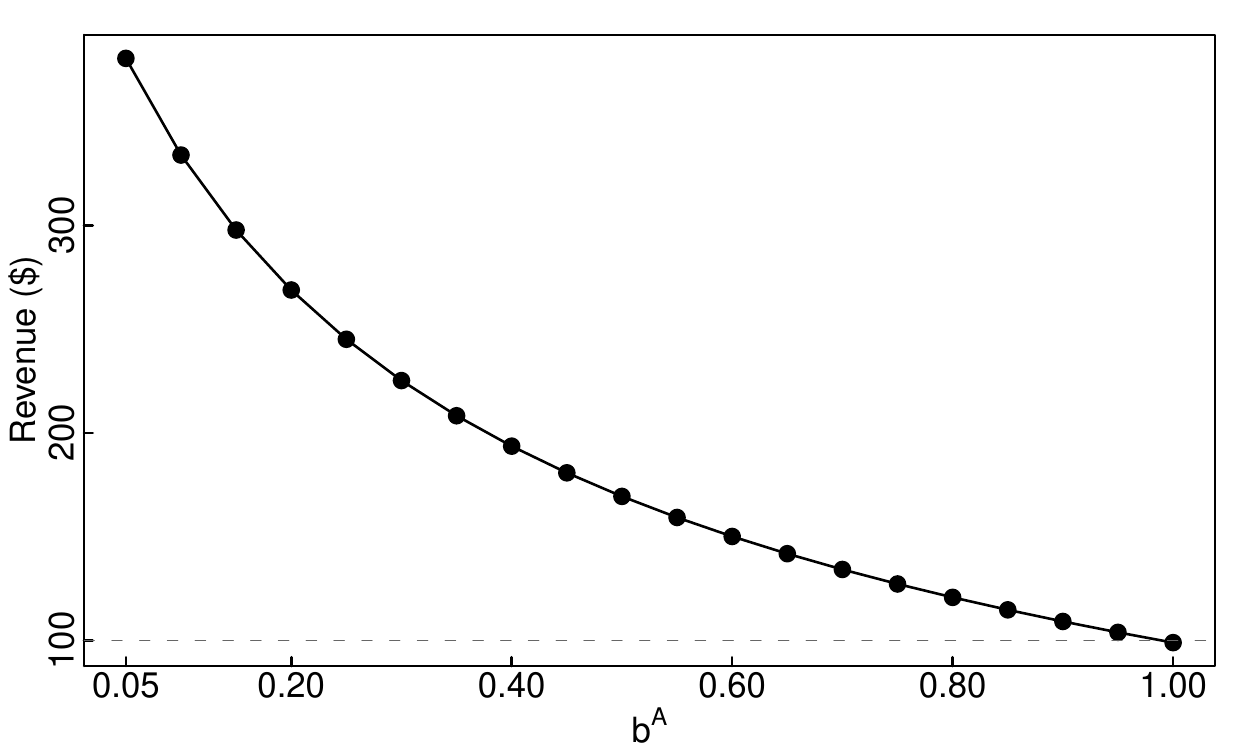} 
   \label{fig:vary-b-k1-revenue}
}
\subfigure[Relative likelihood of the single player in group $A$ winning the auction when compared with a specific player in group $B$.]
{
   \includegraphics[scale=0.55]{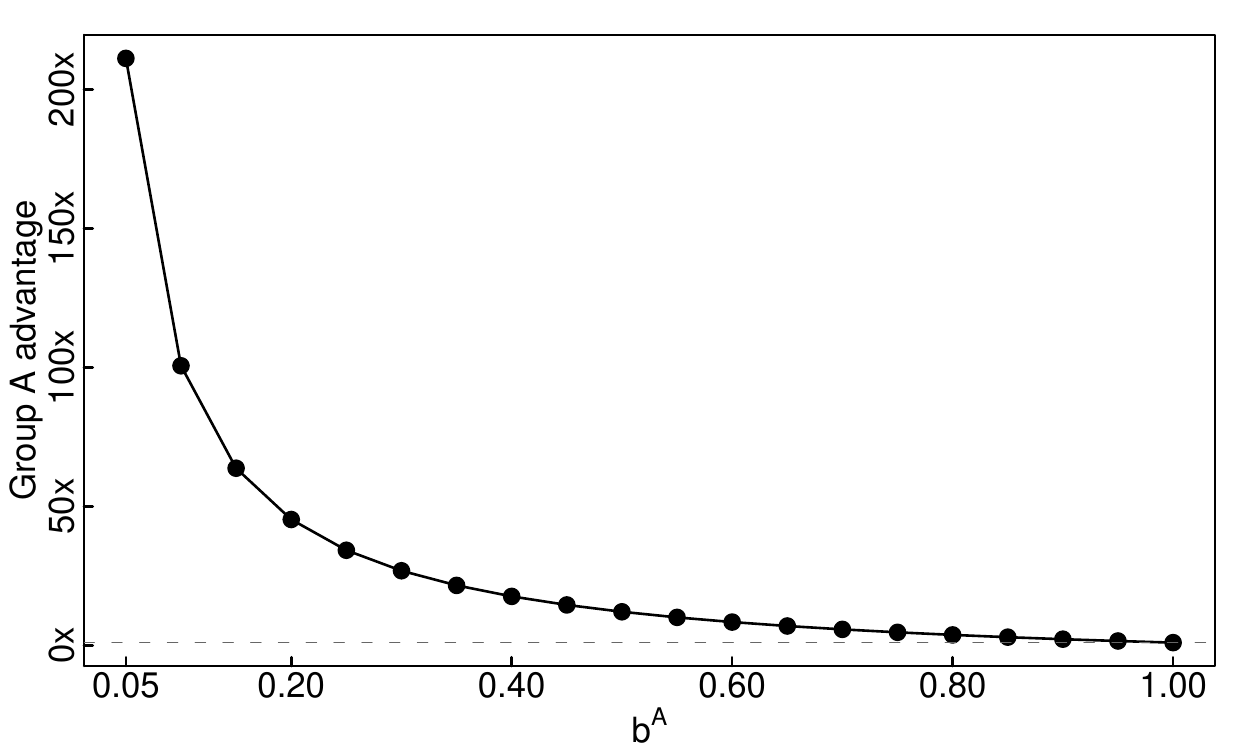} 
   \label{fig:vary-b-k1-relative-p}
}
\caption{A fixed-priced auction with a single player provisioned with cheap bids; $n=50$, $v=100$ and $b^{B}=1$.}
\label{fig:vary-b-k1}
\end{figure}

\fi


\ifnum\tr=1
\subsection{Ascending-Price Auctions}

Using the methods of Section \ref{sec:model-ext2}, we can also compute
the expected revenue of an ascending-price auction with varying bid
fees. Sample results using a price increment of $s = 0.25$ are shown
in Figure \ref{fig:vary-b-ascending-revenue}. Figure
\ref{fig:vary-b-ascending-relative-p} displays the relative likelihood
of a specific player in group $A$ winning the auction compared to a specific player in
group $B$ as the price of the cheaper bid varies.  The results are
similar, although as one might expect, the gains to Swoopo are smaller
in this setting.  

\begin{figure}[htbp]
\centering
\subfigure[Swoopo's revenue as the price of cheap bids varies.]
{
   \includegraphics[scale=0.55]{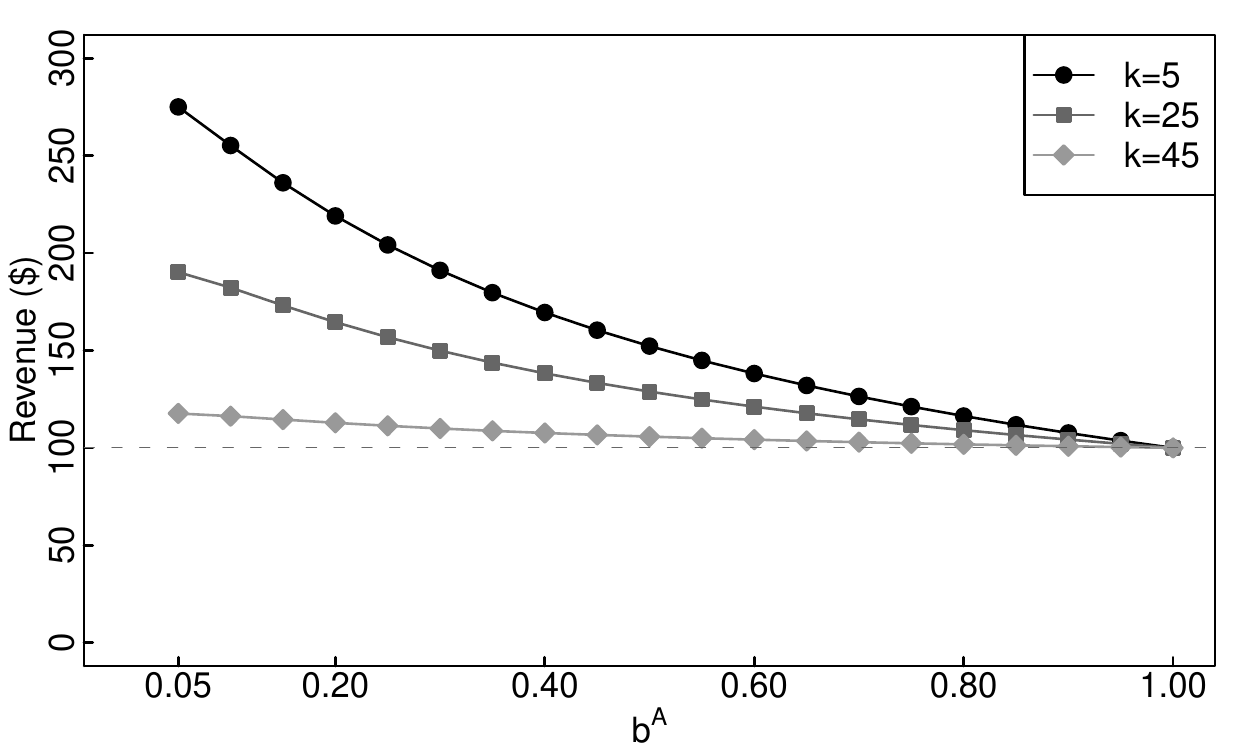} 
   \label{fig:vary-b-ascending-revenue}
}
\subfigure[Relative likelihood of a specific player in group $A$ winning the auction when compared with a specific player in group $B$.]
{
   \includegraphics[scale=0.55]{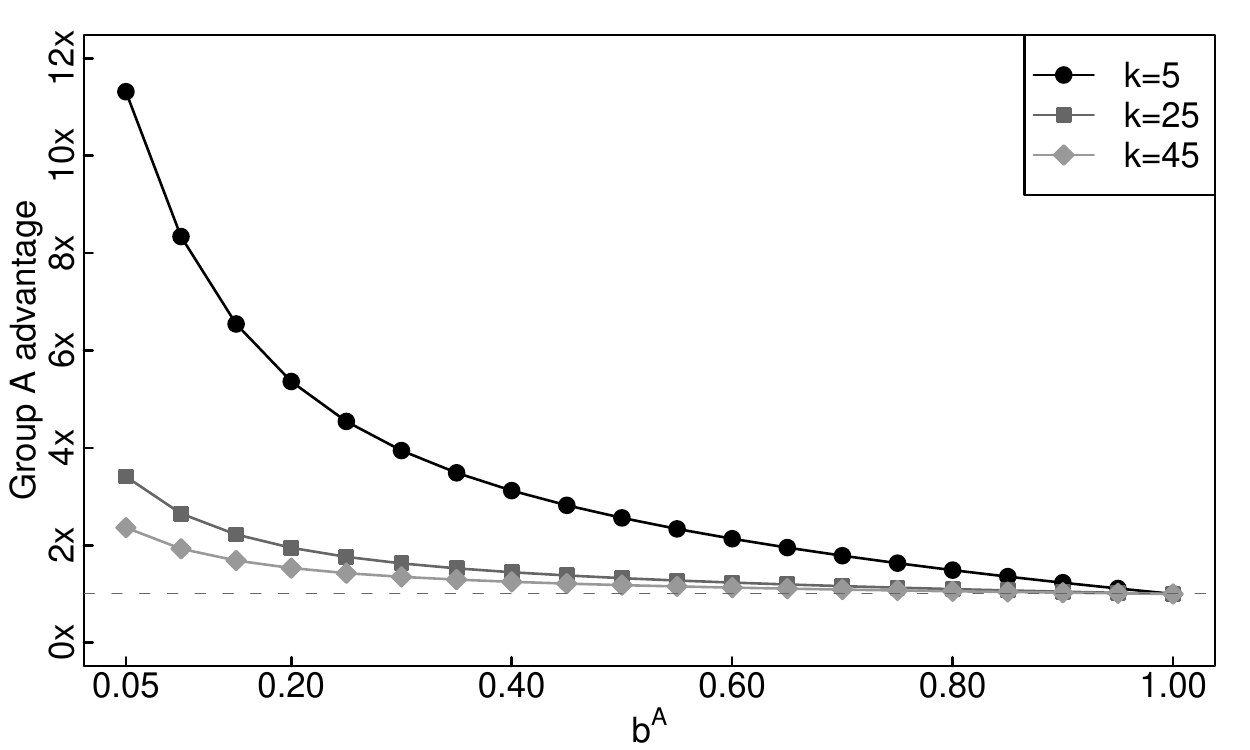} 
   \label{fig:vary-b-ascending-relative-p}
}
\caption{Ascending-price auctions with varying bid fees; $n=50$, $v=100$, $b^{B}=1$ and $s=0.25$.}
\label{fig:vary-b-ascending}
\end{figure}

\fi

\ifnum\ec=1
We point out before continuing that, using the same approach, we can
also analyze the setting where players have differing intrinsic
valuations for the item being auctioned~\cite{bmz2010}. 
\fi

\ifnum\tr=1
\section{Varying Object Valuations}
\label{sec:vary-v}

We now consider the impact of varying the perceived value $v$ of the
auctioned item among the players.  There are multiple motivations for
this consideration.  A first motivation is that people may simply
value the item differently.  In particular, Swoopo provides a nominal
retail value for the auction item, which is generally significantly
higher than the purchase price one could easily obtain elsewhere (such
as on Amazon).  There may therefore be na\"ive players who
base their valuation on the nominal retail price and more
sophisticated players who know the actual retail price of an item.
(This motivation is touched on briefly in \cite{augenblick2009},
although our analysis is markedly different.)

Another more surprising motivation is that Swoopo runs its auctions
simultaneously in multiple countries.  That is, the participants in an
auction often correspond to players in different countries, bidding on
the local version of the Swoopo site.  This necessarily introduces
small inconsistencies due to the use of different currencies, so bids
as well as valuations are likely to differ in some small degree.  (One
dollar is not a fixed whole number of euros.)  But larger variations
often occur because the auction corresponds to different items in
different countries.  This is not without justification -- a certain
type of TV, or monitor, or other electronic device generally would
only work in its country of origin, so close substitutes have to be
found for different countries.  
\ifnum\tr=1
Figure~\ref{fig:auction-261695} shows
an actual example, with screenshots of an auction with the same
auction ID and the same bidders at the US and German Swoopo sites,
with different currencies and different items with different
valuations up for auction.  
\fi
This strongly suggests that the case of different valuations 
is present in real auctions.

\ifnum\tr=1
\begin{figure}[htbp]
\centering
\subfigure[US auction.]
{
   \includegraphics[scale=0.35]{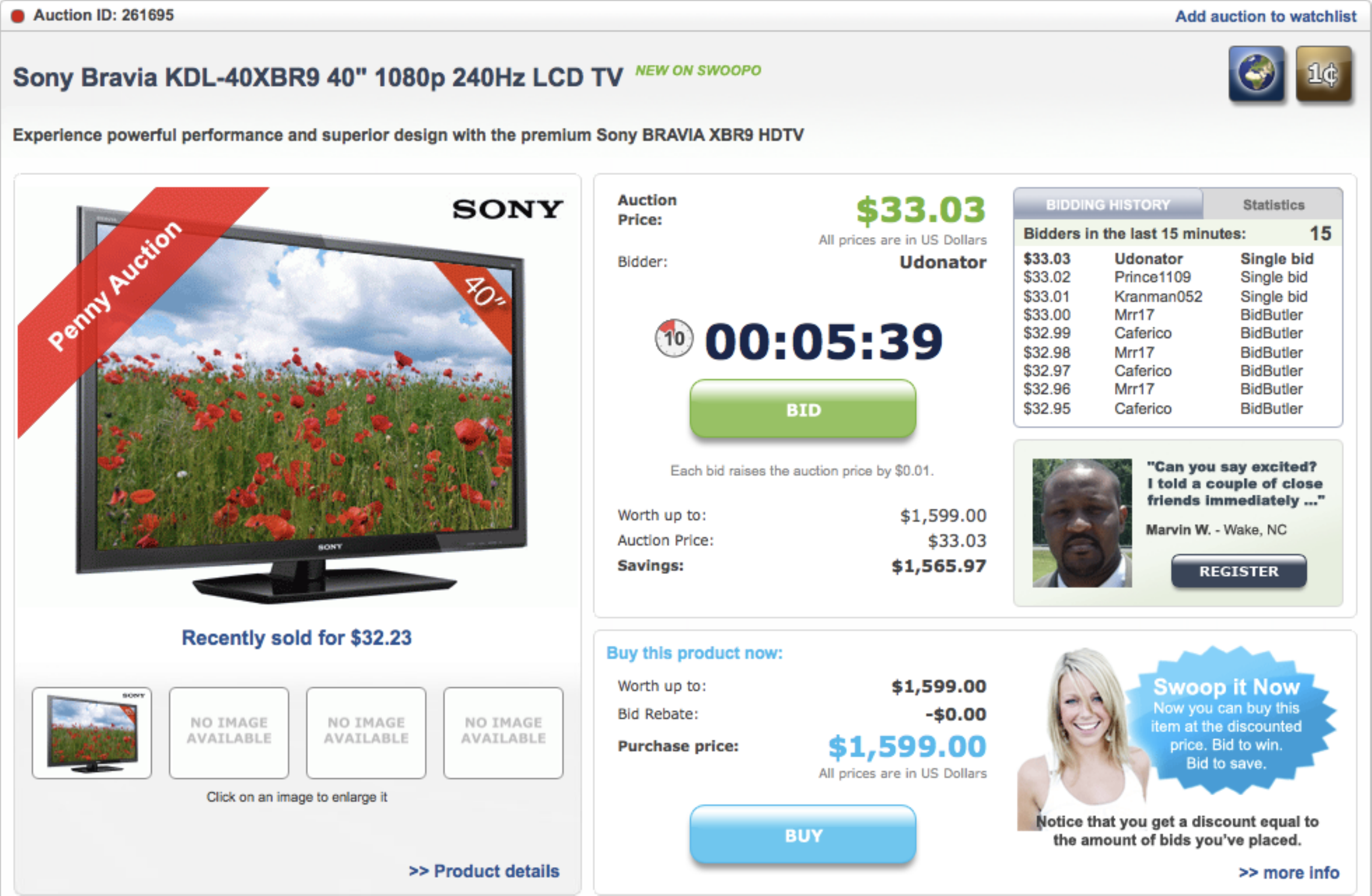} 
   \label{fig:auction-261695-us}
}
\subfigure[German auction.]
{
   \includegraphics[scale=0.35]{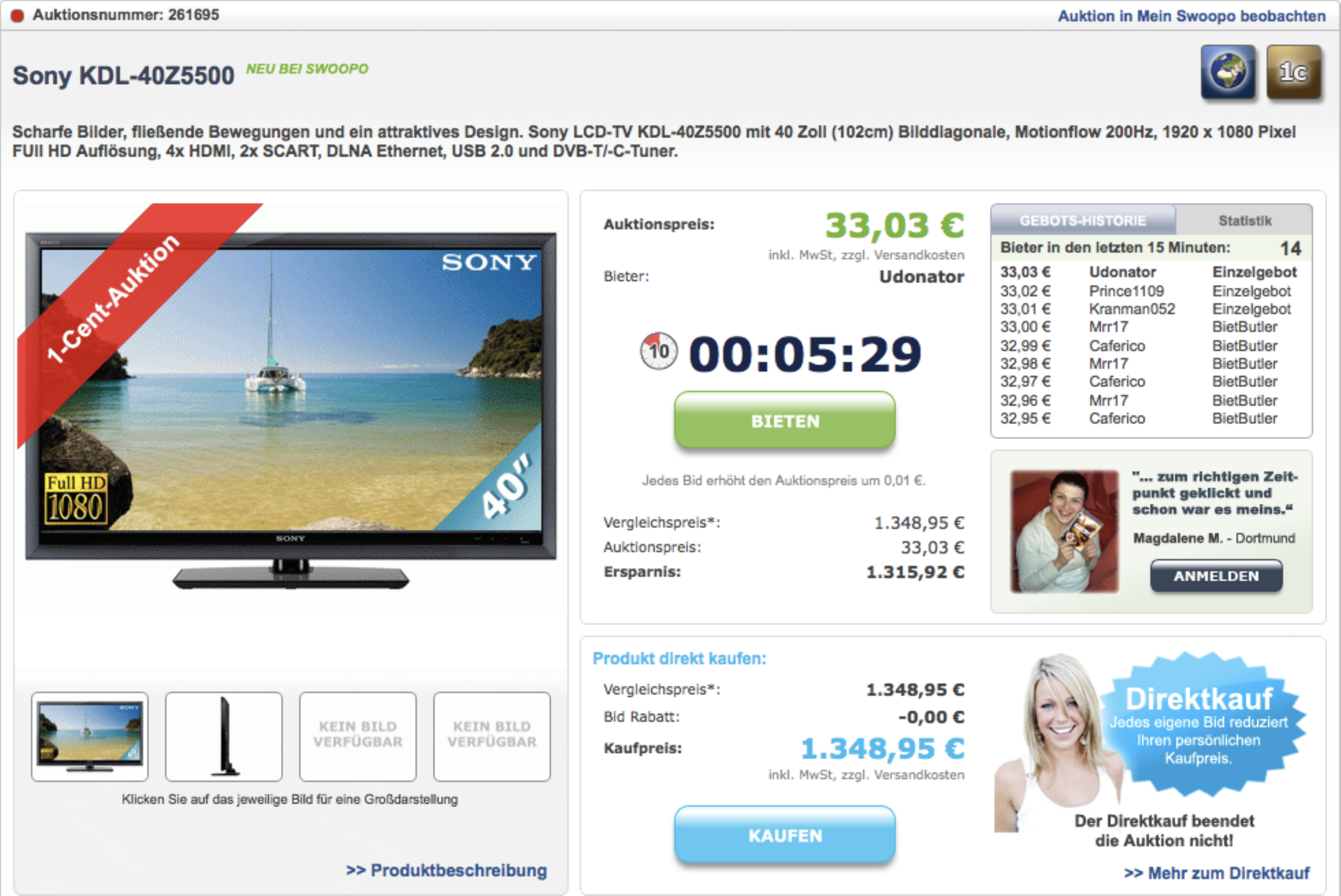} 
   \label{fig:auction-261695-de}
}
\caption{The same auction as seen from the US and German versions of the Swoopo website.}
\label{fig:auction-261695}
\end{figure}
\fi

\ifnum\tr=1
\subsection{Fixed-Price Auctions}
\fi

Motivated by the example of the same auction in different countries,
we consider the following model.  There are two groups of bidders.
Group $A$ of size $k$ values the auctioned object at $\alpha v$ where
$\alpha \in (0,\infty)$. Group $B$ of size $n-k$ values the same object at
$v$. Furthermore, the members of each group perceive everyone as
having the same valuation as themselves. The number of players, $n$, is globally known. 

\ifnum\tr=1
The indifference condition for group $A$ is $(1 - \nu^{A}_{q})(\alpha
v - p) = b$ where $\nu_{q}^{A}$ is again the \emph{perceived} probability
that anyone makes the $q$th bid according to players in
group $A$.
This implies that
\begin{equation}
\nu_{q}^{A} = 1 - \frac{b}{\alpha v-p}.
\end{equation}
Similarly, for group $B$ we have
\begin{equation}
\nu_{q}^{B} = 1 - \frac{b}{v-p}.
\end{equation}

We know that $(1 - \nu^{A}_{q}) = (1 - \beta_{q}^{A})^{n-1}$, and similarly for group $B$.
This implies that the \emph{true} probabilities with which individual players in the two groups bid are
\begin{align}
\beta^{A}_{q} &= 1 - \left(\frac{b}{\alpha v-p}\right)^{\frac{1}{n-1}}, \\
\beta^{B}_{q} &= 1 - \left(\frac{b}{v-p}\right)^{\frac{1}{n-1}} .
\end{align}
Furthermore, the \emph{true} collective probabilities of a bid being placed by either group are
\begin{align}
\mu^{A}_{q} &= 
\begin{cases}
1 - (\frac{b}{\alpha v-p})^{\frac{k}{n-1}} & \quad \textrm{if the current leader is in group $A$,}\\
1 - (\frac{b}{\alpha v-p})^{\frac{k-1}{n-1}} & \quad \textrm{if the current leader is in group $B$,}\\
\end{cases}\\
\mu^{B}_{q} &= 
\begin{cases}
1 - (\frac{b}{v-p})^{\frac{n-k-1}{n-1}} & \quad \textrm{if the current leader is in group $A$,}\\
1 - (\frac{b}{v-p})^{\frac{n-k}{n-1}} & \quad \textrm{if the current leader is in group $B$,}\\
\end{cases}
\end{align}
Then the auction termination probability is $(1-\mu_{q}) = (1-\mu_{q}^{A})(1-\mu_{q}^{B})$.
\fi

\skipthis{
Figure \ref{fig:different-values-revenue-k} displays Swoopo's revenue
for an auction $k$ players value the auctioned item at $\frac{v}{2}$
and $n-k$ players at $v$. The size of the former group of players,
$k$, varies on $x-$axis.
}

Figure \ref{fig:different-values-revenue-alpha} displays Swoopo's
revenue as the valuation parameter $\alpha$ varies from $5\%$ to
$200\%$ for three different values of $k$ with $n=50$ and $b=1$.
The results are natural;  the more players that overvalue an item,
the better for Swoopo.  Unlike some of the other variations
we have studied, the effects on the revenue are naturally bounded:
if the maximum valuation of an item among all players is $2v$, the expected revenue in
this model will not exceed $2v$.  Similar results hold for ascending-price auctions.

\begin{figure}[t]
   \centering
   \includegraphics[scale=0.55]{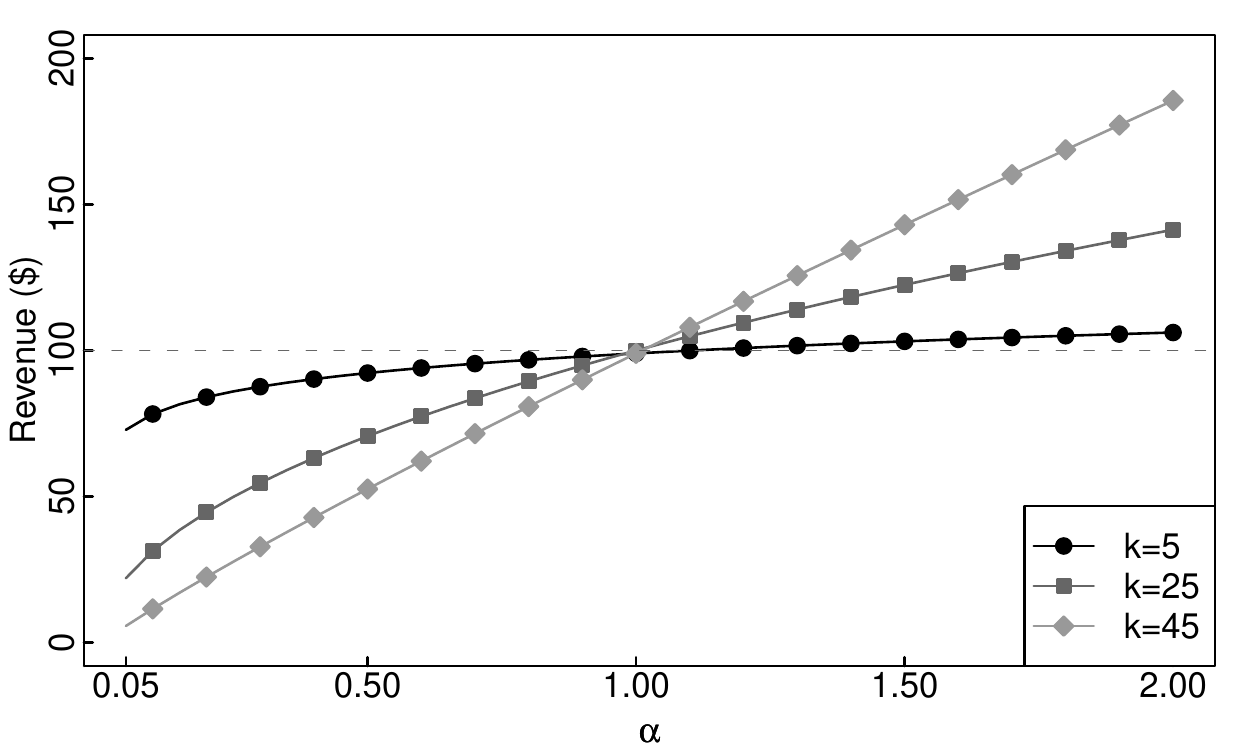} 
   \caption{Revenue for Swoopo as the valuation parameter $\alpha$ varies. Shown for three different values of $k$ and $n=50$, $v=100$, $b=1$ in a fixed-price auction.}
   \label{fig:different-values-revenue-alpha}
\end{figure}

\fi

\ifnum\tr=1
\section{Hidden Information:  Collusion and Shill Bidders}
\fi

\ifnum\ec=1
\vspace{-2em}
\section{Collusion and Shill Bidders}
\fi

\ifnum\tr=1
The analysis for varying the number of players gives us a simple
framework for analyzing two other standard situations with information
asymmetry, namely where certain players have hidden information. 
In the first setting, we consider the case where a subset of players 
collude to form a bidding coalition.  In the second, we consider
{\em shill bidders}, or bidders in the employ of the auctioneer.
\fi

\ifnum\ec=1
Our previous analysis allows us to consider other standard situations with information
asymmetry due to hidden information. 
Here we examine the setting where a subset of players 
collude to form a bidding coalition, and a setting with 
{\em shill bidders}, or bidders in the employ of the auctioneer.
\fi

\subsection{Collusion}
\label{sec:coalitions}

\ifnum\old=1
In the setting of collusion, the natural approach is for the members of the
coalition to agree to not bid against each other, so that if one of them
is currently leading the auction the others bid with zero probability,
and as a group they can choose to make a single bid if they are not the
leader. Conceptually, we can think of the coalition as a single player
or as an agent acting on behalf of the coalition members.  The benefit
of colluding here is that the collusion can potentially intimidate
other players from bidding, by making it appear that there are more
players in the auction than there actually are, thereby reducing the
probability that other players bid.  We wish to quantify the advantage
gained by this form of collusion in terms of the size of the
coalition.

\subsubsection{Fixed-Price Auctions}
\fi

\ifnum\tr=1
In the setting of collusion, the natural approach is for the members
of the coalition to agree to not bid against each other, so that if
one of them is currently leading the auction the others bid with zero
probability.  The agents can either behave independently, or we can
think of the coalition as a single player or agent acting on
behalf of the coalition members.  A benefit of colluding here is
that the coalition can potentially intimidate other players from
bidding, by making it appear that there are more players in the
auction than there actually are, thereby reducing the probability that
other players bid.  We wish to quantify the advantage gained by this
form of collusion in terms of the size of the coalition.

\subsubsection{Fixed-Price Auctions}
\fi

\ifnum\ec=1
A natural approach for collusion is for members of a coalition to
agree to not bid against each other, so that if one of them is
currently leading the auction, the others bid with zero probability.
We wish to quantify the advantage gained by this form of collusion in
terms of the size of the coalition.

\fi

For our analysis, we assume that there is a group $A$ of $k$ players in a coalition,
and a group $B$ of $n-k$ other players not in the coalition.  To these $n-k$ players,
there appear to be $n$ identical players in the auction. 
Again, the coalition players bid as usual, provided a coalition member is not the leader. 
Proceeds and expenses are shared equally between coalition members.

Non-coalition members bid according to their \emph{perceived} indifference condition:
\begin{equation}
\nu_{q}^{B} = 1 - \frac{b}{v-p}.
\end{equation}
This yields 
\begin{align}
\ifnum\tr=1
1 - \nu_{q}^{B} &= (1 - \beta_{q}^{B})^{n-1} \\
\fi
\beta_{q}^{B} &= 1 - \left( \frac{b}{v-p} \right)^{\frac{1}{n-1}}.
\label{eqn:collusion-individual-bid-B}
\end{align}
{F}rom this we can derive the \emph{true} probability of a bid by group $B$:
\begin{align}
\mu_{q}^{B} = 
\begin{cases}
1 - \left(\frac{b}{v-p}\right)^{\frac{n-k}{n-1}} & \quad \textrm{if group $A$ is leading,} \\
1 - \left(\frac{b}{v-p}\right)^{\frac{n-k-1}{n-1}} & \quad \textrm{if group $B$ is leading.} \\
\end{cases}
\end{align}
We observe that players in group $B$, just as a consequence of overestimating the total population, bid less frequently than they should. This fact alone is enough for the coalition of players in group $A$ to gain an edge in winning the auction.

\ifnum\tr=1
Next, we look at the indifference condition for a player in group $A$
when someone from group $B$ is leading the auction, as otherwise group
$A$ players do not bid. Remember that both costs and proceeds are
shared and that players in $A$ have full information:
\begin{align}
\frac{1}{k}b &= \frac{1}{k}(v - p) (1 - \mu_{q}) \\
b &= (v - p) (1 - \mu_{q}) \\
b &= (v - p) (1 - \mu_{q}^{A})(1-\mu_{q}^{B}) \\
\mu_{q}^{A} &= 1 - \frac{b}{(v-p)(1-\mu_{q}^{B})}\\
\mu_{q}^{A} &= 1 - \left( \frac{b}{v-p} \right)^{\frac{k}{n-1}}.
\end{align}
We can derive the individual bidding probabilities for coalition members at equilibrium.  
Recall as we stated earlier when group $B$ is leading the auction group $A$ players act independently.  Hence
\begin{equation}
\beta_{q}^{A} =
\begin{cases}
0 & \quad \textrm{if group $A$ is leading,} \\
1 - \left(\frac{b}{v-p}\right)^{\frac{1}{n-1}} & \quad \textrm{if group $B$ is leading.}
\end{cases}
\label{eqn:collusion-individual-bid-A}
\end{equation}
\fi

\ifnum\ec=1
Next, we look at the indifference condition for a player in group $A$
when someone from group $B$ is leading the auction:
\begin{align}
b &= (v - p) (1 - \mu_{q}^{A})(1-\mu_{q}^{B}) \\
\mu_{q}^{A} &= 1 - \left( \frac{b}{v-p} \right)^{\frac{k}{n-1}}.
\end{align}
Recall as we stated earlier when group $B$ is leading the auction group $A$ players act independently.  Hence
\begin{equation}
\beta_{q}^{A} =
\begin{cases}
0 & \quad \textrm{if group $A$ is leading,} \\
1 - \left(\frac{b}{v-p}\right)^{\frac{1}{n-1}} & \quad \textrm{if group $B$ is leading.}
\end{cases}
\label{eqn:collusion-individual-bid-A}
\end{equation}
\fi

Finally, using the fact that $1-\mu_{q}=(1-\mu_{q}^{A})(1-\mu_{q}^{B})$, we can derive the following expression for the probability of a bid being placed by either group:
\begin{equation}
\mu_{q}=
\begin{cases}
1-\left(\frac{b}{v-p}\right)^{\frac{n-k}{n-1}} & \quad \textrm{if group $A$ is leading,}\\
1-\frac{b}{v-p} & \quad \textrm{if group $B$ is leading.}
\end{cases}
\end{equation}
The increased chances of group $A$ winning the auction are apparent, as the auction is more likely to end when $A$ leads.

\ifnum\tr=1
Equations \ref{eqn:collusion-individual-bid-B} and
\ref{eqn:collusion-individual-bid-A} are nearly sufficient to
determine the probabilities for our Markov chain analysis.  The only
remaining issue regards our choice of tie-breaking rule.  
In our
description thus far, members of the coalition behave independently
when bidding, and hence if several members of the coalition bid, we
would expect each to have a chance to become the leader.  We refer to this
as the coalition having \emph{many bidders}. We could
instead imagine the coalition acting essentially as a single player
controlling many identities and only selecting a single one to use at
each opportunity to bid (albeit with an upwards adjusted probability of bidding,
i.e., $\mu_{q}^{A}$ instead of $\beta_{q}^{A}$).  In this case, 
the coalition would be less likely to win in case of ties.
\fi

\ifnum\ec=1
Equations \ref{eqn:collusion-individual-bid-B} and
\ref{eqn:collusion-individual-bid-A} are nearly sufficient to
determine the probabilities for our Markov chain analysis.  The only
remaining issue regards our choice of tie-breaking rule.  
Notice that a  highly optimized coalition could act as a single player 
  controlling many identities, only selecting a single one to use at 
  each opportunity to bid (albeit with higher probability 
   $\mu_{q}^{A}$ instead of $\beta_{q}^{A}$).  
In this case, the coalition would be less likely to win in case of ties.
We refer to this as a {\em single bidder} coalition, and the
  original, independent case as a {\em many bidder} coalition.
\fi

One would expect two consequences of collusion.  First, 
a coalition of $k$ bidders should have more than $k$ times the
probability of a non-colluding bidder to win. Second, 
the overestimation of the actual player population should negatively 
impact Swoopo's revenues.  We confirm both of these consequences
empirically.

\begin{figure}[t]
\centering
\subfigure[Revenue for Swoopo when collusion occurs.]
{
   \includegraphics[scale=0.55]{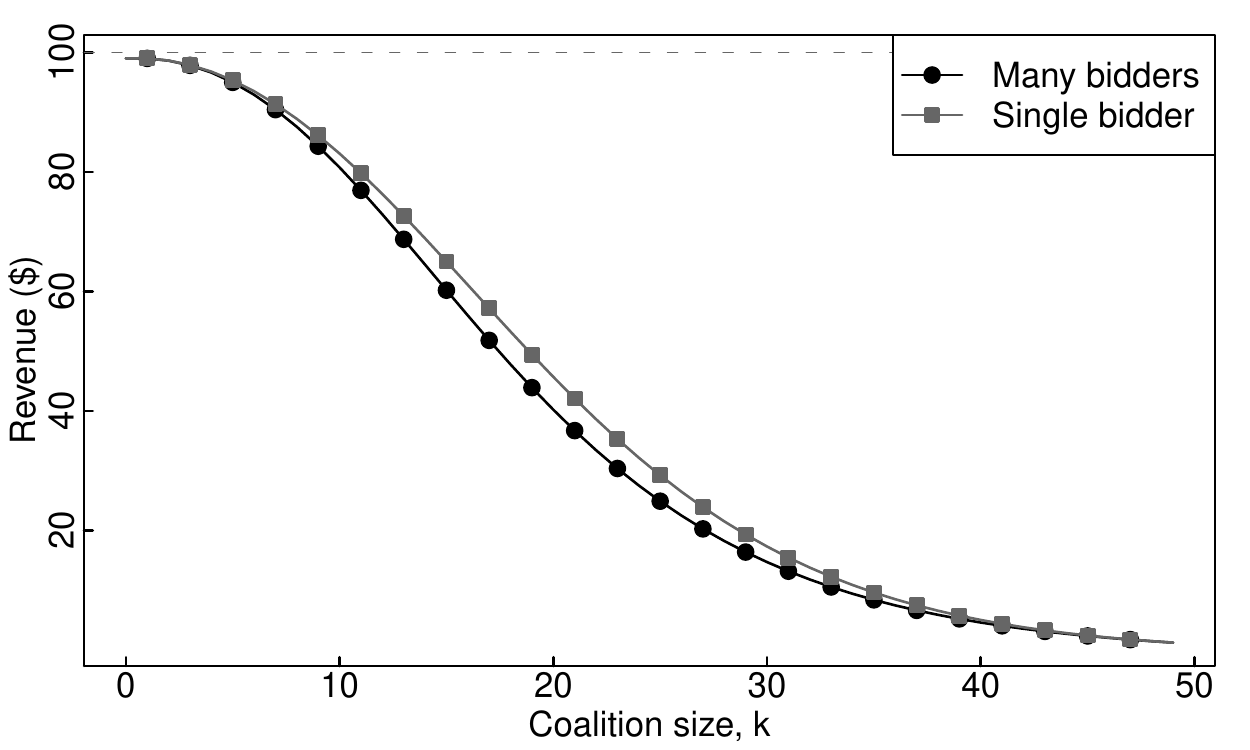} 
   \label{fig:collusion-revenue}
}
\subfigure[Relative likelihood of the colluding group winning the auction vs. any specific outsider.]
{
   \includegraphics[scale=0.55]{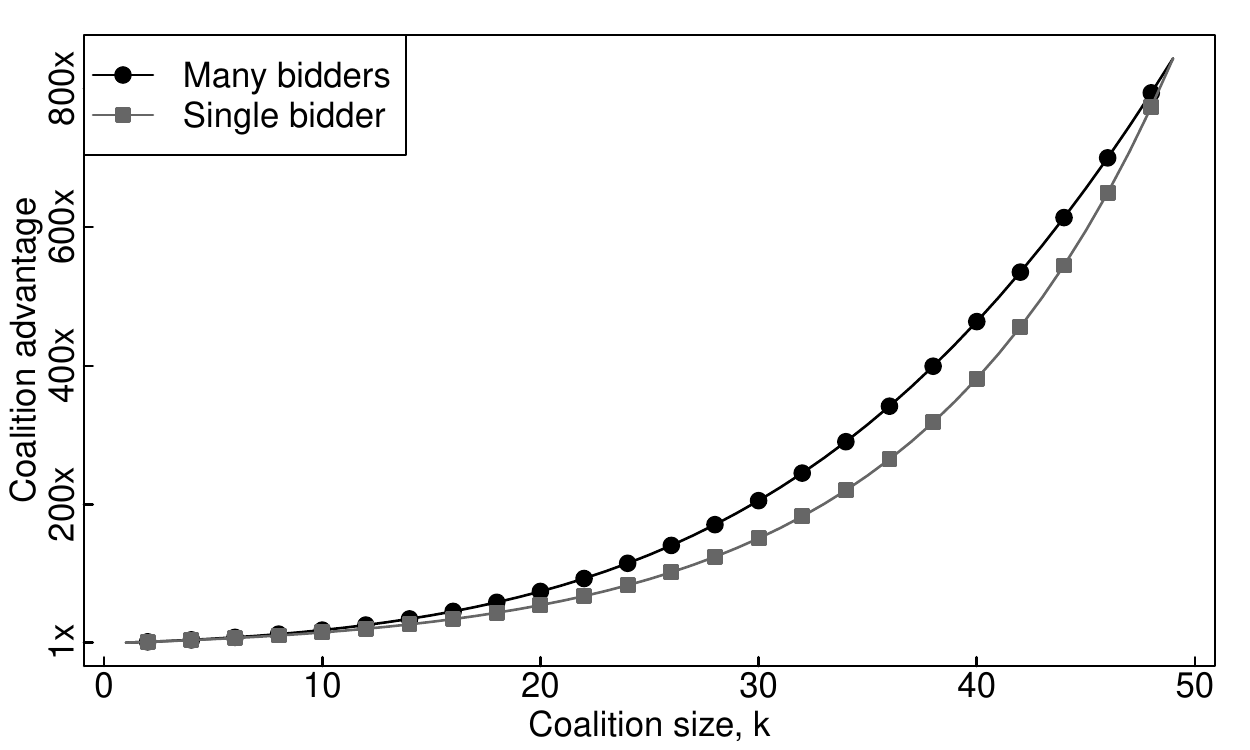} 
   \label{fig:collusion-relative-p}
}
\caption{Fixed-price auctions with a coalition of size $k$; $n=50$, $v=100$ and $b=1$.}
\label{fig:collusion-fixed}
\end{figure}

Figure \ref{fig:collusion-revenue} displays the revenue Swoopo can
expect in the presence of a coalition of size $k$ for both
tie-breaking rules. As can be seen, revenue declines significantly when 
large coalitions are present.  Figure \ref{fig:collusion-relative-p} displays the
relative likelihood of the coalition winning the auction  compared to  
any \emph{particular} outsider.  Even small groups of
colluding players can gain a very large advantage in winning the
auction, superlinear in the size of the coalition, offering a
significant incentive to collude.

\ifnum\tr=1 
\subsubsection{Ascending-Price Auctions}

When considering ascending-price auctions we are faced with a time-inhomogeneous Markov chain as the bid probabilities vary in the price
of the item. A closed-form solution is not as easily attainable as for
time-homogeneous, fixed-price auctions. Instead we resort to the numerical
evaluation methods we have described in Section \ref{sec:model-ext2}. 
\ifnum\tr=1
We consider our usual auction for \$100 in cash with a bid
fee of \$1 and a price increment of 25 cents. Figures
\ref{fig:collusion-ascending-revenue} and
\ref{fig:collusion-ascending-relative-p} respectively display the
the revenue for Swoopo and the relative likelihood of a colluding bidder winning as the size of the coalition grows.
\fi
The results are quite similar in this setting to the fixed-price case;  a coalition can dramatically lower profits,
and have a competitive advantage that grows superlinearly in their size.    

\skipthis{
More specifically, let $p(A, q)$ be the probability that $A$ is leading the auction after $q$ rounds and define $p(B, q)$ similarly. Furthermore, let $p(AB, q)$ be the transition probability from state $A$ to state $B$ in round $q$ and define the other three transition probabilities in the same manner. Then we have:
\begin{equation}
p(A, q+1) = p(A, q) p(AA, q) + p(B, q) p(BA, q)
\end{equation}
Summing over all $Q$ rounds gives us the overall probability that group $A$ is leading. Then the expected revenue of the auction is:
\begin{equation}
Revenue = (b + s) \left( \sum_{i=0}^{Q} p(A, i) + \sum_{i=0}^{Q} p(B, i) \right)
\end{equation}

The probability that $A$ wins the auction after exactly $q$ rounds is given by
\begin{equation}
p(W_{A}, q) = p(A, q) (1 - p(AA, q) - p(AB, q)
\end{equation}
and similarly for group $B$. Again, summing over all $Q$ rounds will give us the total probability of $A$ winning the auction. Then we can use these probabilities are used to compute the relative likelihood of bidders in each of the groups winning the auction.
}
\fi

\ifnum\tr=1
\begin{figure}[t]
\centering
\subfigure[Revenue for Swoopo as the size $k$ of the coalition grows in an ascending price auction.]
{
   \includegraphics[scale=0.55]{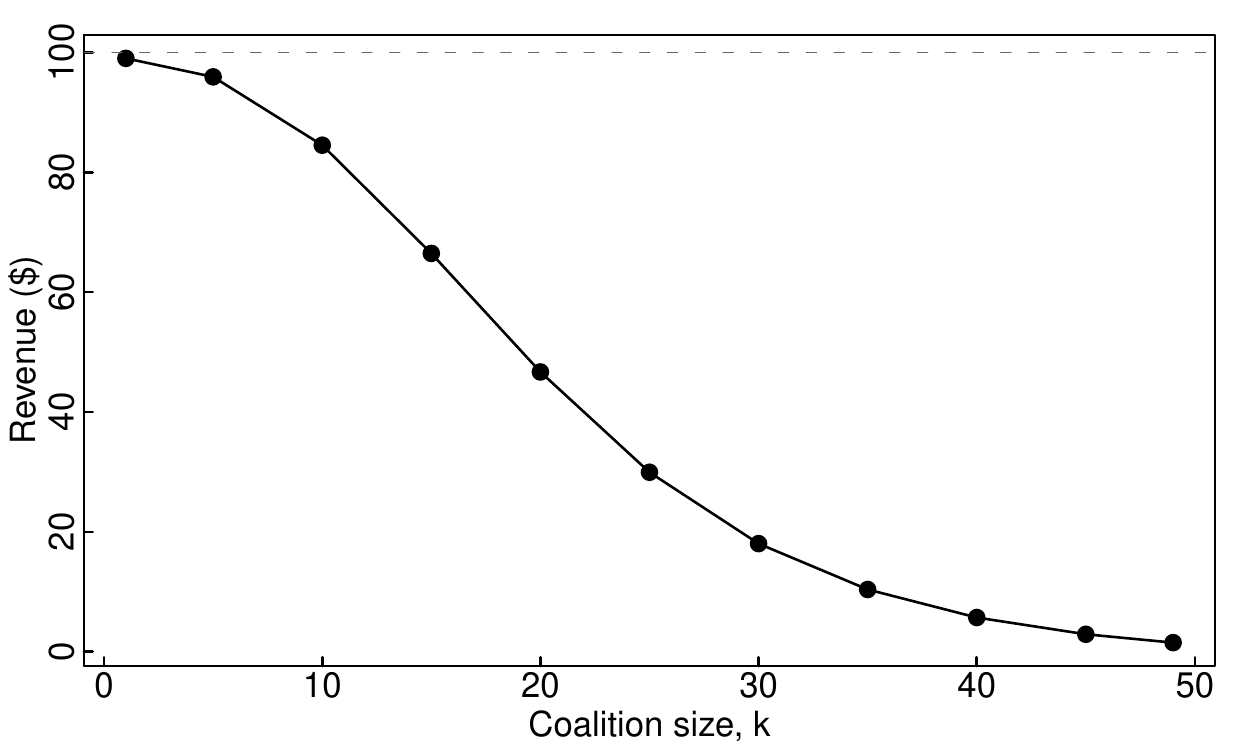} 
   \label{fig:collusion-ascending-revenue}
}
\subfigure[Relative likelihood of the colluding group winning the auction vs. any specific outsider.]
{
   \includegraphics[scale=0.55]{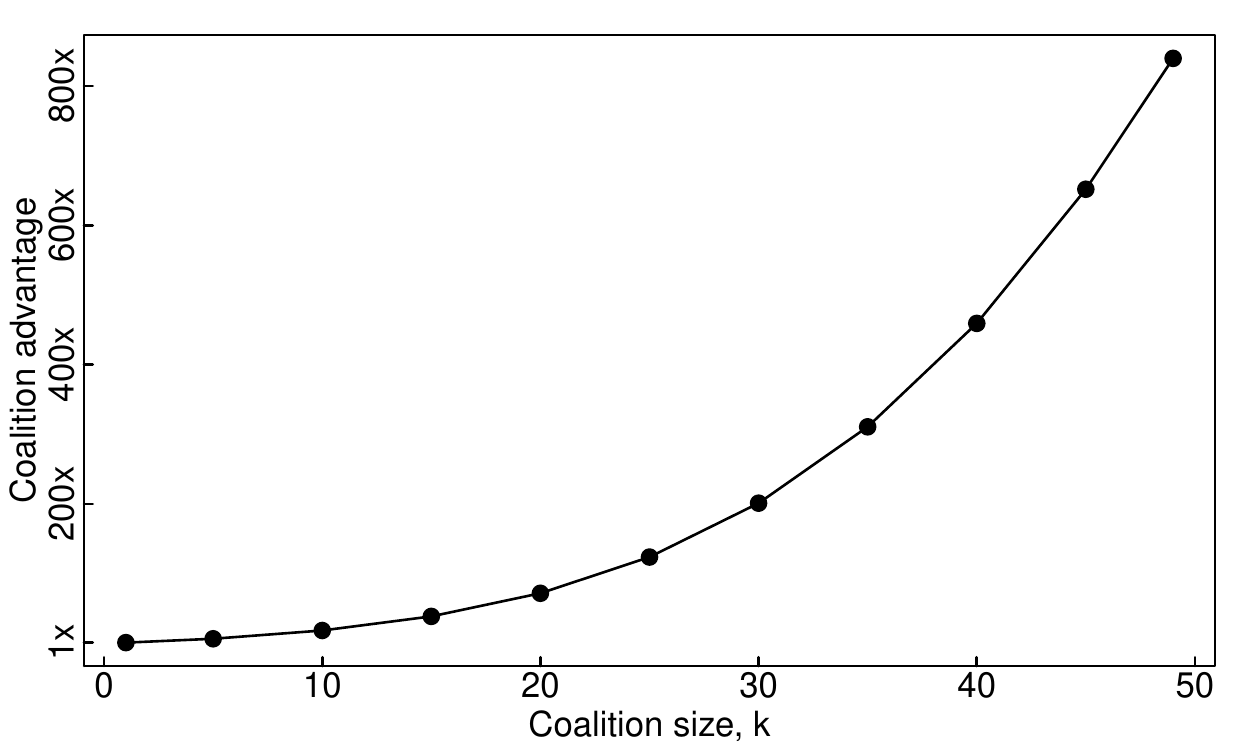} 
   \label{fig:collusion-ascending-relative-p}
}
\caption{Ascending-price auctions with a coalition of size $k$; $n=50$, $v=100$, $b=1$ and $s=0.25$.}
\label{fig:collusion-ascending}
\end{figure}
\fi

\subsection{Shill Bidding}
\label{sec:shill}

\skipthis{ The trustworthiness of the auctioneer is a common concern
of bidders in pay-per-bid and regular auctions alike. One method the
auctioneer can use to defraud players and artificially inflate profits
is shill bidding which refers to the participation of the auctioneer
in the auction under a false identity. Evidence of shill bidding
\cite{} is sparse at best and in no way do we make claims of providing
more. However, considering the relatively short track-record of many
per-per-bid auction websites the question of how much the auctioneer
can profit from employing shill bidding becomes more pertinent.

In \cite{platt2009} Platt et al. find that shill bidding of the form
they define doesn't lead to increased profits for the auctioneer. In
their version a shill bidder bids randomly so as to maintain the
indifference condition for the other players. Arguably, this
definition makes a shill bidder very hard to detect which should be
one of the auctioneer's primary concerns; but at the same time it
provides zero incentive to use a shill in the first place. }

A further consideration is the effect of shill bidders,
or bidders under the employ of the auction site who attempt to drive
up revenue by bidding in order to prevent auctions from terminating early.
This is not a theoretical problem; pay-per-bid auction sites other than Swoopo
have been accused of using shill bidding \cite{pennyauctionwatchshills}.  
In the working paper
\cite{platt2009}, shill bidders were considered, but it was assumed
that they would behave equivalently to other players in the auction.
This assumption was necessary to maintain the symmetry of the
analysis, and was justified by the argument that if shill bidders
behave exactly as other players, they would be more difficult to
detect.  We argue that sites employing shill bidding may be willing to shoulder 
the increased detection risk associated with increased shill bidding
as long as it is accompanied by increased
profit. 

There are several possible ways of introducing shill bidders.  Here we
focus on the following natural one: we define a $(\rho,L)$-shill as one
that enters the auction with probability $\rho$ and bids with probability
one at each opportunity when they are not the leader until $L$ bids
have been made (in total, by all players), at which stage he drops out
of the auction.  Such an approach provides useful tradeoffs;
increasing $\rho$ or $L$ increases the probability of detection, but
offers the potential for increased profit.

\ifnum\tr=1
To analyze shill bidding we employ our usual framework.  When there is
no shill (with probability $1-\rho$), we assume we have a standard auction
with $n$ players; with probability $\rho$ the shill enters and the auction
has $n+1$ players.  To analyze auctions with a shill, we separate
the bidders into two groups: group $A$ consists of the lone shill,
and group $B$ consists of the $n$ legitimate players who are not
informed of the shill's presence. 

This information is sufficient to determine the transition
probabilities in order to use our Markov chain analysis.  Recall 
that shill bidders produce no revenue for the auctioneer, so the expected revenue is
determined by the expected number of times a legitimate player is the
leader.  For convenience we adopt our usual tie-breaking rule, so the
leader is picked uniformly at random from the players,
including the shill, who decide to make a bid at each step.  
(One could
imagine other rules -- such as the shill is never picked in case of a tie, to maximize bid
fees obtained.)  
\fi

\ifnum\ec=1
To analyze shill bidding we employ our usual framework.  We have
a standard auction with $n$ players with probability $1-\rho$.
With probability $\rho$ the shill enters and the auction
has $n+1$ players.  In this case, we separate
the bidders into two groups: group $A$ consists of the lone shill,
and group $B$ consists of the $n$ legitimate players who are not
informed of the shill's presence. 
As before we can determine the transition
probabilities and use our Markov chain analysis.  Recall 
that shill bidders produce no revenue for the auctioneer, so the expected revenue is
determined by the expected number of times a legitimate player is the
leader.  For convenience we adopt our usual tie-breaking rule, so the
leader is picked uniformly at random from the players
who bid at each step.  
\fi

\begin{figure}[t]
\centering

\ifnum\tr=1
\subfigure[Fixed-price auction, $p=0$.]
{
   \includegraphics[scale=0.55]{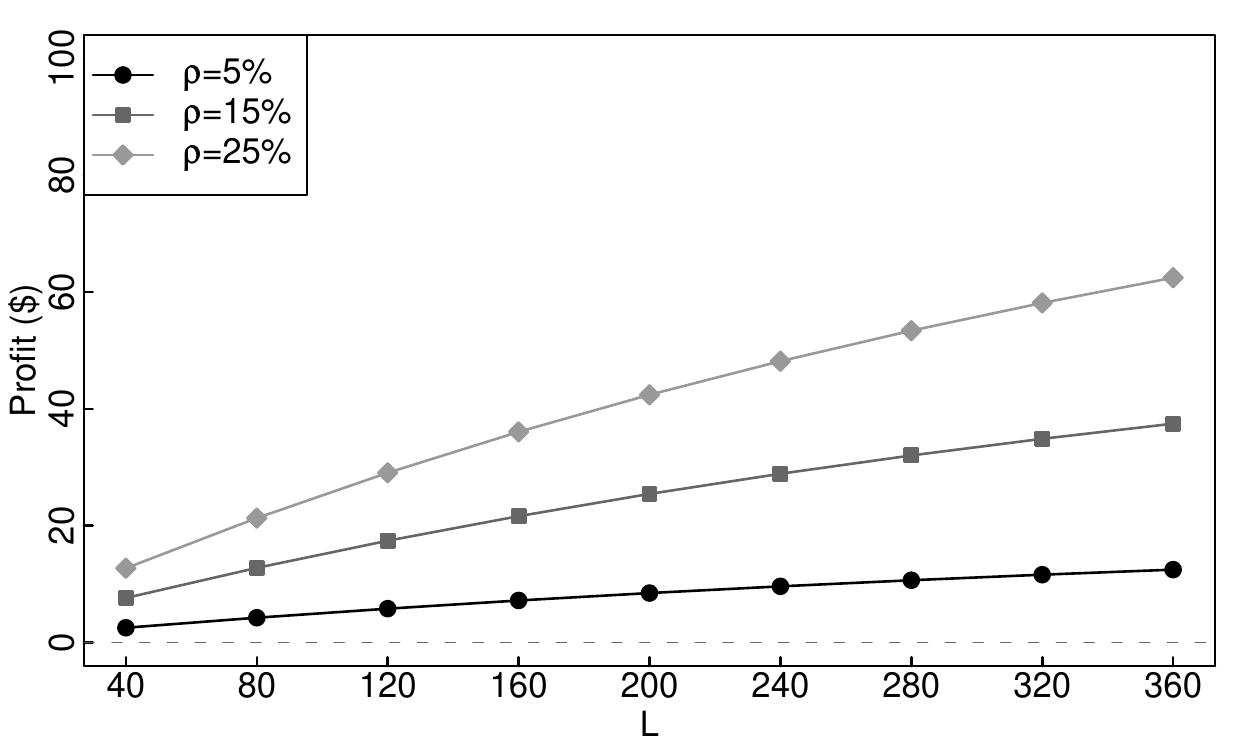} 
   \label{fig:shill-fixed-profit-1}
}
\subfigure[Ascending-price auction.]
{
   \includegraphics[scale=0.55]{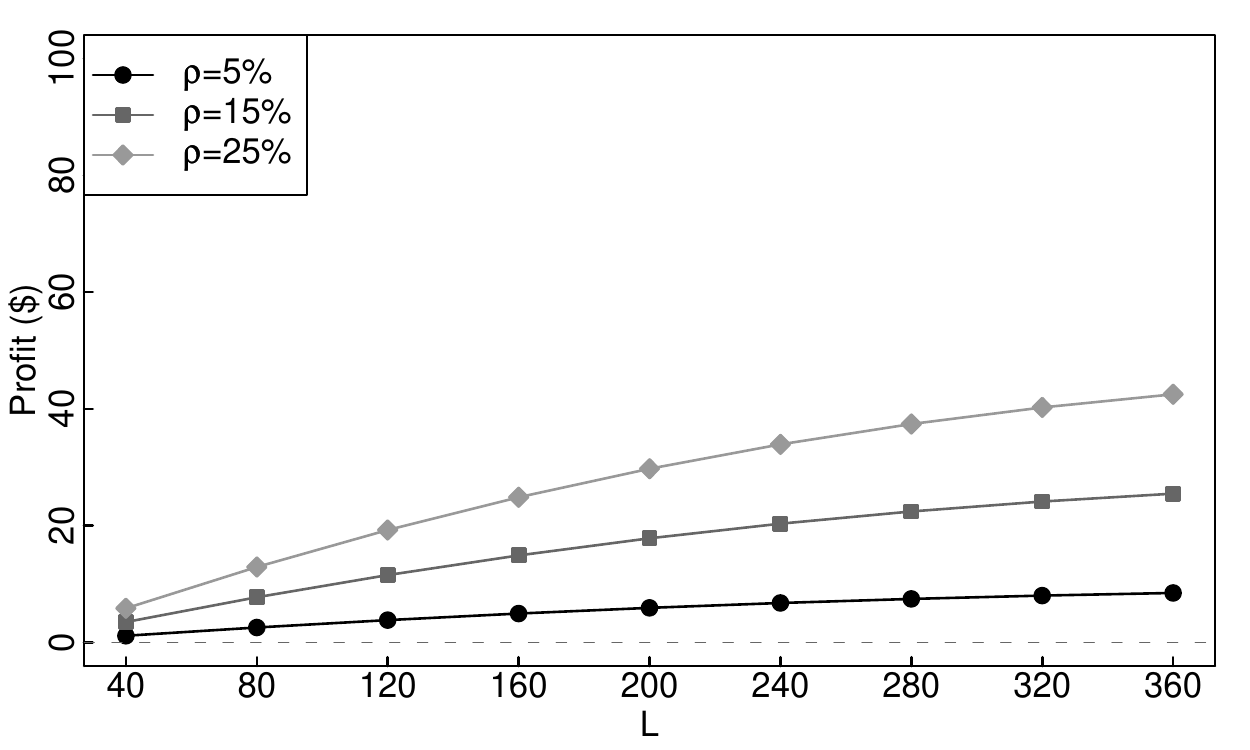} 
   \label{fig:shill-ascending-profit-1}
}
\fi

\ifnum\ec=1
   \includegraphics[scale=0.55]{figures/single-shill-ascending-profit.pdf} 
   \label{fig:shill-ascending-profit-1}
\fi

\ifnum\tr=1
\caption{Expected profit for Swoopo in the presence of a single-identity $(\rho, L)$-shill; $n=50$, $v=100$, $b=1$ and $s=0.25$.}
\label{fig:shill-profit-1}
\fi
\ifnum\ec=1
\caption{Expected profit for Swoopo with a $(\rho, L)$-shill; $n=50$, $v=100$, $b=1$ and $s=0.25$.}
\label{fig:shill-profit-1}
\fi
\end{figure}

\ifnum\tr=1
Before presenting some example data, we provide some intuition.  We
first consider a fixed-price auction with a shill bidder.  The auction
can be thought of as proceeding in two stages: in the first stage that
lasts for up to $L$ bids the shill participates, and in the second
stage the shill abstains. It should be clear that if the auction
reaches the second stage the expected revenue is nearly $v$; the only
issue is that there are only $n$ legitimate players, but they each
behave as though there are $n+1$ players.  For large enough $n$ the
revenue remains close to $v$ in the second stage, and the auctioneer's
marginal profit from using a shill is therefore essentially equal to
the bid fees from other players in the first stage of the auction.  
The case of ascending-price auctions is slightly more interesting due to the 
fact that legitimate players bid with decreasing probability as the price
of the item goes up.  As a consequence, it is more likely for the shill
to win the item as $L$ increases, in which case the auctioneer only
earns revenue from the bid fees.

In order to see the effect of shills more clearly, rather than plot 
\fi
\ifnum\ec=1
Rather than plot
\fi
the per-auction revenue with shill bidders, we instead plot the per-auction
{\em profit}.  We do this for two reasons.  
First, since a symmetric, full-information auction results in 
  zero expected profit for the auctioneer in our model, all profit in our plots can be 
  attributed to the presence of the shill. 
Second, in this setting, there is some chance 
that the shill will win the auction, in which case the auctioneer's revenue
is all profit, a fact not well captured by a revenue plot.
\ifnum\tr=1
Figures \ref{fig:shill-fixed-profit-1} and \ref{fig:shill-ascending-profit-1} display
the expected profit for Swoopo in the presence of a $(\rho, L)$-shill
for fixed and ascending-price auctions respectively. 
\fi
\ifnum\ec=1
Figure 
\ref{fig:shill-profit-1} displays
the expected profit for Swoopo in the presence of a $(\rho, L)$-shill
for an ascending-price auction. 
\fi
Shill participation in ascending-price auctions has diminishing returns with $L$,
which is to be expected; even though the shill is forcibly extending the expected
length of the auction, as the price of the item goes up,
legitimate players become less willing to participate.

\ifnum\tr=1
As an example of another possible shill model, we have also considered
a variation where the shill player uses two distinct identities, so
that he can bid with a second identity when his first identity is the leader,
and vice versa.  In this way, the shill can guarantee that $L$ bids will occur before
dropping out of the auction, at the expense of introducing another
perceived player.  This approach seems riskier in terms of likelihood of 
detection, but leads to additional revenue, since it prevents the
auction from ending early with the shill winning and thereby
effectively provides more opportunities for other players to bid.

While the results for the double-identity shill plotted in Figure~\ref{fig:shill-profit-2} 
  are similar to those depicted for the single-identity shill, the profit
  gains due to extra bids are more substantial and the plots exhibit only minimal 
  diminishing returns. 

\begin{figure}[t]
\centering
\subfigure[Fixed-price auction, $p=0$.]
{
   \includegraphics[scale=0.55]{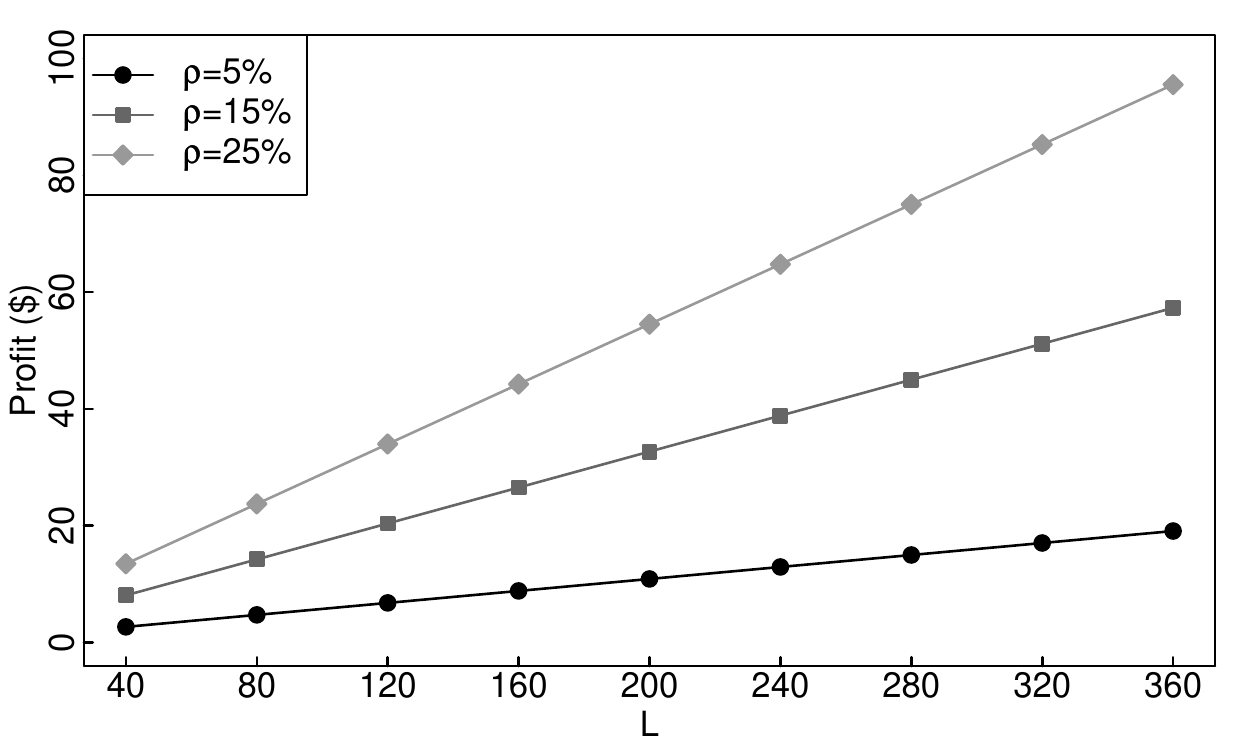} 
   \label{fig:shill-fixed-profit-2}
}
\subfigure[Ascending-price auction.]
{
   \includegraphics[scale=0.55]{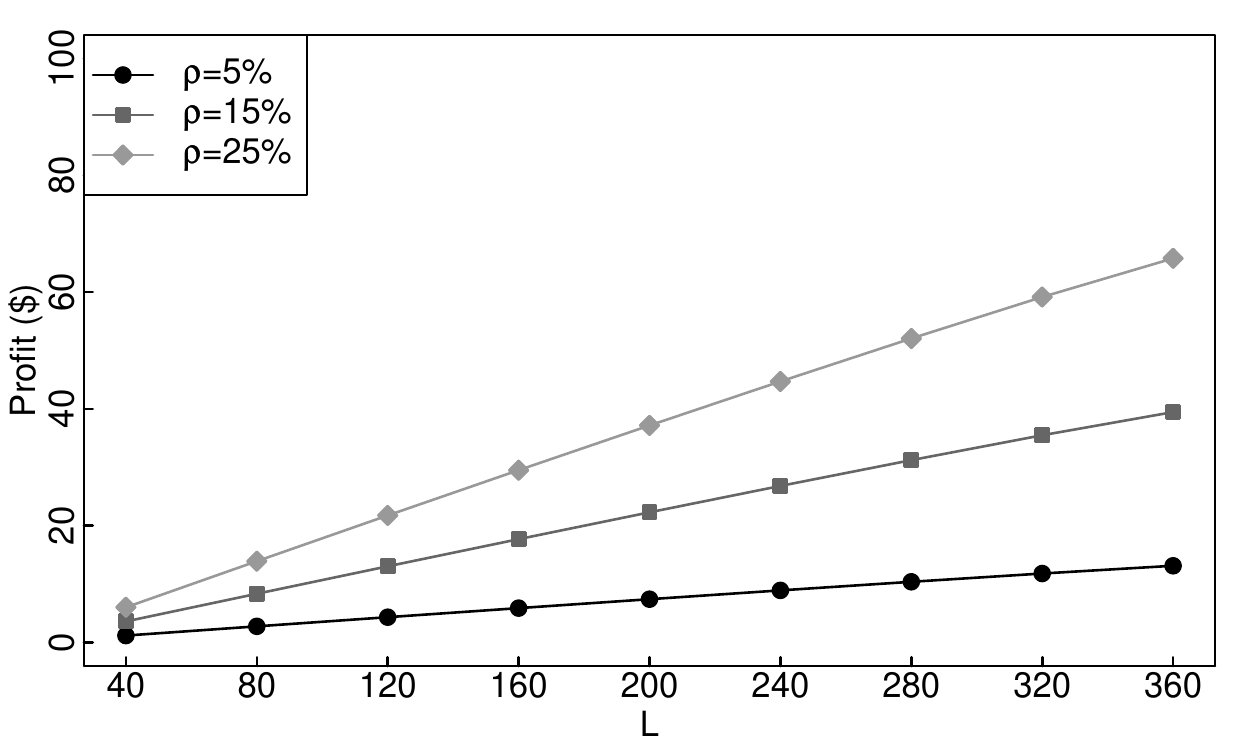} 
   \label{fig:shill-ascending-profit-2}
}
\caption{Expected profit for Swoopo in the presence of a double-identity $(\rho, L)$-shill; $n=50$, $v=100$, $b=1$ and $s=0.25$.}
\label{fig:shill-profit-2}
\end{figure}
\fi

\ifnum\tr=1
\section{Playing Chicken and the Impact of Aggression}
\fi
\ifnum\ec=1
\section{Chicken and Aggression}
\fi
\label{sec:chicken}

In this section we address a recently added feature to Swoopo's
interface, Swoop It Now, that appears to have not been analyzed
previously.  This feature has the potential to significantly change
the dynamics of Swoopo auctions. Our suggestion is that this feature
may lead to a subclass of players whose strategy makes some Swoopo auctions resemble
the game of chicken \cite{rapoport1966}, in contrast to 
the Markovian games we have modeled in previous sections.\footnote{One might argue
that the resemblance is more to a war of attrition auction \cite{krishnamorgan} than to chicken;  we find
the chicken nomenclature easier to use, but the underlying idea is the same.}
In games of 
chicken, it is generally understood that it can be useful for players to 
signal their intentions,
explicitly or implicitly, to other players, in order to cause them to
give up and allow the signaling player to win.  A natural signaling
approach in the timed auction context is to bid both frequently and
quickly after another player bids.  This bidding strategy has been
noted previously, in the work of \cite{augenblick2009}, where the author dubs
this ``bidding aggressively'' and finds that aggressive bids have higher
expected profit.  Here we undertake an independent study, making
several new contributions.  Besides presenting how this behavior can
be viewed as a signaling mechanism for a game of chicken embedded in
Swoopo, we provide a novel and natural definition of aggression for
pay-per-bid auctions. Then, using our trace data, we analyze auctions
for signs of aggressiveness, and estimate how aggressiveness
correlates with winning auctions and profitability for players.  A
surprising finding is that both too little and too much aggression
appear to be losing strategies.  
\ifnum\tr=1
We emphasize that here our analysis
is based less on formal models of player behavior than in previous
sections; our analysis is therefore necessarily more speculative and
worthy of future study.
\fi

\ifnum\ec=1
\vspace{-0.5em}
\fi
\subsection{Swoop It Now and Chicken}

\ifnum\tr=1
Swoopo recently added the Swoop It Now option to auctions on its site, 
which gives each player the ability to purchase the item at a
given price even if one loses the auction.\footnote{While we do not
know the exact date of its introduction, based on various Web
postings, deployment on the US site appears to have occurred around
July 2009, before we began taking traces of auctions.}  That is, in
many auctions, Swoopo provides a nominal retail value for the auction
item, call it $r$.  At the end of the auction, a player who has spent
a total of $\delta$ on bids during the course of the auction can buy the
item at a price of $r - \delta$; that is, the bids are transformed from
otherwise unrecoverable sunk costs to a partial payment for the auction item.
%
%
As argued in previous work \cite{augenblick2009,platt2009}, the nominal retail price provided by
Swoopo is generally significantly higher than the price for which one
could buy the item online, and is surely well above Swoopo's price for
most items.

Unfortunately we do not currently know how often Swoop It Now is used;
to our knowledge such information is neither given by Swoopo, nor derivable
from any data Swoopo makes available.  
We suspect the feature is often overlooked, or that players are not interested, 
  given the high nominal retail value.
On the other hand, 
  after a large losing investment in an auction, this option may become 
  attractive to certain players.
\fi

\ifnum\ec=1
Swoopo recently added the Swoop It Now option to auctions on its site, 
which gives each player the ability to purchase the item at a
given price even if one loses the auction. (Deployment on the US site appears to have occurred around
July 2009, before we began taking traces of auctions.)  That is, in
many auctions, Swoopo provides a nominal retail value for the auction
item, call it $r$.  At the end of the auction, a player who has 
incurred a total bid cost of $\delta$ 
can purchase the item at a price of $r - \delta$, effectively transferring  
otherwise unrecoverable sunk costs to a partial payment for the auction item.
%
%
The retail value $r$
given by Swoopo is generally significantly higher than the lowest 
online retail price \cite{augenblick2009,platt2009}.

Unfortunately we do not currently know how often Swoop It Now is used;
to our knowledge such information is neither given by Swoopo, nor derivable
from any data Swoopo makes available.  
While the high nominal retail value is unattractive,
  after a large losing investment in an auction, this option may become 
  attractive to certain players.
\fi

Let us consider the behavior this additional feature introduces and
its consequences in two settings: the case where only one {\em committed} player 
  is willing to exercise this option, and the case where
multiple players potentially are.  Our assumption here is that $r = \alpha v$,
where $v$ is a common value of the item for all players and $\alpha >
1$.  Our key finding is that when multiple players 
consider taking advantage of this option as a backstop, the game becomes a variant
of chicken.

\ifnum\tr=1
\begin{figure}[t]
\centering
\subfigure[Fixed-price auction.]
{
   \includegraphics[scale=0.55]{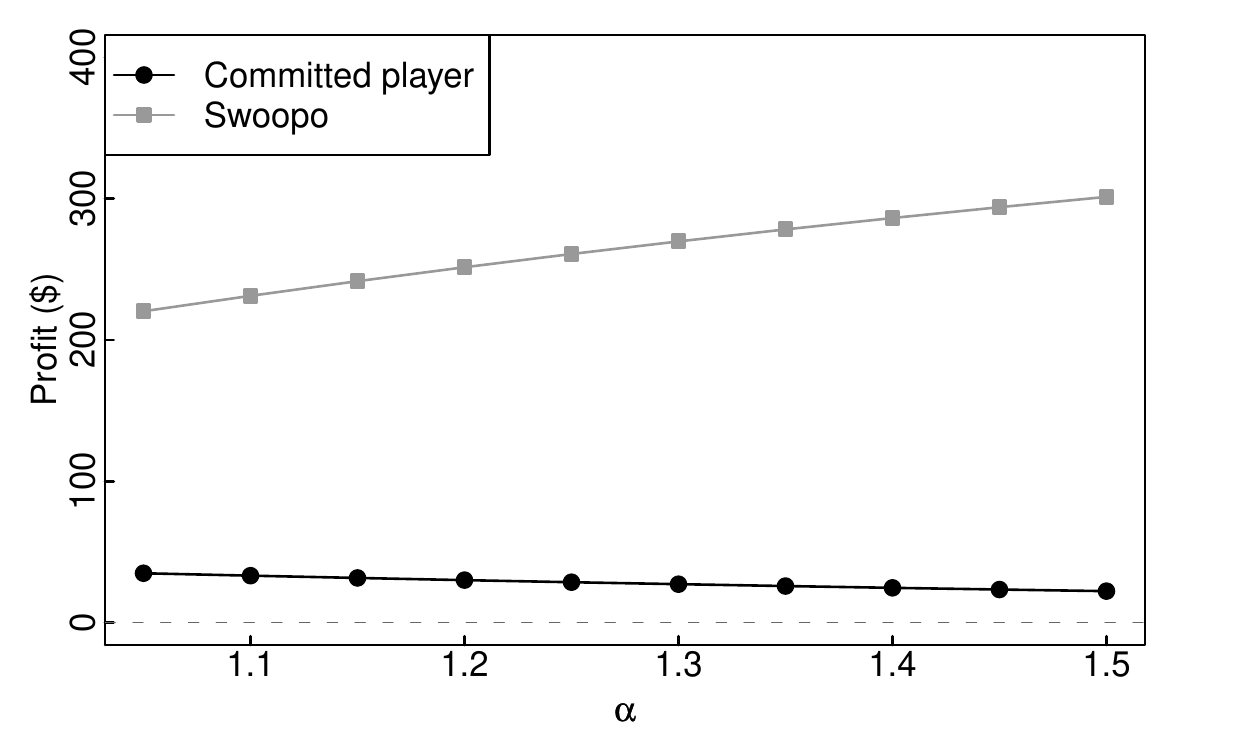} 
   \label{fig:single-aggressive-fixed}
}
\subfigure[Ascending-price auction.] 
{
   \includegraphics[scale=0.55]{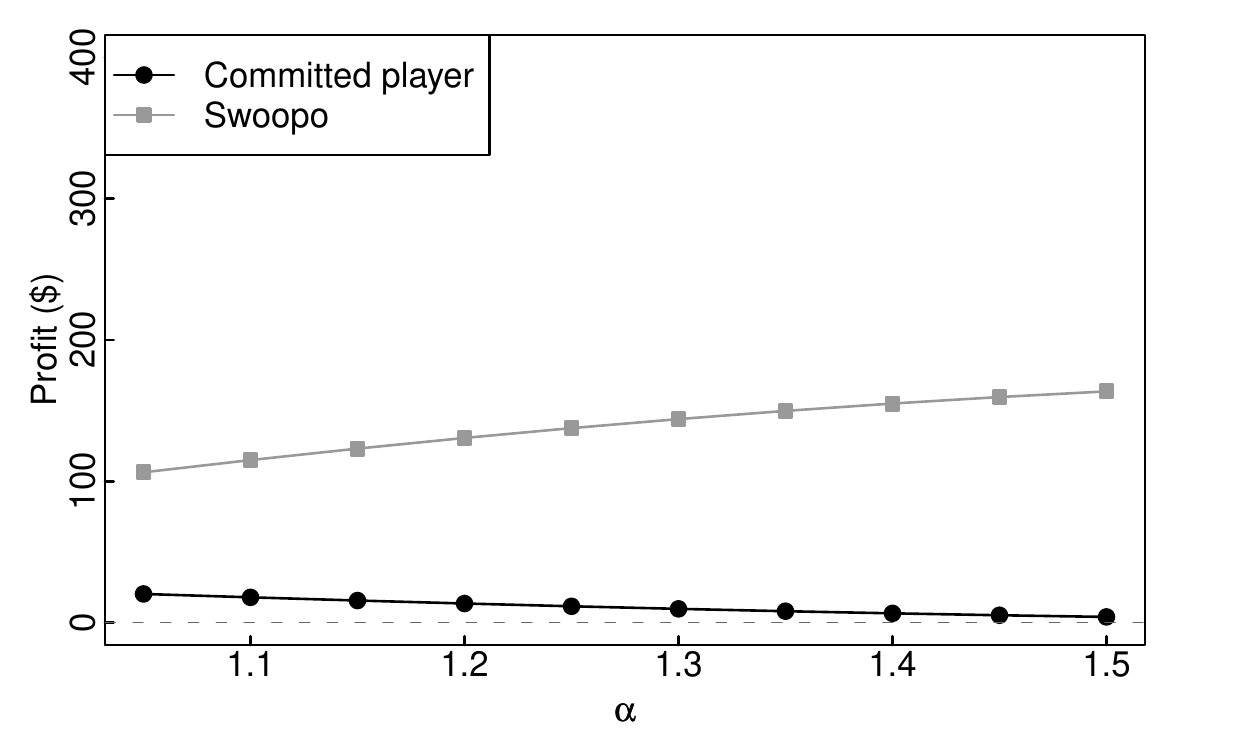} 
   \label{fig:single-aggressive-ascending}
}
\caption{Profit for Swoopo and a committed player as $\alpha$ varies; $n=50$, $v=100$, $b=1$ and $s=0.25$.}
\label{fig:single-committed}
\end{figure}
\fi

\ifnum\ec=1
\begin{figure}[t]
\centering
{
   \includegraphics[scale=0.55]{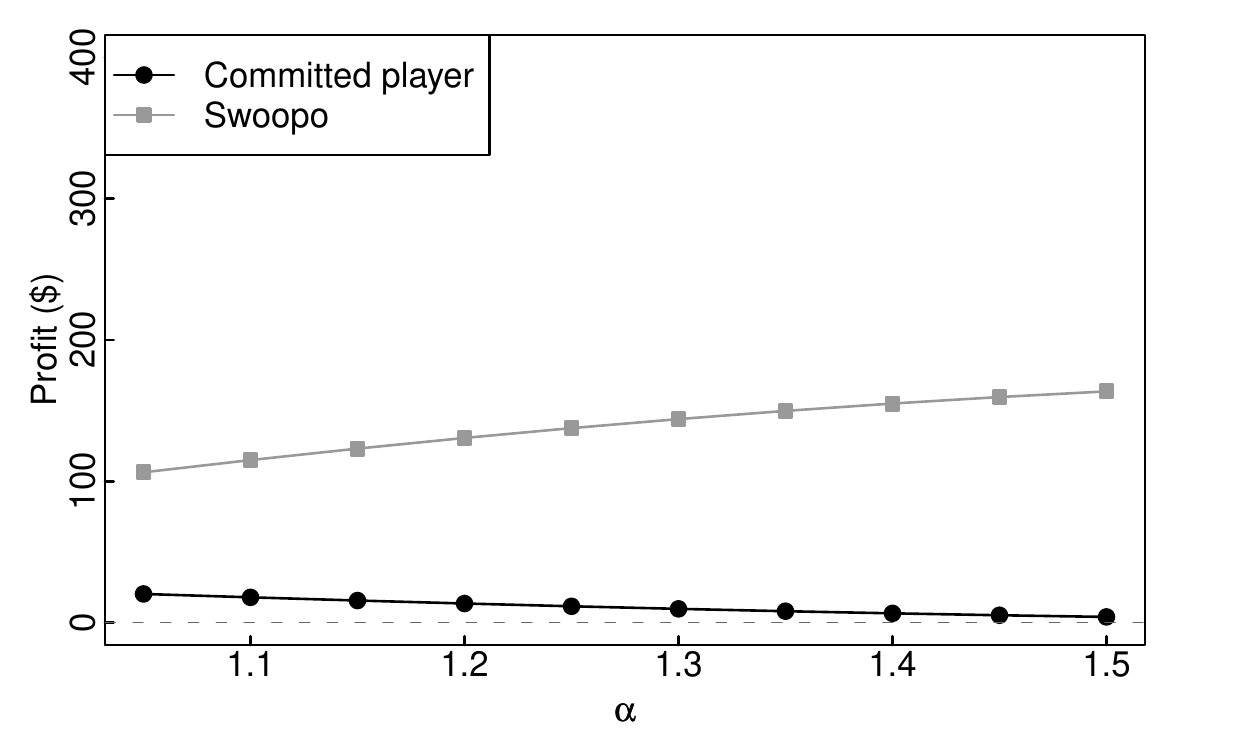} 
   \label{fig:single-aggressive-ascending}
}
\caption{Profit for Swoopo and a committed player as $\alpha$ varies; $n=50$, $v=100$, $b=1$ and $s=0.25$.}
\label{fig:single-committed}
\end{figure}
\fi

\ifnum\tr=1
Let us first suppose that a single player has the opportunity to buy
at the price $r$, including the amount they spend on bids.  This
player may believe the odds of winning the auction early are
sufficient to keep bidding at every possible step, finding that the
expected gain from winning early dominates the maximum possible loss
of $(\alpha-1)v$.  This player will therefore continue to bid until
either winning the auction or spending enough so that it is cheaper
to buy the item using Swoop It Now than to win it at the current auction price; the
other $n-1$ players will play as usual.  In effect, in this situation,
the player is essentially equivalent to a shill bidder, except here the
player keeps bidding until spending a certain amount, rather than until
a certain number of bids have been made.  The approach of
Section~\ref{sec:shill} can be applied with minor variations.  (In
such settings, the Markov chain state space must be expanded, so this
player can keep the amount spent thus far as part of the state; this
is easy to accomplish.)  Also, note in this case Swoopo always
sells at least one copy of the item, as opposed to the setting with
a shill bidder.  The main outcome, naturally, is that having a player
who intends to use the Swoop It Now feature increases Swoopo's profit
by prolonging the auction, assuming the presence of this player  does not
change other players' strategies. Figure \ref{fig:single-committed}
displays the profit earned by the single committed participant as well
as Swoopo.  Here, as when studying shill bidders, it makes more sense
to consider the profit for Swoopo.  In the case where the player uses the
Swoop It Now feature, if they have bid $\delta$ so far, they will have
to pay an additional side payment of $r-\delta$ to complete the purchase.
Also, we decrement Swoopo's profit by an additional $v$ to account for
the transfer of a second item to auction winner.
The higher $\alpha$ is the longer the committed player stays in the game.
As a result, profit for the committed player is decreasing in $\alpha$ while
the reverse holds for Swoopo. As usual, for ascending-price auctions Swoopo's
profit is bounded - no one will bid after $Q+1$ rounds.
On the other hand, if the player who is committed to the Swoop It Now purchase can
signal their intention through aggressive bidding so that other players
drop out of the auction, the end result is significantly less profit
for Swoopo, as the auction will end early.  
\fi

\ifnum\ec=1
Let us first suppose that a single player has the opportunity to buy
at the price $r$, including the amount they spend on bids.  This
player may attempt to win the auction early 
by bidding at every possible step, believing that the
expected gain dominates the maximum possible loss
of $(\alpha-1)v$.  This player will therefore bid until
either winning the auction or spending enough so that it is cheaper
to buy the item using Swoop It Now than to win it at the current auction price; the
other $n-1$ players will play as usual.  The player is essentially equivalent to a shill bidder, except they bid until spending a certain amount, rather than until
a certain number of bids have been made, and they actually
purchase the item if they win.  The approach of
Section~\ref{sec:shill} can be applied with minor variations.  
The main outcome, naturally, is that such a player
increases Swoopo's profit
by prolonging the auction, assuming their presence does not
change other players' strategies. Figure \ref{fig:single-committed}
displays the profit earned by the single committed participant as well
as Swoopo for an ascending-price auction.  
When a player uses the Swoop It Now feature, if they have bid $\delta$ so far, they will have
   to pay an additional side payment of $r-\delta$ to complete the purchase.
Also, we decrement Swoopo's profit by an additional $v$ to account for
the transfer of a second item to the auction winner.
Profit for the committed player decreases in $\alpha$ while
the reverse holds for Swoopo.  This model ignores the possibility that a player
might signal their intention through aggressive bidding so that other players
drop out of the auction, 
resulting in less profit for Swoopo.
\fi

In the case of two (or more) players who are interested in using the
Swoop It Now feature, the resulting game instead resembles the
game-theoretic version of chicken.  For convenience we consider a
fixed-price auction with a price of $0$.  Suppose that two players
plan to continue to bid until either obtaining the item or spending
$r$ in bids and then using the Swoop It Now feature.  
If both exhaust their bids, they will both lose
$(\alpha-1)v$ in value.  But if, instead, one of them
backs off, allowing the other player to win, that player will lose
only what they have bid so far -- call this $\beta$ -- and the other
player will 
purchase the item at a discount -- call their
gain $\gamma v$, on average.  
Table \ref{tab:chicken-payouts} displays
the chicken game in the standard payoff matrix notation.


\ifnum\tr=1
Obviously, we have simplified things considerably in this description;
there may be more than two players willing to use the Swoop It Now
feature, there may be other players involved, and it may be unclear
which players intend to treat the auction as a game of chicken.  This
is clearly a subject in need of further study.  However, we do show
that the Swoop It Now feature, by keeping individual losses bounded,
does embed the potential for games of chicken to erupt within Swoopo
auctions.  As aggressive bidding is a natural way to signal intent in
this setting, (and may be a sound tactic in its own right), we turn to
a study of aggression, making use of our \textsc{Trace} dataset.
\fi

\ifnum\ec=1
Obviously, we have simplified things considerably in this description;
for example, there may be more than two players willing to play chicken
in this setting.  This is clearly a subject in need of
further study.  However, the Swoop It Now feature,
by keeping individual losses bounded, does appear to embed the potential for 
games of chicken to erupt within Swoopo auctions.  As aggressive bidding 
is a natural way to signal intent in this setting, (and may be a sound 
tactic in its own right), we turn to a study of aggression, making use of 
our \textsc{Trace} dataset.
\fi

\begin{table}[t]
\centering
  \begin{tabular}{|c|cc|}
    \hline
     & Quit & Play Till End \\
    \hline
    \hline
    Quit & $-\beta,-\beta$ & $-\beta,\gamma v$  \\
    Play Till End & $\gamma v,-\beta$ & $-\alpha v, - \alpha v$ \\
    \hline
  \end{tabular}
  \label{tab:chicken-payouts}
\caption{Payouts for chicken strategies}
\end{table}

\subsection{Aggression}
\label{sec:aggression}

In earlier work \cite{augenblick2009}, Augenblick has suggested that 
aggressive bidding, including bidding immediately after another player has bid
and bidding frequently in the same auction, leads to higher expected 
value for a player.  His analysis is based on individual bids rather than
bidders;
that is, he considers for each bid how
the time since the previous bid and the number of bids by the bidder
for that item correlate to the expected profit, using regression techniques.

\ifnum\tr=1
We adopt a different approach, by looking instead at how aggressive
bidding affects the profitability of a {\em player} within an auction,
and by providing a single aggression metric to measure the
aggressiveness of a strategy.  As a bidder may vary his strategy
across auctions, we define aggressiveness in the context of a given
auction.  We believe that considering the effects of aggressiveness at
the level of player profitability offers important insights as it
views the merits of an aggressive strategy holistically.  (Also, it is
clear that having bid many times previously will affect the expected
profit of a single bid, simply because having bid many times means the
auction has gone on longer, which affects the current probability the
auction terminates.)
\fi

\ifnum\ec=1
We adopt a different approach, by looking instead at how aggressive
bidding affects the profitability of a {\em player} within an auction,
and by providing a single aggression metric to measure the
aggressiveness of a strategy.  As a bidder may vary his strategy
across auctions, we define aggressiveness in the context of a given
auction.  We believe that considering the effects of aggressiveness at
the level of player profitability offers important insights as it
views the merits of an aggressive strategy holistically.  
\fi

We define an aggressive strategy as one which consists of placing many 
bids in rapid succession to preceding bids.  Specifically, let 
the {\em response time} for a bid be the number of seconds since the preceding bid.  
Aggression should be inversely proportional to response time and proportional to 
the number of bids a bidder places within an auction.  
Hence we define the aggression of a bidder in a given auction as:
\begin{equation}
{\mbox{Aggression}} = \frac{{\mbox{Number of bids}}}{{\mbox{Average response time (seconds/bid)}}}.
\end{equation}

\begin{figure}[t]
\centering
\subfigure[Empirical CCDF of aggression.]
{
   \includegraphics[scale=0.55]{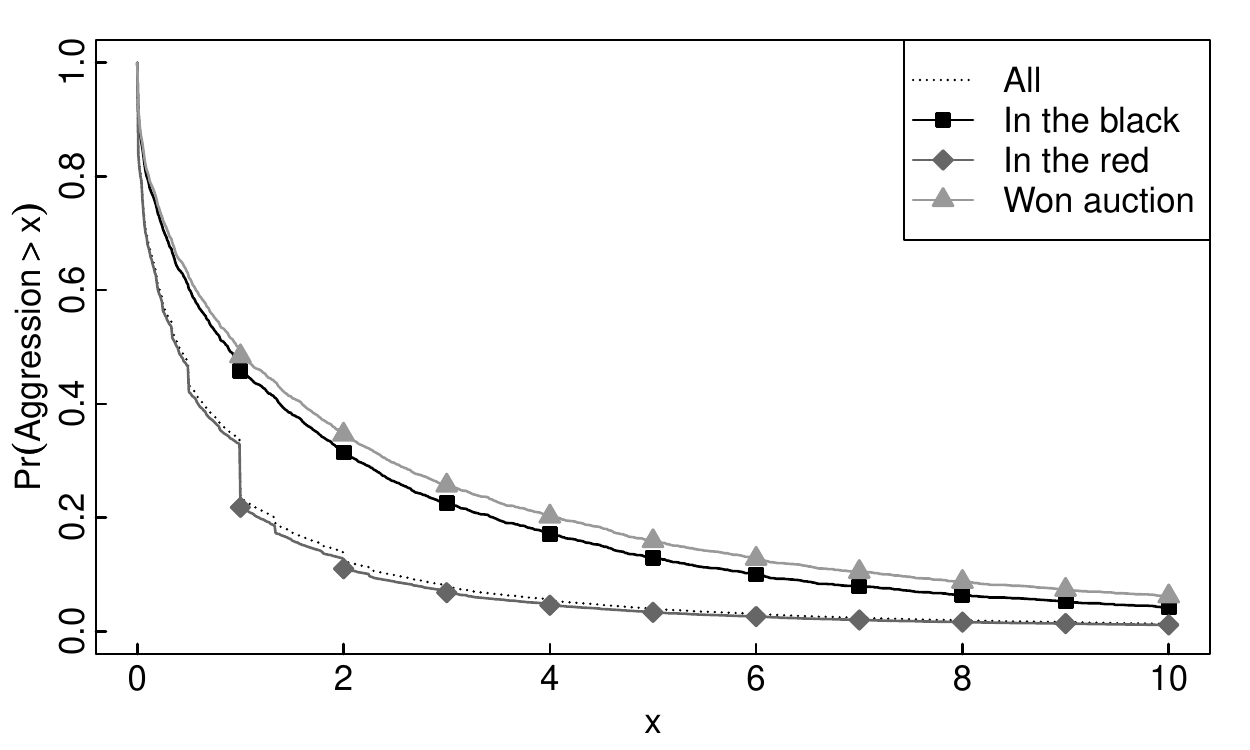} 
   \label{fig:aggression-eccdf}
}
\subfigure[Cumulative profit vs. aggression rank.]
{
   \includegraphics[scale=0.55]{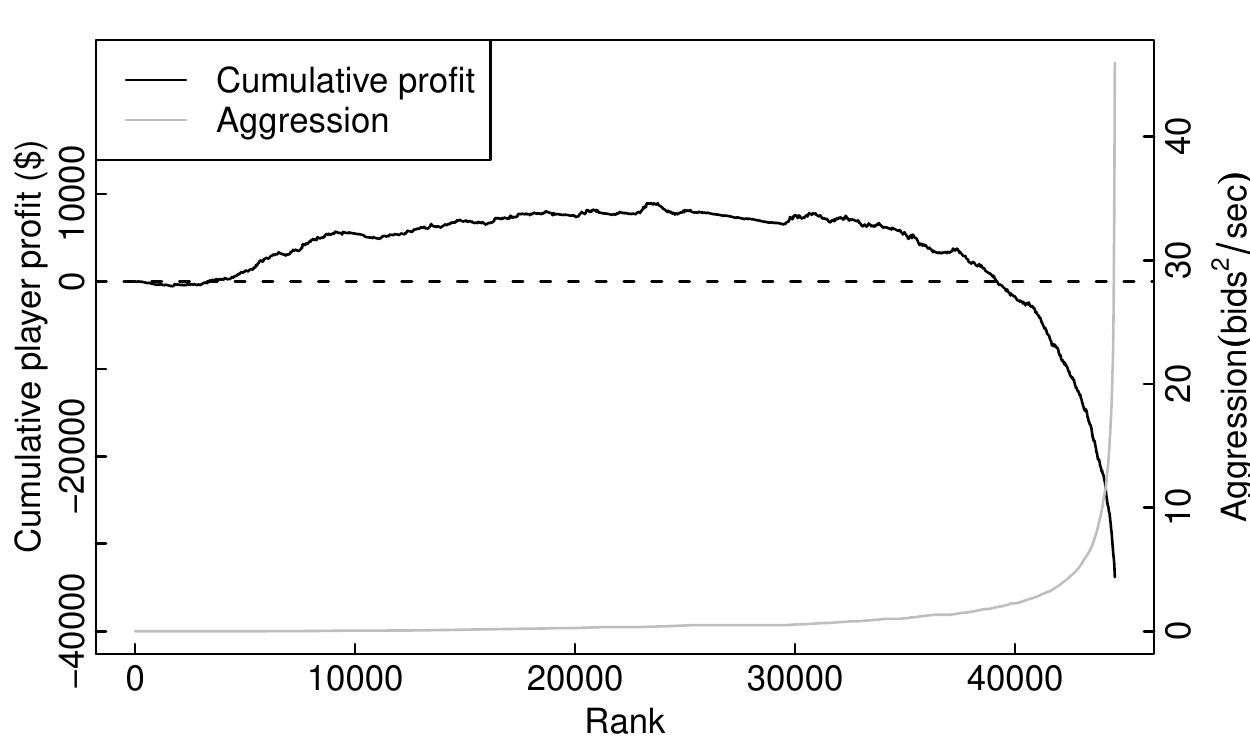} 
   \label{fig:aggression-cumulative-profit}
}
\ifnum\tr=1
\caption{Characterizing he profitability of aggressive strategies.}
\fi
\ifnum\ec=1
\caption{Aggression and profitability.}
\fi
\label{fig:aggressive-strategies}
\end{figure}

To investigate whether aggressive bidding is a successful strategy we
look at the traces of 3,026 complete (no missing bids) ``NailBiter'' auctions (Swoopo auctions which do not
permit the use of automated bids by a ``BidButler'') in our \textsc{Trace} dataset. Figure
\ref{fig:aggression-eccdf} displays the empirical CCDF of aggression
\ifnum\tr=1
across all bidders and for three classes of strategies: 
\begin{itemize}
\item ``Won auction'': strategies that resulted in winning the auction.
\item ``In the black'': strategies which resulted in winning the auction profitably.
\item ``In the red'': strategies which resulted in losing money irrespective of whether the auction was won or lost.
\end{itemize}
\fi
\ifnum\ec=1
for bidders who ended up 1) ``in the black'', by winning the auction profitably;  2) winning, but perhaps
not profitably; and 3) ``in the red'', by incurring bid fees in excess of the value of any item won. 
Note that the classes are not exclusive.
\fi
(For Figure \ref{fig:aggression-eccdf}, when calculating the profit of a bidder, we assume a fixed bid cost of 60 cents, as we cannot determine the true cost. This could affect our interpretation of the results.)

Our first observation is that aggression follows a highly skewed distribution: the 
majority of players display little aggression, while a small number of players are
highly aggressive. Also, not surprisingly, those players winning the auction were
bidding much more aggressively than others.
More interestingly, we see that successful players, i.e., those who not
only won the auction, but did so profitably, are more aggressive than average, but
less aggressive than those who win auctions.
Arguably, aggression is successful in moderation.

Figure \ref{fig:aggression-cumulative-profit} provides more insight
into this latter point.  For all bidder-auction pairs in our dataset,
we compute the aggression and profitability of each outcome, and rank these 
outcomes by aggression (least aggressive first).  We then plot the {\em cumulative} profit
for all outcomes through a given rank with dark shading.  For reference, we also plot the aggression
of bidders at a given rank using light shading and the scale depicted on the right-hand side of the plot.  
We see that successful strategies are mostly concentrated at  aggression
ranks lower than average. More interestingly, a fact not evident in Figure \ref{fig:aggression-eccdf}, the highly 
aggressive players are responsible for most of Swoopo's profits. 

\ifnum\tr=1

\begin{table}[t]
\centering
  \begin{tabular}{|r|r|r|r|}
    \hline
  	\multicolumn{1}{|c}{Aggressive} & \multicolumn{1}{|c}{Number of} & \multicolumn{1}{|c|}{Auction revenue} & \multicolumn{1}{|c|}{Mean winner}\\
    \multicolumn{1}{|c}{bidders} & \multicolumn{1}{|c}{auctions} & \multicolumn{1}{|c|}{(as \% of retail price)} & \multicolumn{1}{|c|}{profit margin}\\
	\hline
	\hline
	$0$ & 1,699 & 62\% & 77\%\\
	$1$ & 493 & 135\% & 51\%\\
	$\ge 2$ & 834 & 246\% & 26\%\\
	\hline
  \end{tabular}
  \label{tab:chicken-in-the-wild}
  \caption{Evidence of chicken}
\end{table}

In Table \ref{tab:chicken-in-the-wild}, we give empirical evidence that Swoopo
  profits are strongly and positively impacted by the presence of multiple
  aggressive players, defined as players with an aggression level of at 
  least 3 bids$^2$/sec.
Strikingly, when two or more aggressive players are present, Swoopo's per-auction
   revenues are in excess of 2.4 times the stated retail price of the good.
At the same time, and as expected, as the number of aggressive players in an auction increases the profit
margin of the auction winner decreases. The fourth column of Table \ref{tab:chicken-in-the-wild}
displays the mean of winner profit margins.

Finally, having provided empirical evidence regarding aggressive behavior, we
\fi
\ifnum\ec=1
We 
\fi
now revisit the question of whether games of chicken
  are also taking place within Swoopo.  To do that we turn to the {\sc
  Trace} dataset and look at the 3,026 ``NailBiter'' auctions for
  evidence of \emph{duels}: auctions culminating in long sequences of
  back-and-forth bidding between two opponents.  We find that 9\% of
  all auctions culminated in a duel lasting at least 10 bids, 
  5\% lasted at least 20 bids, and 1\% lasted at least 50 bids. The
  longest duel we observed was 201 bids long and somewhat humorously took place
  between users \texttt{Cikcik} and \texttt{Thedduell}.  We believe
  this provides further evidence that at least some auctions are
  becoming essentially games of chicken, and reiterate our supposition
  that aggressive bidding is used as a signaling method in such settings.

\ifnum\ec=1
\vspace{-0.8em}
\fi
\section{Conclusions and Future Work}

\ifnum\tr=1
Swoopo provides a fascinating case study in how new, non-trivial auction
mechanisms perform in real-world situations, and it also provides a focal
point for developing the theory of auction mechanisms in the context
of human behaviors.  Here we have focused on the key issue of 
asymmetry, and in particular, how various
manifestations of information asymmetry may be responsible 
in large part 
for the significant profits Swoopo appears to enjoy today.  
At the same time, we have also shown that the profitability of these auctions 
is potentially fragile, especially in cases where signaling by committed
players willing to play a game of chicken or collusion between players can end 
the auction early.  In these settings,
information asymmetry can potentially reduce the profits enjoyed by the auctioneer.

There are clearly many interesting directions to follow from here.
One area we have started to examine is asymmetric models of pay-per-bid
auctions with full information.  For example, players could have differing 
bid fees or valuations of the item, but with these fees and valuations known 
in advance to all players.  Although a full information setting may not be 
as immediately practical as that considered in our work, asymmetries expose a 
richer set of issues than the symmetric, full information case, and have not yet 
been considered in any depth in previous work.
\fi

\ifnum\ec=1
Swoopo provides a fascinating case study in how new, non-trivial auction
mechanisms perform in real-world situations.  Here we have focused on the key issue of 
asymmetry, and in particular, how various
manifestations of information asymmetry may be responsible in large part 
for the significant profits Swoopo appears to enjoy today.  
At the same time, we have also shown that the profitability of these auctions 
is potentially fragile, especially in cases where signaling by committed
players willing to play a game of chicken or collusion between players can end 
the auction early.  

There are clearly many interesting directions to follow from here.
One area we have started to examine is asymmetric models of pay-per-bid
auctions with full information.  Players could have differing 
bid fees or valuations of the item, but with these fees and valuations known 
in advance to all players.  Interestingly, such models have not yet 
been considered in any depth in previous work.
\fi

\ifnum\tr=1
Let us consider what such a model might look like.  
For convenience, let us consider a fixed-price
auction.  Suppose that there are players $A_1,A_2,\ldots,A_n$, with
player $A_i$ having bid fee $b^{A_i}$ and value $v^{A_i}$.  
Let $\beta_{A_i}$ be the probability that $A_i$ should bid when $A_i$ 
is not in the lead in equilibrium.  The
indifference condition for player $A_i$ is now given simply by:
$$b^{A_i} = (v^{A_i}-p) \left (1 - \prod_{j\neq i} \beta_{A_j} \right),$$
or 
$$\prod_{j\neq i} \beta_{A_j} = 1 - \frac{b^{A_i}}{(v^{A_i}-p)}.$$
It is helpful to substitute $\eta_i = \ln \beta_{A_i}$ and $\zeta_i =
\ln \left (1 - \frac{b^{A_i}}{(v^{A_i}-p)} \right)$.  Then we can
readily solve the family of simple linear equations:
$$\sum_{j\neq i} \eta_{A_j} = \zeta_i.$$
This assumes an equilibrium based on the indifference condition (and, for example, that $\beta_{A_i} \neq 0$).  
It would be interesting to consider under what conditions such equilibria exist, as 
well as if they are the only equilibria.  In some cases, the auction might for example reduce to a game of chicken,
in which case there may be multiple equilibria.  For the case of ascending-price auctions, the calculations are more
challenging: one can use backward induction, finding the probabilities
for the $Q$th bid and then calculating downward.

It is worth noting that with varying bids, in a full information
auction where all players ascribe the same value $v$ to an item and
use a strategy based on the indifference condition, Swoopo's expected
revenue for successful auctions remains $v$ following the
same argument as given in Section~\ref{sec:model}.  This highlights
the key role of information asymmetry in our result showing that
varying bids can lead to large profits for Swoopo.  Further comparisons
with a full information model should be similarly enlightening.  
\fi

\ifnum\tr=1
Another broad topic for future work regards more extensive study of user
  behavior on Swoopo.
While our study considers the impact of one natural tactic, aggression,
  in Swoopo auctions, there are many others that could also be considered.
The impact of timing has generally been abstracted away, but it is probably
  important to user behavior, as our study of aggression suggests. 
More understanding of the impact of timing considerations appears important
  for further study. 
In particular, perhaps there are
   timing-based or other active signaling mechanisms that a strategic player 
  could leverage to maximize expected return in these auctions.
A possible direction related to signaling is the issue of learning.
Can players dynamically change their beliefs about underlying auction parameters, such
  as the bid fee other players are charged or the intentions of other players, based on 
  how the auction proceeds in order to improve their performance?
As another example, one feature that we have not yet studied is the impact of bidders 
  who use automatic bidding agents, such as BidButlers:  how do they influence the auction, and can such bidders be
  detected?
And finally, there remains the thorny problem of attempting to quantify
the impact of how specific auction characteristics we have considered 
-- misestimates of the
number of players, varying bid fees, varying valuations of items, the
ability for players to use the BidButler and the Swoop It Now features -- 
on real-world profitability.  Such an effort would likely 
require more detailed information regarding bids currently available 
only to Swoopo, and better models of user behavior.
\fi

\ifnum\ec=1
Other broad topics for future work include more extensive study of user
  behavior on Swoopo, the impact of timing (that we and others have abstracted away),
  models where users can dynamically change their beliefs and strategies, and the impact
  of automatic bidding agents such as BidButlers. 
Finally, there remains the thorny problem of attempting to quantify
  directly the impact of specific auction characteristics 
  on real-world profitability.  
\fi

\ifnum\tr=1
\section{Acknowledgments}

Michael Mitzenmacher would like to thank Maher Said for several helpful discussions
regarding Swoopo. Georgios Zervas would like to thank Azer Bestavros for helpful 
discussions regarding collusion.
\fi

\ifnum\tr=1
\bibliography{swoopo}
\bibliographystyle{abbrv}{}
\fi

\ifnum\ec=1
\bibliographystyle{abbrv}{}
\bibliography{swoopo}

\begin{thebibliography}{10}

\bibitem{akerlof1970}
G.~A. Akerlof.
\newblock The market for ``lemons'': Quality uncertainty and the market
  mechanism.
\newblock {\em The Quarterly Journal of Economics}, 84(3):488--500, 1970.

\bibitem{augenblick2009}
N.~Augenblick.
\newblock {Consumer and Producer Behavior in the Market for Penny Auctions: A
  Theoretical and Empirical Analysis}.
\newblock Unpublished manuscript. Available at \url{www.stanford.edu/~ned789},
  2009.

\bibitem{grinstead1997}
C.~M. Grinstead and J.~L. Snell.
\newblock {\em Introduction to Probability}.
\newblock American Mathematical Society, 1997.

\bibitem{hinnosaar2009}
T.~Hinnosaar.
\newblock {Penny Auctions}.
\newblock Unpublished manuscript at \url{http://toomas.hinnosaar.net/}, 2009.

\bibitem{krishnamorgan}
V.~Krishna and J.~Morgan.
\newblock {An analysis of the war of attrition and the all-pay auction}.
\newblock {\em Journal of Economic Theory}, 72(2):343--362, 1997.

\bibitem{mitzenmacher2004brief}
M.~Mitzenmacher.
\newblock {A brief history of generative models for power law and lognormal
  distributions}.
\newblock {\em Internet mathematics}, 1(2):226--251, 2004.

\bibitem{mitzenmacher2005editorial}
M.~Mitzenmacher.
\newblock {Editorial: The future of power law research}.
\newblock {\em Internet Mathematics}, 2(4):525--534, 2005.

\bibitem{pennyauctionwatchshills}
\url{http://www.pennyauctionwatch.com/tag/shills/}.

\bibitem{platt2009}
B.~C. Platt, J.~Price, and H.~Tappen.
\newblock {Pay-to-Bid Auctions}.
\newblock Unpublished manuscript. Available at
  \url{http://econ.byu.edu/Faculty/Platt}, 2009.

\bibitem{rapoport1966}
A.~Rapoport and A.~Chammah.
\newblock The game of chicken.
\newblock {\em American Behavioral Scientist}, 10, 1966.

\bibitem{rothschild1976}
M.~Rothschild and J.~Stiglitz.
\newblock {Equilibrium in competitive insurance markets: An essay on the
  economics of imperfect information}.
\newblock {\em The Quarterly Journal of Economics}, 90(4):629--649, 1976.

\bibitem{shubik1971}
M.~Shubik.
\newblock The dollar auction game: a paradox in noncooperative behavior and
  escalation.
\newblock {\em Journal of Conflict Resolution}, 15(1):109--111, March 1971.

\bibitem{spence1973}
M.~Spence.
\newblock Job market signaling.
\newblock {\em The Quarterly Journal of Economics}, 87(3):355--374, 1973.

\bibitem{nyt1}
B.~Stone.
\newblock Sites ask users to spend to save.
\newblock {\em New York Times}, August 17, 2009.

\bibitem{nyt2}
R.~H. Thaler.
\newblock Paying a price for the thrill of the hunt.
\newblock {\em New York Times}, November 15, 2009.

\bibitem{willinger2009mathematics}
W.~Willinger, D.~Alderson, and J.~Doyle.
\newblock {Mathematics and the internet: A source of enormous confusion and
  great potential}.
\newblock {\em Notices of the American Mathematical Society}, 56(5):586--599,
  2009.

\end{thebibliography}
\fi

\ifnum\tr=1
\appendix
\section{Computing the Steady State of a Markov Chain}
\label{sec:appendix-model-ext}

Let $f(k;n,p)$ be the \emph{probability mass function} of the binomial distribution where $n$ is the number of trials, $k$ is the number of successes and $p$ it the probability of success. The transition probabilities in Figure \ref{fig:two-groups-markov}, presented in Section \ref{sec:asymmetric-markov},
 are given by:
\begin{align}
p_{AA} &= 
\sum_{i=1}^{k-1} 
\sum_{j=0}^{n-k}
f(i;k-1,\beta_{q}^{A})
f(j;n-k,\beta_{q}^{B})
\frac{i}{i+j} \\
p_{AB} &= 
\sum_{i=0}^{k-1} 
\sum_{j=1}^{n-k}
f(i;k-1,\beta_{q}^{A})
f(j;k-1,\beta_{q}^{B})
\frac{j}{i+j} \\
p_{BB} &= 
\sum_{i=0}^{k} 
\sum_{j=1}^{n-k-1}
f(i;k,\beta_{q}^{A})
f(j;n-k-1,\beta_{q}^{B})
\frac{j}{i+j} \\
p_{BA} &= 
\sum_{i=1}^{k} 
\sum_{j=0}^{n-k-1}
f(i;k,\beta_{q}^{A})
f(j;n-k-1,\beta_{q}^{B})
\frac{i}{i+j}
\end{align}
where $\beta_{q}^{A}$ and $\beta_{q}^{B}$ are the individual bidding probabilities of players in group $A$ and group $B$ respectively. Their exact values will depend on the specific asymmetry we are considering. 

We will be interested in computing the steady state probability
distribution of the Markov chain. The goal of finding closed-form
solutions for time-inhomogeneous Markov chains for ascending price
auctions is beyond the scope of our work; hence for these chains we resort to numerical
methods, simply calculating the probability of being in each state in each stage,
as described in Section~\ref{sec:asymmetric-markov}. But for fixed-price auctions the analysis is far more
straightforward and we present it here following the framework of
\cite{grinstead1997}. First define the labeled transition matrix $P$
as:

\begin{equation}
P=
\left(
\begin{array}{ccccc}
& \mathbf{A} & \mathbf{B} & \mathbf{W_{A}} & \mathbf{W_{B}} \\
\mathbf{A} & p_{AA} & p_{AB} & 1-p_{AA}-p_{AB} & 0\\
\mathbf{B} & p_{BA} & p_{BB} & 0 & 1-p_{BB}-p_{BA}\\
\mathbf{W_{A}} & 0 & 0 & 1 & 0\\
\mathbf{W_{B}} & 0 & 0 & 0 & 1\\
\end{array}
\right)
\end{equation}
$P$ is in \emph{canonical} form with the transient states coming before the absorbing ones. Let $Q$ be the submatrix of $P$ containing solely the transient states $A$ and $B$. For an absorbing Markov chain $P$ the matrix $N = (I - Q)^{-1}$ is called its \emph{fundamental} matrix. The entry $n_{ij}$ of $N$ gives the expected number of times the process is in state $j$ given that it started in state $i$. Notice that $N$ exists and is equal to $I + Q^{2} + Q^{3} + \cdots$. We can express $N$ in terms of the transition probabilities:
\begin{equation}
N=\left(
\begin{array}{cc}
 \frac{1-p_{BB}}{p_{BB} p_{AA}-p_{AA}-p_{BB}-p_{AB} p_{BA}+1} & \frac{p_{AB}}{p_{BB} p_{AA}-p_{AA}-p_{BB}-p_{AB} p_{BA}+1} \\
 \frac{p_{BA}}{p_{BB} p_{AA}-p_{AA}-p_{BB}-p_{AB} p_{BA}+1} & \frac{1-p_{AA}}{p_{BB} p_{AA}-p_{AA}-p_{BB}-p_{AB} p_{BA}+1}
\end{array}
\right).
\end{equation}
For any given starting state we can compute the time to absorption by computing the vector $t=N c$, where $c$ is a column vector whose entries are $1$. Effectively, by adding together the elements of $N$ row-wise we are summing up the time spent in each of the transient states. In particular we have
\begin{equation}
t=
\left(
\begin{array}{c}
 \frac{-p_{AB}+p_{BB}-1}{p_{AA} (-p_{BB})+p_{AA}+p_{AB} p_{BA}+p_{BB}-1} \\
 \frac{-p_{AA}+p_{BA}+1}{(p_{AA}-1) (p_{BB}-1)-p_{AB} p_{BA}}
\end{array}
\right).
\end{equation}
Let $p_{A}$ and $p_{B}$ be the probabilities of the game starting in states $A$ and $B$ respectively and let $p_{0}$ be a vector containing them. It is worth highlighting here that $p_{A} + p_{B} < 1$ as there is always the probability that nobody will place the first bid. We can compute the expected number of bids the game will last by calculating $t p_{0}$. Alternatively, as we often do, if we wish to assume that first bid always occurs $p_{0}$ has to be scaled by $1/(p_{A}+p_{B})$ so that $p_{A} + p_{B} = 1$.

We will also be interested in computing the revenue of the auction. When players are charged a single bid fee the revenue can be easily obtained from expected number of bids. If the two groups of players are charged differently  the revenue of is given by $p_{0} N b$ where $b$ is a vector of bid fees.

Next, define $R$ as the upper-right, two-by-two submatrix of $P$. Let $S$ be a matrix whose $(i, j)^{th}$ element is the probability that the chain will be absorbed in state $j$ given that it started in state $i$. Then
\begin{equation}
S = N R = 
\left(
\begin{array}{cc}
\frac{\left(p_{AA}+p_{AB}-1\right) \left(p_{BB}-1\right)}{\left(p_{AA}-1\right) \left(p_{BB}-1\right)-p_{AB} p_{BA}} & \frac{p_{AB} \left(p_{BB}+p_{BA}-1\right)}{p_{AB} p_{BA}-\left(p_{AA}-1\right)
   \left(p_{BB}-1\right)} \\
-\frac{\left(p_{AA}+p_{AB}-1\right) p_{BA}}{\left(p_{AA}-1\right) \left(p_{BB}-1\right)-p_{AB} p_{BA}} & \frac{\left(p_{AA}-1\right) \left(p_{BB}+p_{BA}-1\right)}{\left(p_{AA}-1\right)
   \left(p_{BB}-1\right)-p_{AB} p_{BA}}
\end{array}
\right).
\end{equation}
We can also compute the unconditional probabilities $w$ of the chain being absorbed in each of the sink states assuming the first bid is placed
(using the scaled version of $p_0$):
\begin{equation}
w = p_{0} S = 
\left(
\begin{array}{c}
\frac{\left(p_{AA}+p_{AB}-1\right) \left(p_{BB}-1\right) p_{A}-p_{AB} \left(p_{BB}+p_{BA}-1\right) p_{B}}{\left(p_{AA}-1\right) \left(p_{BB}-1\right)-p_{AB} p_{BA}}\\
\frac{\left(p_{AA}-1\right) \left(p_{BB}+p_{BA}-1\right) p_{B}-\left(p_{AA}+p_{AB}-1\right) p_{BA} p_{A}}{\left(p_{AA}-1\right) \left(p_{BB}-1\right)-p_{AB} p_{BA}}
\end{array}
\right)'.
\end{equation}
This framework will allow us to characterize a series of asymmetries by computing the vectors $t$ and $w$, analytically where possible and computationally otherwise.

\section{Dataset Description}
\label{sec:appendix-dataset}

In this section we provide more information about the two datasets we collected and used throughout this paper. The dataset is publicly available 
at \texttt{http://cs-people.bu.edu/zg/swoopo-dataset.tar.gz}.

\subsection{Outcomes Dataset}

Our \textsc{Outcomes} dataset comprises 121,419 auctions conducted between September 8, 2008 and December 12, 2009. For each auction we collected the following information:

\vskip 0.1cm
\begin{tabular}{lp{7cm}r}
\hline
Field name & Description & Example value\\
\hline\hline
\texttt{auction\_id}	& A unique numerical id for the auction. & \texttt{259070} \\
\texttt{product\_id}	& A unique product id. & \texttt{10011706} \\
\texttt{item} & A text string describing the product. & \texttt{300-bids-voucher} \\
\texttt{desc}	 & More information about the product. & \texttt{300 Bids Voucher} \\
\texttt{retail} & The stated retail value of the item, in dollars. & \texttt{180} \\
\texttt{price} & The price the auction reached, in dollars. & \texttt{31.26} \\
\texttt{finalprice} & The price charged to the winner in dollars.\footnotemark & \texttt{31.26} \\
\texttt{bidincrement} & The price increment of a bid, in cents. & \texttt{6} \\
\texttt{bidfee} & The cost incurred to make a bid, in cents. & \texttt{60} \\
\texttt{winner} & The winner's username. & \texttt{Schonmir1500} \\
\texttt{placedbids} & The number of paid bids placed by the winner. & \texttt{106} \\
\texttt{freebids} & The number of free bids place by the winner. & \texttt{0} \\
\texttt{endtime\_str} & The auction's end time. & \texttt{13:29 PDT 12-12-2009} \\
\texttt{flg\_click\_only} & A binary flag indicating a ``NailBiter'' auction. & \texttt{1} \\
\texttt{flg\_beginnerauction}  & A binary flag indicating a beginner auction. & \texttt{0} \\
\texttt{flg\_fixedprice}	& A binary flag indicating a fixed-price auction. & \texttt{0} \\
\texttt{flg\_endprice} & A binary flag indicating a 100\%-off auction. & \texttt{0} \\
\hline
\end{tabular}
\footnotetext{This can differ from the \texttt{price} field in the case of fixed-price auctions.}
\vskip 0.1cm

\subsection{Trace Dataset}

Our \textsc{Trace} dataset comprises 7,352 auctions conducted between
October 1, 2009 and December 12, 2009.  In addition to the information
contained in the \textsc{Outcomes} dataset, our \textsc{Trace} dataset
also contains information about the actual bids that were placed
during the auction. We probed Swoopo at semi-regular time-intervals -
more frequently near the end of the auction when bids are placed in
rapid succession, less frequently near the beginning of the auction
when bidding is sparse. Specifically, to avoid inundating Swoopo with
useless requests we implemented the following back-off procedure.
Our initial probing interval is set to one second.
When the countdown clock had more than 10 minutes left on it, and there
were no bids reported in our last request, we would increase our
probing interval by half a second, up to a maximum of a minute. A new
request including at least one bid would reset the probing interval to
one second.  When the countdown clock had at least 2 and less than 10
minutes left on it, our maximum probing interval was 10
seconds. Finally, when the countdown clock had fewer than 2 minutes
left on it we probed at the rate Swoopo suggested but at least once every second (every probe response
was associated with a Swoopo defined update-interval that Swoopo uses to instruct the user's browser
when to next update).
Swoopo would respond with a line of the following format:

\begin{verbatim}
ct=15|cs=1|ra=0|cw=Schonmir1500|cp=3126|bh=521:Schonmir1500:1:3126:0:#|lui=4#1#0#0
\end{verbatim}
Each line was timestamped on our end to provide the actual time of the bid(s). In plain English the above line tells us that the 521st bid of the auction was placed by user \texttt{Schonmir1500}. It raised the price of the item to \$31.26 and it was not placed by a BidButler. Finally, when the bid was placed another 4 seconds were added to countdown clock to reset it to its current value of 15. Specifically, the fields, which are separated using vertical bars, carry the following information:

\vskip 0.1cm
\begin{tabular}{lp{12cm}}
\hline
Field name & Description \\
\hline\hline
\texttt{ct} & Current time on the countdown clock. \\ 
\texttt{cs} & Current state, 1 means the auction is still active, 20 that is has ended.  Swoopo defines more status codes which we did not encounter.
 \\
\texttt{ra} & Technical use. When set to 1 forces the browser to reload the page, irrelevant to the auction. \\
\texttt{cw} & Username of current winner. \\
\texttt{cp} & Current price in cents. \\
\texttt{bh} & Bid history, \texttt{\#}-separated tuples of the form: \texttt{bidnumber:username:bidtype: price:yourbid}. A \texttt{bidtype} of 1 indicates a player bid whereas a \texttt{bidtype} of 2 indicates a BidButler bid. The field \texttt{yourbid} is set to 1 when the observer has placed the corresponding bid.\\
\texttt{lui} & Last update index, four \texttt{\#}-separated numbers: seconds added to clock from player bids, number of player bids, seconds added to clock by BidButler bids, number of BidButler bids.\\
\hline
\end{tabular}
\vskip 0.1cm

\fi

\end{document}